\documentclass{emulateapj}

\shorttitle{Bias of Line-Ratio AGN Selection}
\shortauthors{Trump et al.}

\begin{document}

\title{The Biases of Optical Line-Ratio Selection for Active Galactic
  Nuclei, and the Intrinsic Relationship between Black Hole Accretion
  and Galaxy Star Formation}

\author{
  Jonathan R. Trump,\altaffilmark{1,$\dagger$}
  Mouyuan Sun,\altaffilmark{1,2}
  Gregory R. Zeimann,\altaffilmark{1}
  Cuyler Luck,\altaffilmark{3}
  Joanna S. Bridge,\altaffilmark{1}
  Catherine J. Grier,\altaffilmark{1}
  Alex Hagen,\altaffilmark{1}
  Stephanie Juneau,\altaffilmark{4}
  Antonio Montero-Dorta,\altaffilmark{5}
  David J. Rosario,\altaffilmark{6}
  W. Niel Brandt,\altaffilmark{1}
  Robin Ciardullo,\altaffilmark{1}
  and Donald P. Schneider\altaffilmark{1}
}

\altaffiltext{1}{
  Department of Astronomy and Astrophysics and Institute for
  Gravitation and the Cosmos, 525 Davey Lab, The Pennsylvania State
  University, University Park, PA 16802, USA
\label{PSU}}

\altaffiltext{2}{
  Department of Astronomy and Institute of Theoretical Physics and
  Astrophysics, Xiamen University, Xiamen, Fujian 361005, China
\label{Xiamen}}

\altaffiltext{3}{
  State College High School, 650 Westerly Parkway, State College, PA
  16801, USA
\label{State High}}

\altaffiltext{4}{
  Irfu/Service d'Astrophysique, CEA-Saclay, Orme des Merisiers, 91191
  Gif-sur-Yvette Cedex, France
\label{Saclay}}

\altaffiltext{5}{
  Department of Physics and Astronomy, The University of Utah, 115
  South 1400 East, Salt Lake City, UT 84112, USA
\label{Utah}}

\altaffiltext{6}{
  Max-Planck-Institut f\"{u}r extraterrestrische Physik (MPE),
  Giessenbachstrasse 1, D-85748 Garching bei M\"{u}nchen, Germany
\label{MPE}}



\altaffiltext{$\dagger$}{
  Hubble Fellow
\label{HF}}

\def\etal{et al.}
\newcommand{\Ha}{\hbox{{\rm H}$\alpha$}}
\newcommand{\Hb}{\hbox{{\rm H}$\beta$}}
\newcommand{\OII}{\hbox{{\rm [O}\kern 0.1em{\sc ii}{\rm ]}}}
\newcommand{\NeIII}{\hbox{{\rm [Ne}\kern 0.1em{\sc iii}{\rm ]}}}
\newcommand{\NeIV}{\hbox{{\rm [Ne}\kern 0.1em{\sc iv}{\rm ]}}}
\newcommand{\NeV}{\hbox{{\rm [Ne}\kern 0.1em{\sc v}{\rm ]}}}
\newcommand{\HeII}{\hbox{{\rm He}\kern 0.1em{\sc ii}}}
\newcommand{\OIII}{\hbox{{\rm [O}\kern 0.1em{\sc iii}{\rm ]}}}
\newcommand{\NII}{\hbox{{\rm [N}\kern 0.1em{\sc ii}{\rm ]}}}
\newcommand{\SII}{\hbox{{\rm [S}\kern 0.1em{\sc ii}{\rm ]}}}
\newcommand{\CIV}{\hbox{{\rm C}\kern 0.1em{\sc iv}}}
\newcommand{\CIII}{\hbox{{\rm C}\kern 0.1em{\sc iii}{\rm ]}}}
\newcommand{\HII}{\hbox{{\rm H}\kern 0.1em{\sc ii}}}
\newcommand{\Mbh}{M_{\rm BH}}
\newcommand{\Lbol}{L_{\rm bol}}
\newcommand{\eddratio}{L_{\rm bol}/L_{\rm Edd}}

\begin{abstract}

  We use 317,000 emission-line galaxies from the Sloan Digital Sky
  Survey to investigate line-ratio selection of active galactic nuclei
  (AGNs).  In particular, we demonstrate that ``star formation
  dilution'' by $\HII$ regions causes a significant bias against AGN
  selection in low-mass, blue, star-forming, disk-dominated galaxies.
  This bias is responsible for the observed preference of AGNs among
  high-mass, green, moderately star-forming, bulge-dominated hosts.
  We account for the bias and simulate the intrinsic population of
  emission-line AGNs using a physically-motivated Eddington ratio
  distribution, intrinsic AGN narrow line region line ratios, a
  luminosity-dependent $\Lbol/L\OIII$ bolometric correction, and the
  observed $\Mbh-\sigma$ relation.  These simulations indicate that,
  in massive ($\log(M_*/M_\odot) \gtrsim 10$) galaxies, AGN accretion
  is correlated with specific star formation rate but is otherwise
  uniform with stellar mass.  There is some hint of lower black hole
  occupation in low-mass ($\log(M_*/M_\odot) \lesssim 10$) hosts,
  although our modeling is limited by uncertainties in measuring and
  interpreting the velocity dispersions of low-mass galaxies.  The
  presence of star formation dilution means that AGNs contribute
  little to the observed strong optical emission lines (e.g., $\OIII$
  and $\Ha$) in low-mass and star-forming hosts.  However the AGN
  population recovered by our modeling indicates that feedback by
  typical (low- to moderate-accretion) low-redshift AGNs has nearly
  uniform efficiency at all stellar masses, star formation rates, and
  morphologies.  Taken together, our characterization of the
  observational bias and resultant AGN occupation function suggest
  that AGNs are unlikely to be the dominant source of star formation
  quenching in galaxies, but instead are fueled by the same gas which
  drives star formation activity.

\end{abstract}


\section{Introduction}

The observed correlations between the mass of a galaxy's bulge and the
mass of its supermassive black hole (SMBH) \citep[e.g.,][]{mag98,
  geb00, fer00, mar03, kor13} imply that galaxy growth via star
formation (SF) must have a corresponding period of SMBH growth in the
active galactic nucleus (AGN) phase.  Yet the details that couple
AGN-galaxy coevolution remain mysterious.  Theoretical frameworks
invoke mergers \citep[e.g.,][]{san88,dim05,hop06,hop08}, violent disk
instabilities \citep{dek09,bou11,gab13}, or secular processes in
gas-rich disks \citep[e.g.,][]{shlos89,hh09,hop14} to simultaneously
(or near-simultaneously, e.g., \citealp{wild10}) fuel rapid star
formation and drive gas inward to power an AGN.  Many simulations also
invoke AGN ``feedback'' which quenches star formation through gas
blowout by radiative winds \citep{sil98,fab02,dim05} or gas heating by
radio jets \citep{cro06}.  Both AGN feedback and coupled AGN-SF
fueling leave an imprint on the observed properties (e.g., color, star
formation rate, and morphology) of AGN host galaxies.  This scenario
means the physical processes behind AGN-galaxy coevolution can be
revealed by the simple question: \textit{Are AGN host galaxies
  special?}

In seminal work \citet{kau03} used $\sim$120,000 galaxies from the
Sloan Digital Sky Survey \citep[SDSS,][]{sdss} to demonstrate that
AGNs are most frequently observed in massive bulge-dominated galaxies,
with more luminous AGNs found in host galaxies with more recent star
formation.  \citet{kau03}, along with subsequent work reaching similar
conclusions \citep{hec04,schaw10}, selected AGNs using the
characteristic optical signature of the AGN narrow line region (NLR):
AGN emission produces higher ratios of partially ionized forbidden
lines compared to Balmer recombination lines \citep{bpt81,vo87}.
Studies of AGN host galaxies using large samples with X-ray
\citep{nan07, sal07, geo08, sil08, schaw09, hic09, koc09, hag10},
infrared \citep{hic09}, and radio \citep{best05} AGN selection found a
similar observed preference for massive and bulge-dominated hosts,
with infrared $\Rightarrow$ X-ray $\Rightarrow$ radio corresponding to
a sequence in declining star formation rate (SFR) in the host galaxy
\citep{hic09}.  Taken together these observations implied a critical
stellar mass for accretion onto an AGN, with indirect evidence for AGN
feedback in the green/red colors of X-ray/radio AGN host galaxies.

More recent studies, however, have suggested that the apparent
green-valley preference of X-ray AGNs is purely a relic of selection
effects \citep{sil09,xue10}.  There is also a mass bias for AGN
selection at a fixed accretion rate \citep{aird12}, most easily
understand with the unitless Eddington ratio, $\lambda_{\rm Edd}
\equiv \eddratio$.  The bolometric luminosity is driven by mass
accretion, $\Lbol=\eta\dot{M}c^2$, and the Eddington luminosity is
defined as $L_{\rm Edd}=(1.3 \times 10^{38}~{\rm
  erg~s}^{-1})\Mbh/M_\odot$, so that $\eddratio \simeq 0.44
\eta(\dot{M}/{\rm Gyr^{-1}})/\Mbh$.  The radiative efficiency $\eta
\sim 0.1$ at $\dot{m} \equiv \dot{M}/[L_{\rm Edd}/(\eta c^2)] \gtrsim
10^{-2}$, while lower accretion rates lead to radiatively inefficient
accretion and $\log(\eta) \sim -1+0.5\log(\dot{m}/10^{-2})$
\citep{xie12}.  The Eddington ratio better describes the mode of AGN
accretion and feedback than does luminosity, as the AGN accretion flow
changes from a geometrically thin accretion disk with radiative winds
at $\lambda_{\rm Edd} \gtrsim 0.01$ to a radiatively inefficient
accretion flow with stronger radio jets at $\lambda_{\rm Edd} \lesssim
0.01$ \citep{nar08,ho08,tru11a,hec14}.  Because $\Mbh$ is correlated
with galaxy bulge mass $M_{\rm bulge}$, which is in turn correlated
with total stellar mass $M_*$, a given AGN flux limit means it is
easier to find lower-$\lambda_{\rm Edd}$ AGNs in high-mass galaxies
than in low-mass galaxies.  The $\lambda_{\rm Edd}$ function for AGNs
is likely to steeply decline with increasing accretion rate, since
there are many more weakly-accreting than rapidly-accreting AGNs in
the local universe \citep[e.g.,][]{ho08}.  In a flux-limited sample
this results in a steep bias towards finding more AGNs in massive
hosts.

\citet{aird12} reported that the X-ray AGN host occupation fraction is
consistent with a \textit{uniform AGN Eddington ratio distribution
  across all host galaxy stellar masses}.  Thus, after controlling for
selection bias, AGN host galaxies are not special in terms of stellar
mass.  It remains unclear if AGN host galaxies are special in SFR or
morphology.  Recent work combining far-infrared observations with
X-ray selection demonstrates that AGNs are most common in star-forming
galaxies \citep{mul12, har12, jun13, ros13a, ros13b, chen13}.  This
result is generally similar to the connection between AGN luminosity
and host galaxy SFR seen in the initial line-ratio
AGN\footnote{Throughout the paper we refer to AGNs selected by
  Seyfert-like optical line ratios as ``line-ratio AGNs.''} sample of
\citet{kau03} and in subsequent studies of broad-line AGN hosts
\citep{tru13a,mat14}.  In terms of host morphology, most recent
observations demonstrate that AGNs are not preferentially fueled by
major mergers \citep{gro05, gab09, cis11, koc12}, although mergers may
play a role in fueling nearby AGNs that are very luminous and/or
obscured \citep{koss10, ell11, tre12, tru13b}.  Observations are mixed
on whether violent disk instabilities play a significant role in AGN
fueling, with evidence both for \citep{bou12} and against
\citep{tru14}.  So while X-ray AGN hosts do not appear to be special
in their stellar mass, it remains under debate if they are special in
SFR or morphology.  And it is still unclear how the larger population
of line-ratio AGNs connect to this X-ray picture.


In this work we use line-ratio AGN selection to recover the intrinsic
AGN occupation fraction with host galaxy properties.  For this purpose
we use a very large sample of $>$300,000 galaxies from the SDSS,
described in Section 2.  Just as \citet{kau03} originally found,
Section 3 demonstrates that line-ratio AGNs are observed to be most
common in massive, concentrated galaxies with intermediate colors and
specific star formation rates (sSFRs).  The selection bias for
line-ratio AGNs is a function of the contrast between AGN and galaxy
emission lines rather than a simple flux limit, and in Section 4 we
quantify this bias with host galaxy properties.  In Section 5 we model
the intrinsic AGN population as a function of galaxy properties,
beginning with a uniform Eddington ratio distribution, then allowing
Eddington ratio and black hole occupation to vary with galaxy
properties.  Section 6 describes the implications of the modeled AGN
occupation function for galaxy emission-line measurements, the
connection between AGN accretion and galaxy SFR, AGN feedback, and
black hole seed formation.  Throughout the paper we use a basic
$\Lambda$CDM cosmology with $h=0.70$, $\Omega_M=0.3$,
$\Omega_{\Lambda}=0.7$.

\section{Observational Data}

Our galaxy sample is drawn from the Sloan Digital Sky Survey
\citep[SDSS,][]{sdss}.  SDSS data include $ugriz$ broad-band
photometry \citep{fuku96} and spectroscopy with $R\sim2000$ over
$3800<\lambda<9200$\AA \citep{smee13}, taken using a 2.5-m telescope
at Apache Point Observatory \citep{gunn06}.  The galaxy sample in this
work is drawn from the SDSS Data Release 10 \citep{sdssdr10}, which
covers a total sky area of 14,555 deg$^2$.  We used the public
SkyServer SQL
server\footnote{http://skyserver.sdss3.org/dr10/en/tools/search/sql.aspx}
to select a parent sample of 317,192 galaxies ({\tt class=GALAXY})
with $r<17.77$, spectroscopic redshifts in the range $0.01<z<0.1$, and
line flux errors of $0<\sigma_{\rm
  line}<10^{-15}$~erg~s$^{-1}$~cm$^{-2}$ (to remove bad spectra).  The
$r$-magnitude limit represents the main spectroscopy limit of the SDSS
galaxy survey \citep{strauss02}.  The limits in the line flux errors
effectively remove objects with problematic spectra (artifacts and bad
pixels) in the line regions.  These spectral issues are uncorrelated
with galaxy properties and are essentially random occurrences.

The redshift range of $0.01<z<0.1$ is designed to include a broad
range of galaxy stellar masses without significant 
evolution in galaxy properties.  The upper limit in redshift
corresponds roughly to a mass limit of $\log(M_*/M_{\odot}) \gtrsim
10.7$ with the SDSS magnitude limit, and the lower redshift limit adds
lower-mass galaxies to $\log(M_*/M_{\odot}) \gtrsim 8$.  In Section
4.1 we demonstrate that the fractional coverage of the SDSS fiber
varies by $\lesssim$20\% over the stellar mass range of the full
sample.  There is also minimal galaxy evolution within this redshift
range: for example, the cosmic SFR evolves by $\lesssim$25\%
\citep{wyd05,cuc12,mad14}.  Thus we do not expect a significant change
in AGN fraction with redshift driven by changing galaxy properties.

SDSS galaxies have redshifts and line measurements computed using the
{\tt idlspec2d} software: details of this automated processing are
given by \citet{bol12}.  In brief, galaxy classifications and
redshifts are computed by finding the best-fit (minimum-$\chi^2$)
template, with templates constructed from sets of eigenspectra derived
from a principal component analysis of training spectra.  Emission
line fluxes are computed using the best-fit velocity dispersion
template as the continuum, which implicitly corrects for stellar
absorption.  Each line is fit as a Gaussian, with line width fixed
across the Balmer lines, and a separate fixed line width for the group
of all other emission lines visible in our redshift range.

Stellar velocity dispersion is computed by fitting a set of stellar
templates from the library of \citet{pru01}, degrading the templates
to match the instrumental broadening ($\sim$70~km~s$^{-1}$) and
minimizing $\chi^2$ over a grid of $0<\sigma<850$~km~s$^{-1}$ with
25~km~s$^{-1}$ bins.  Since these velocity dispersions are fit using
the stellar absorption lines, they are unaffected by AGN emission.
Low-mass galaxies ($\log(M_*)<10^{10}M_\odot$) tend to have measured
dispersions less than the instrumental resolution: we include
dispersions for these galaxies using survival analysis
\citep[e.g.,][]{fei85} on their 1-$\sigma$ upper limits.  The lack of
reliable velocity dispersions in low-mass galaxies is an issue we
return to in Sections 5 and 6.


We create samples of galaxies with well-measured line ratios,
following \citet{jun14} and requiring the signal-to-noise ratio
($S/N$) of the line ratios to have $(S/N)_{\rm ratio}>3/\sqrt{2}$.
This is the equivalent of each line detected at $(S/N)_{\rm line}>3$,
but has the advantage of including well-constrained line ratios where
one line is bright and the other faint (for example, $\Hb$ is
frequently bright and $\OIII$ weak in massive galaxies, and $\Ha$ is
bright and $\NII$ weak in low-metallicity galaxies).  From the parent
sample of 317,192 galaxies with $r<17.77$ and $0.01<z<0.1$, 223,448
galaxies possess each of the $\OIII\lambda5007/\Hb$ and
$\NII\lambda6584/\Ha$ ratios measured at $(S/N)_{\rm
  ratio}>3/\sqrt{2}$: these lines form the ``BPT'' \citep{bpt81}
diagram and so we designate this galaxy set as the ``BPT well-measured
sample.''  Similarly the $\OIII\lambda5007/\Hb$ and
$(\SII\lambda6717+6731)/\Ha$ ratios define the ``VO87'' \citep{vo87}
diagram, and the ``VO87 well-measured sample'' includes the 215,242
galaxies with these ratios measured at $(S/N)_{\rm ratio}>3/\sqrt{2}$.
For most galaxies the $(S/N)_{\rm ratio}$ threshold corresponds to a
line flux limit of $f \gtrsim 1 \times
10^{-16}$~erg~s$^{-1}$~cm$^{-2}$.  Our line-ratio classifications
generally use only these well-measured samples, although we also
discuss the likely AGN fraction among the full parent sample of
galaxies (including those with poorly-measured line ratios) in Section
3.3.

\begin{figure*}[ht]   
\epsscale{1.15}
{\plotone{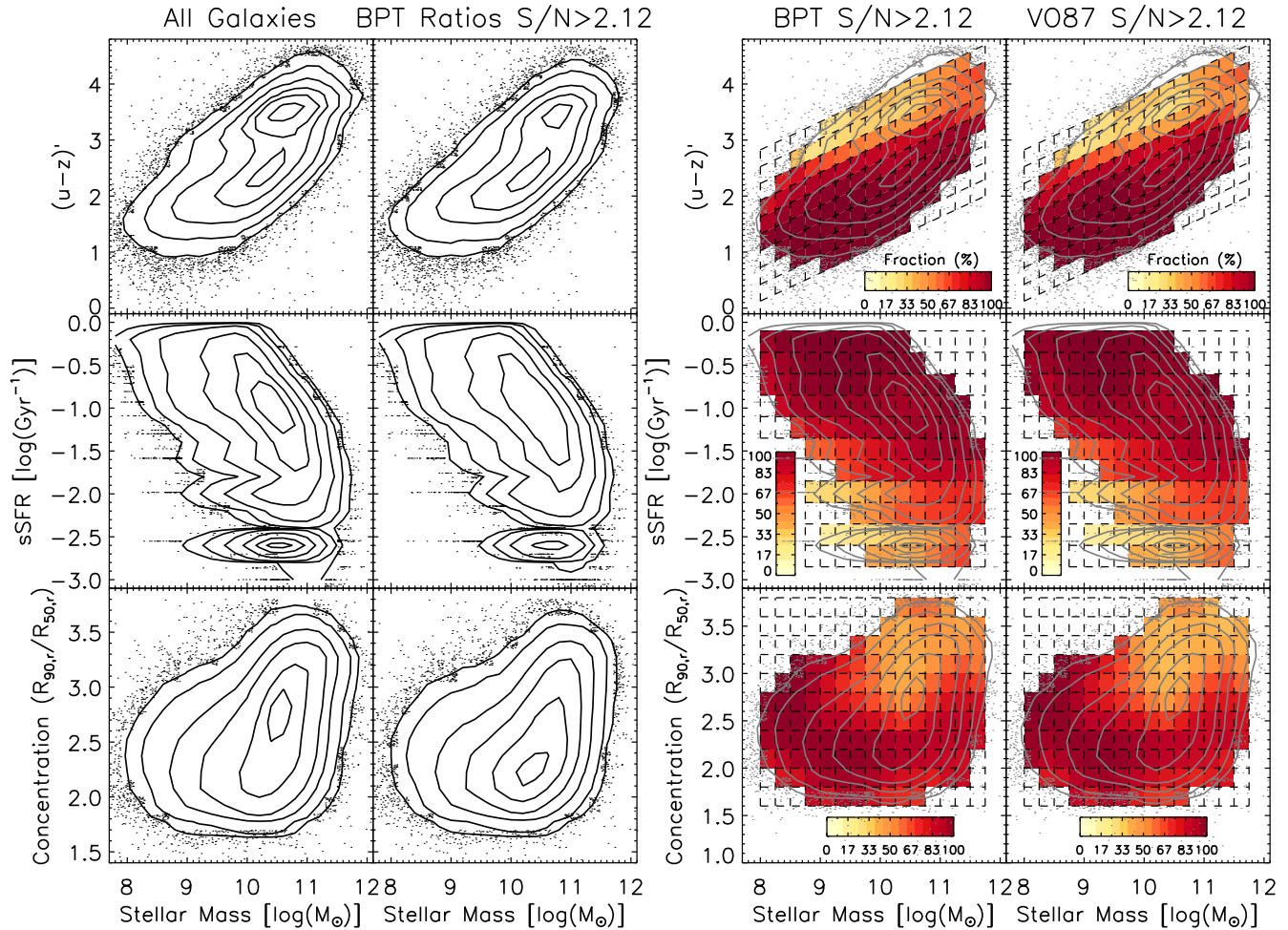}}
\figcaption{The distribution of galaxy properties for the full parent
  sample (left column, 317,192 galaxies) with the well-measured sample
  of galaxies having both BPT line ratios measured at $(S/N)_{\rm
    ratio}>3/\sqrt{2}$ (second column, 223,448 galaxies).  The two
  columns at right also display the fraction of the full sample with
  well-measured ($(S/N)_{\rm ratio}>3/\sqrt{2}$) BPT or VO87 line
  ratios.  The emission-line ratio $S/N$ requirements result in fewer
  red, low-sSFR, concentrated galaxies.
\label{fig:comparecat}}
\end{figure*}   

Unless otherwise specified, the emission-line luminosities in each
galaxy are dust-corrected using the measured Balmer decrement and a
\citet{ccm89} attenuation curve with \citet{odonnell94} coefficients.
The extinction corrections use an intrinsic (dust-free) Balmer
decrement of $\Ha/\Hb=2.86$ for SF-dominated emission-lines (i.e.,
classified as $\HII$-dominated on the BPT or VO87 diagrams) and
$\Ha/\Hb=3.1$ for AGN emission \citep[appropriate for narrow line
region gas conditions, e.g.,][]{ost06}.  Galaxies with observed Balmer
decrements less than the assumed intrinsic decrements were assigned
zero extinction.

We characterize basic galaxy properties using stellar mass, color,
specific star formation rate (sSFR), and concentration.  Stellar
masses and specific star formation rates are estimated by
\citet{mon14}, fitting the broad-band $ugriz$ SDSS photometry with a
grid of templates from the flexible stellar population synthesis code
\citep[FSPS,][]{con09}.  The templates assume a \citet{krou01} initial
mass function, allowing a wide range of galaxy formation times $2 <
t_{\rm age}/{\rm Gyr} < T_U(z)$, where $T_U$ is the age of the
universe at redshift $z$.  The stellar mass and sSFR determination
uses a broad-band fit for the dust attenuation (separate from the
Balmer decrements used for dust-correcting the emission lines),
following \citet{cha00} and \citet{calz00} by assuming $\tau_1$ for
birth clouds and $\tau_2$ for old stars, with $\tau_1=3\tau_2$, and
fitting a grid of values spanning $0<\tau_2<0.75$.  The use of
broad-band photometry to estimate stellar mass and sSFR minimizes
potential AGN contamination.  Our study uses only narrow-line AGNs
(not quasars), which generally have weak continuum emission.  The AGN
emission lines also do not significantly affect the broad-band
photometry, since almost all AGNs in the sample have emission line
equivalent widths $<$100\AA.

All galaxy colors are k-corrected to $z=0.05$ using the {\tt kcorrect}
IDL software \citep{kcorrect}.  We characterize galaxy morphology
using a concentration parameter $C_r$ defined as the ratio of the
radii containing 90\% and 50\% of the galaxy's $r$-band light,
$C_r=R_{90,r}/R_{50,r}$.  Details on the measurement of these
\citet{pet76} radii are given by \citet{bla01}.  Galaxies with high
concentration ($R_{90,r}/R_{50,r} \gtrsim 2.5$) tend to be
bulge-dominated, while galaxies with $R_{90,r}/R_{50,r} \lesssim 2.5$
are typically disks.

It is important to note that the sample of galaxies with well-detected
emission lines occupies a slightly different parameter space in galaxy
properties compared to the full SDSS parent sample.  In Figure
\ref{fig:comparecat} we compare the color--mass, sSFR--mass, and
concentration--mass diagrams for the full parent sample with the
well-measured BPT and VO87 samples.  The restrictions on BPT and VO87
line ratios result in nearly identical well-measured samples, with
fewer red, concentrated, and low-sSFR galaxies.  Since the galaxies
removed from the well-measured sample typically have $S/N_{\rm
  line}<3$, they tend to have weaker emission lines and so are
significantly less likely to have AGNs.  In Section 3.3 we devise a
strategy that accounts for the small AGN fraction likely present in
galaxies with poorly-measured line ratios.

\section{Observed Host Properties of Line-Ratio AGNs}

We begin by examining the observed AGN fractions among host galaxies
of different mass, color, SFR, and morphology.  Section 3.1 introduces
our adopted methods of line-ratio selection, including a method to
separate AGNs and low-ionization narrow emission regions (LINERs).  We
use the well-measured sample in Section 3.2 to determine the fraction
of BPT and VO87 AGNs with galaxy color, specific star formation rate,
morphology, and stellar mass.  In Section 3.3 we additionally provide
a correction for the (small) fraction of AGNs missed among galaxies
with poorly-measured line ratios, concluding with an estimate of the
overall AGN fraction among the entire parent sample.

\subsection{Line-Ratio AGN Selection}

The most widely used methods for line-ratio classification of AGNs use
the ``BPT'' \citep{bpt81} and ``VO87'' \citep{vo87} diagrams.  Each
compares the ratios of forbidden lines of partially ionized metals
with Balmer recombination lines of hydrogen: $\OIII\lambda5007/\Hb$ vs
$\NII\lambda6584/\Ha$ for the BPT, and $\OIII\lambda5007/\Hb$ vs
$\SII\lambda(6717+6731)/\Ha$ for the VO87 diagram.  These lines are
also typically among the brightest features in rest-frame optical
spectra of galaxies, and the close wavelength range of each pair means
that the ratios are largely insensitive to dust (assuming that
forbidden metal lines and hydrogen recombination lines are emitted in
the same galaxy regions).  Note that BPT and VO87 classification is
appropriate for use on narrow (i.e. $<$1000~km/s) emission lines only.
Our sample does not include broad-line AGNs, which also exhibit broad
($>$1000~km/s) $\Ha$ and $\Hb$ emission lines in their spectra.

\begin{figure}[t]
\epsscale{1.15}
{\plotone{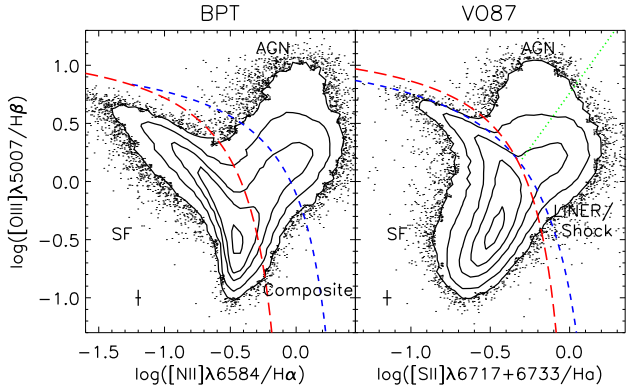}}
\figcaption{The BPT and VO87 diagrams for line-ratio classification of
  AGNs and star-forming galaxies.  Median line ratio errors are shown
  by the error bars in the lower left of each panel.  We adopt the
  long-dashed red lines for AGN/SF classification: in the BPT diagram
  this is the empirical line of \citet{kau03}, and in the VO87 diagram
  this is an empirical line of our own construction.  Also shown in
  each diagram are the \citet{kew01} maximal starburst lines
  (short-dashed blue), and the green dotted line in the VO87 diagram
  is the AGN/LINER division of \citet{kew06}.
\label{fig:bpt}}
\end{figure}

Figure \ref{fig:bpt} presents our well-measured galaxy samples in the
BPT and VO87 diagrams.  Each galaxy in this sample has both of the
relevant line ratios measured at $(S/N)_{\rm ratio}>3/\sqrt{2}$ (at
left, $\OIII/\Hb$ and $\NII/\Ha$, and at right, $\OIII/\Hb$ and
$\SII/\Ha$).  Also shown in Figure \ref{fig:bpt} are several
demarcation lines for classifying galaxies:

\begin{enumerate}
  \item BPT \citet{kau03} empirical AGN/SF division line (long-dashed
    red),
    \begin{equation}
      \log(\OIII/\Hb)=1.3+0.61/[\log(\NII/\Ha)-0.05]
    \end{equation}
  \item VO87 empirical AGN/SF division line defined for this work
    (long-dashed red),
    \begin{equation}
      \log(\OIII/\Hb)=1.3+0.48/[\log(\SII/\Ha)-0.10]
    \end{equation}
  \item BPT \citet{kew01} maximal starburst line (dashed blue),
    \begin{equation}
      \log(\OIII/\Hb)=1.19+0.61/[\log(\NII/\Ha)-0.47]
    \end{equation}
  \item VO87 \citet{kew01} maximal starburst line (dashed blue),
    \begin{equation}
      \log(\OIII/\Hb)=1.3+0.72/[\log(\SII/\Ha)-0.32]
    \end{equation}
  \item VO87 \citet{kew06} division between AGNs and LINERs (dotted
    green),
    \begin{equation}
      \log(\OIII/\Hb)=0.76+1.89\log(\SII/\Ha)
    \end{equation}
\end{enumerate}

We adopt Equations 1 and 2 as our lines for AGN/SF classification in
this work, along with Equation 5 for AGN/LINER separation.

Accurate AGN selection lines are subject to considerable uncertainty,
largely due to differences in interpretation.  For example, the
\citet{kew01} lines (Equations 3 and 4) are the maximal line ratios
for starbursts in photoionization models, and so can be treated as
minimum AGN lines.  Galaxies between the \citet{kau03} and
\citet{kew01} lines (Equations 1 and 3) in the BPT diagram are often
classified as ``composite galaxies'' with emission-line contribution
from both SF and AGN.  In general, composites have higher SFRs than
AGN-dominated galaxies \citep{sal07}, and the two categories exhibit
little difference in X-ray properties \citep{jun11,tro11,tru11b}.
Therefore we assume that composite galaxies host similar AGNs but with
more dilution from $\HII$ regions compare to galaxies with
AGN-dominated line ratios.  For this reason we include both composite
and AGN-dominated galaxies lying above the \citet{kau03} empirical
AGN/SF line (Equation 1) in our initial BPT AGN selection.

We create a new empirical line in the VO87 diagram (Equation 2) which
is designed to be parallel to the star-forming galaxy sequence in a
similar fashion to the \citet{kau03} empirical BPT line.  In
particular, our new VO87 line contrasts with the \citet{kew01} maximal
starburst line in the treatment of low-metallicity galaxies (the upper
left in each panel of Figure \ref{fig:bpt}).  Equation 4 selects
several AGNs from the symmetric tail of the low-metallicity
star-forming galaxy sequence, while our Equation 2 avoids classifying
these galaxies as AGNs.

In the VO87 diagram, the \citet{kew06} line (Equation 5) additionally
separates AGNs from low-ionization narrow emission-line regions
\citep[LINERs,][]{hec80}.  X-ray and radio observations suggest that
LINER galaxies are likely to host nuclear AGNs \citep[][and references
therein]{ho08}.  However LINERs have much lower accretion rates than
stronger-lined AGNs \citep{ho09} and much of their line emission is
probably powered by extended ionization sources rather than nuclear
activity \citep{yan12}.  For these reasons we avoid LINERs in our
sample and define ``VO87 AGNs'' as lying above the new empirical
AGN/SF line (Equation 2) and the \citet{kew06} AGN/LINER line
(Equation 5) in the VO87 diagram.

\begin{figure}[t]
\epsscale{1.15}
{\plotone{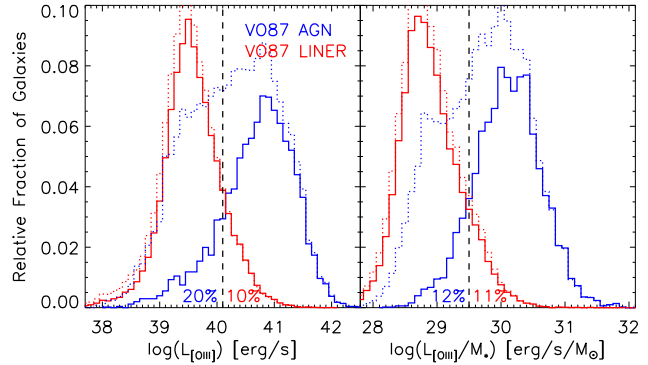}}
\figcaption{Distributions of dust-corrected $L\OIII$ (left) and
  $L\OIII/M_*$ (right) for AGNs and LINERs classified on the VO87
  diagram.  The dotted lines show all objects, while the solid lines
  show only ``well-separated'' AGNs and LINERs with line ratios at
  least $1\sigma$ from the AGN/LINER classification line (Equation 5).
  Vertical lines show empirical divisions between the two categories,
  with percentages of outliers indicated for the well-separated AGN
  (blue) and LINER (red) samples.  The $L\OIII/M_*$ ratio, a proxy for
  Eddington ratio, better distinguishes the two categories than
  $\OIII$ luminosity alone, and we adopt this ratio to separate AGNs
  and LINERs identified by the BPT diagram.
\label{fig:vo87lum}}
\end{figure}

Unlike the ``VO87 AGN'' category, the BPT diagram does not distinguish
between AGNs and LINERs.  Indeed, the BPT diagram does not lend itself
to a similar LINER/AGN separation as the \citet{kew06} line in the
VO87 diagram: while the $\SII$ critical density relates to the
LINER/AGN transition, the $\NII$ critical density is a factor of
$\sim$30 higher and does not.  (\citet{vo87} originally suggested a
simple AGN/LINER separation of $\OIII/\Hb>3$, but AGNs and LINERs form
a continuous rather than bimodal distribution in $\OIII/\Hb$,
e.g. \citealp{ho03}.)

AGNs typically have stronger emission lines than LINERs, and so we
investigate the use of $\OIII$ luminosity and the ratio $L\OIII/M_*$
to separate the two classes in Figure \ref{fig:vo87lum}.  The
$L\OIII/M_*$ ratio is related to the Eddington ratio $\eddratio$: for
a bolometric correction $k_{\rm [OIII]}=\Lbol/L\OIII$ (the inverse of
the fraction of total AGN light reprocessed as $\OIII$ emission) and a
mass ratio $\gamma=\Mbh/M_*$, $L\OIII/M_*=\frac{\gamma}{k_{\rm
    [OIII]}}\frac{\Lbol}{1.3 \times 10^{38}\Mbh}$.  (The variables
$k_{\rm [OIII]}$ and $\gamma$ are not actually constants: more
accurate relationships between $L\OIII/M_*$ and Eddington ratio are
discussed in Section 5.  However $\log(L\OIII/M_*)$ does monotonically
increase with $\log(\eddratio)$ with a nearly linear slope of
$\sim$0.5, making it a useful simple proxy.)

Figure \ref{fig:vo87lum} compares $L\OIII$ and $L\OIII/M_*$ of
``well-separated'' AGNs and LINERs with measured line ratios at least
$1\sigma$ from the AGN/LINER classification line.  The left panel
shows a significant tail of low-luminosity VO87 AGNs overlapping with
the LINERs.  This overlap is reduced when the two populations are
instead compared in $L\OIII/M_*$.  In the right panel of Figure
\ref{fig:vo87lum}, AGNs and LINERs with line ratios well-separated
from the classification line are divided by:
\begin{equation}
  \log(L\OIII/M_*) \ge 29.5~[\log({\rm erg~s}^{-1}~M_\odot^{-1})]
\end{equation}
Due to both the physical motivation and the good empirical separation,
we use Equation 6 to define a ``BPT AGN'' category which is mostly
free of LINERs.

In summary, we identify two categories of AGNs using emission-line
properties:

\begin{itemize}

  \item BPT AGN: Above the \citet{kau03} line
    $\log(\OIII/\Hb)>1.3+0.61/[\log(\NII/\Ha)-0.05]$, and with
    $\log(L\OIII/M_*) \ge 29.5$ (units of erg~s$^{-1}$ and
    $M_\odot$): Equations 1 and 6.
  \item VO87 AGN: Above our empirical AGN/SF line and the
    \citet{kew06} AGN/LINER division,
    $\log(\OIII/\Hb)>1.3+0.48/[\log(\SII/\Ha)-0.10]$ and
    $\log(\OIII/\Hb)>0.76+1.89\log(\SII/\Ha)$: Equations 2 and 5.
\end{itemize}

Among each well-measured sample, 6.3\% (14,175/223,448) of galaxies
are classified as BPT AGNs and 4.3\% (9210/215,242) are classified as
VO87 AGNs.  In the next section we demonstrate that both methods
identify AGNs in similar kinds of galaxies, and the lower AGN fraction
identified by the VO87 is due to slightly lower overall selection
efficiency.

\subsection{Observed AGN Fractions in Well-Measured Galaxies}


\begin{figure*}[ht]   
\epsscale{1.15}
{\plotone{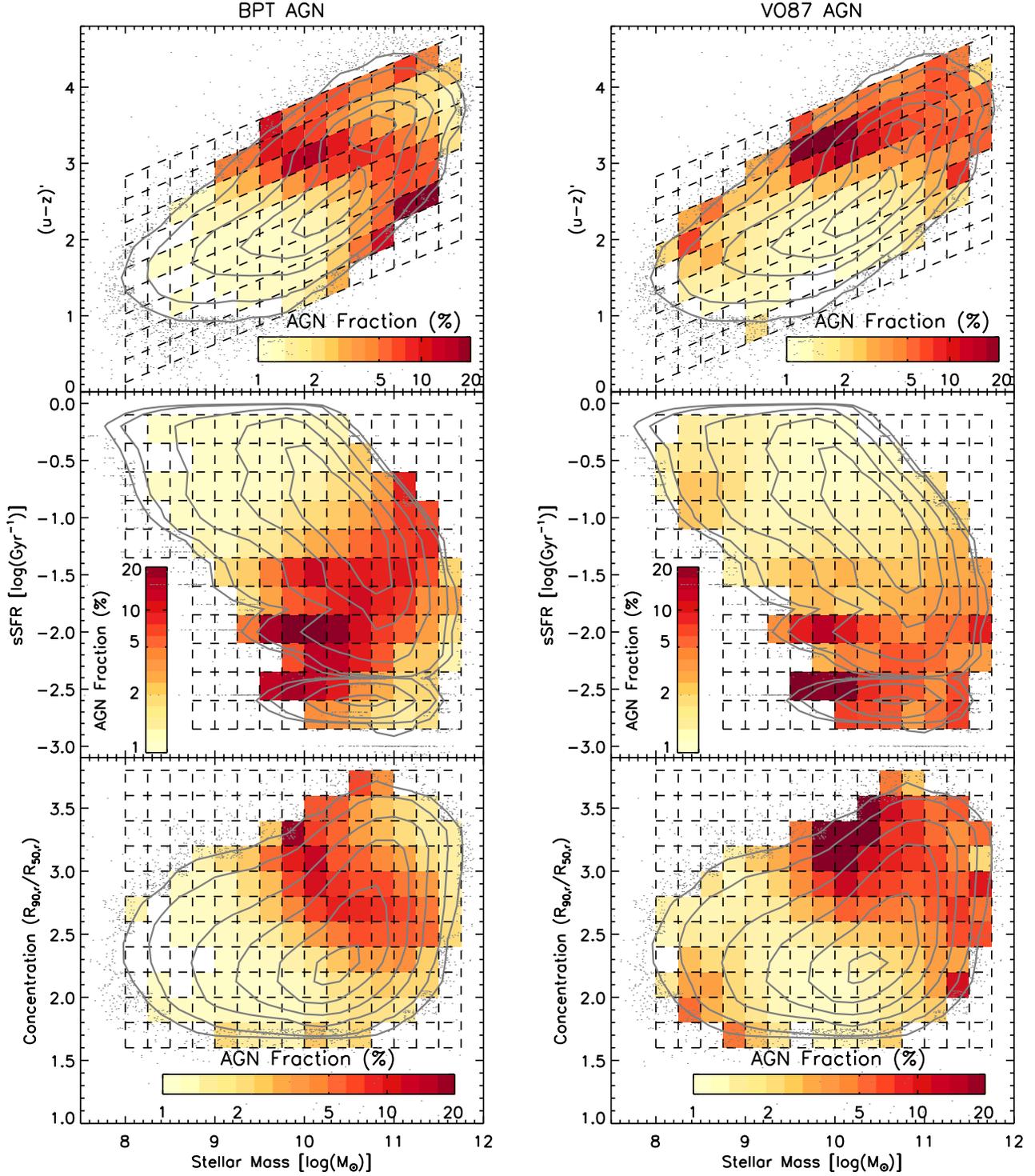}}
\figcaption{The observed fractions of BPT AGNs and VO87 AGNs among
  galaxies in the well-measured samples (with $(S/N)_{\rm
    ratio}>3/\sqrt{2}$), in bins of galaxy color, specific star
  formation rate, and concentration versus stellar mass.  Color is
  k-corrected to be observed at $z=0.05$ and color--mass bins are
  tilted (with a slope of $0.5$) to match the slope of the blue cloud.
  Bins with less than 50 galaxies are left blank (white).  Gray
  contours represent the overall population of well-measured galaxies.
  The observed (non-LINER) AGN fraction is highest in massive green,
  moderate-sSFR, and concentrated galaxies and lower on both the
  low-mass and red (low-sSFR) extremes of the galaxy population.
\label{fig:obsfrac}}
\end{figure*}   

Figure \ref{fig:obsfrac} shows the observed BPT and VO87 AGN fractions
in bins of galaxy color, specific star formation rate, and
concentration versus stellar mass among galaxies of the well-measured
sample.  The AGN fraction is defined as the fraction of galaxies (with
line ratio $(S/N)_{\rm ratio}>3/\sqrt{2}$) in each bin that meet the
criteria of the BPT AGN or VO87 AGN categories defined in Section 3.1,
and bins with less than 50 galaxies are left blank.  In both the BPT
and VO87 diagrams, the AGN fraction is observed to be highest in
massive ($M_*\sim10^{10.5}M_\odot$), green, moderate-sSFR,
bulge-dominated galaxies with $R_{90,r}/R_{50,r}>2.5$.  There is also
a small population of VO87 AGNs in low-mass hosts which are not
identified by the BPT AGN classification.  We demonstrate in Section 5
that this result may be due to a bias of the BPT against AGNs with
low-metallicity NLRs which remain detected in the VO87 diagram.

The maximal AGN fraction in massive green, moderate-sSFR,
bulge-dominated galaxies led to previous conclusions that AGN require
massive host galaxies and cause feedback that quenches star formation
\citep[e.g.,][]{kau03, schaw10}.  \citet{xue10} demonstrated that the
similar massive green-valley host preference observed for X-ray AGNs
is caused solely by selection effects, and \citet{aird12} additionally
showed that the intrinsic AGN distribution is actually uniform at all
stellar masses.  In Sections 4 and 5 we similarly investigate the
selection biases affecting line-ration selection and demonstrate that
the intrinsic AGN occupation fraction differs from the the simplest
interpretation of the observations, especially for low-mass galaxies.

As discussed in Section 2 and shown in Figure \ref{fig:comparecat},
the well-measured galaxy samples (with line ratio $(S/N)_{\rm
  ratio}>3/\sqrt{2}$) include fewer massive, red, concentrated, and
low-sSFR galaxies than the parent population.  In the next subsection
we account for AGNs in poorly-measured galaxies, showing that the
green-valley (moderate-sSFR, concentrated) AGN preference is even more
apparent among the full parent sample of galaxies.

\subsection{The Line-Ratio AGN Fraction Among All Galaxies}

It is only possible to select line-ratio AGNs in galaxies with
well-measured emission-line ratios.  However Figure
\ref{fig:comparecat} demonstrates that the well-measured galaxy sample
results in a biased set of galaxy properties.  Measuring an unbiased
AGN fraction requires understanding the likely AGN occupation among
galaxies with poorly-measured ($S/N<3/\sqrt{2}$) emission-line ratios.

\begin{figure}[t]
\epsscale{1.15}
{\plotone{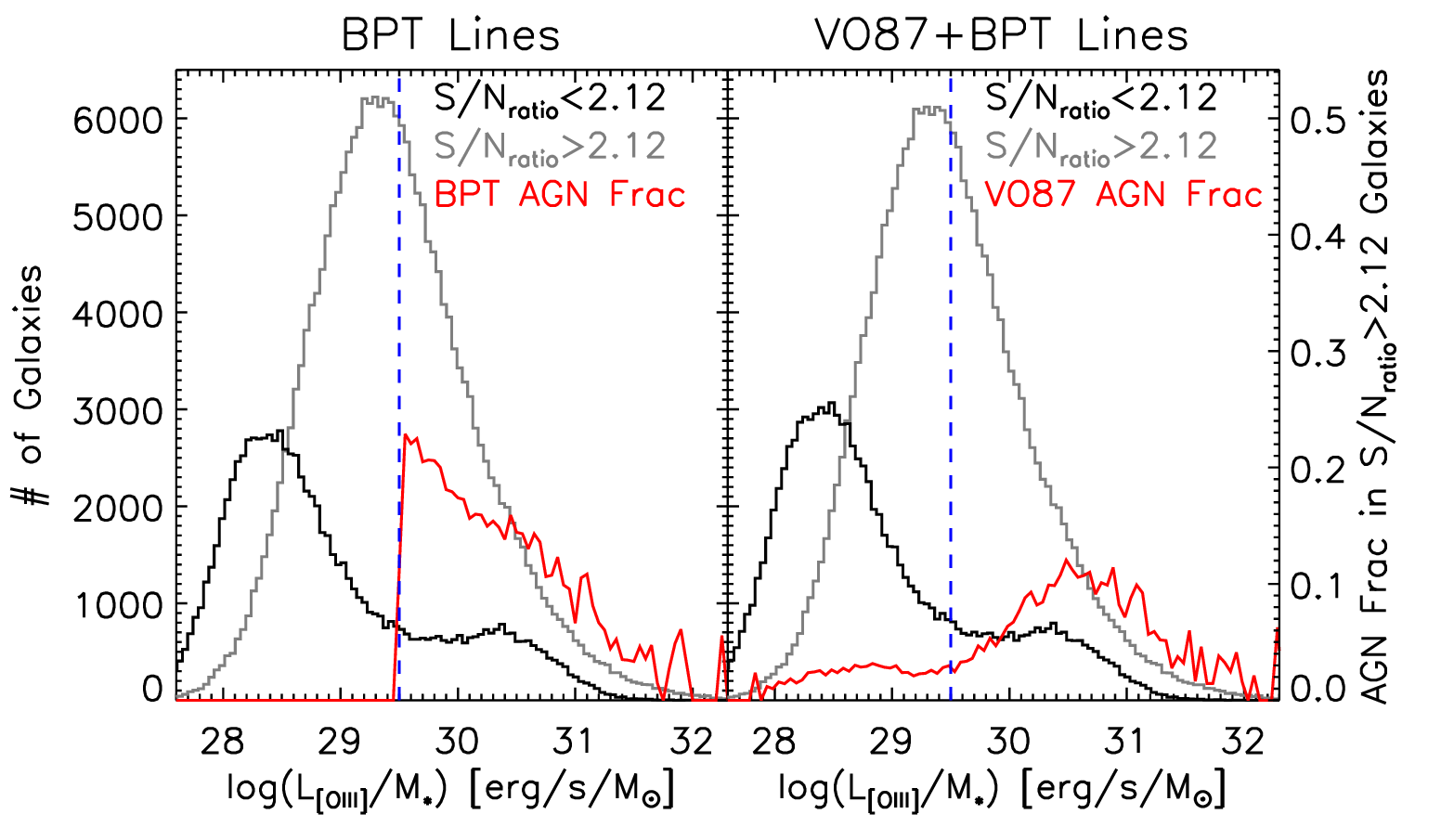}}
\figcaption{The distributions of dust-corrected $L\OIII/M_*$ for
  weak-lined ($S/N_{\rm ratio}<3/\sqrt{2}$) galaxies compared to the
  AGN fraction as a function of $L\OIII/M_*$ measured from
  strong-lined ($S/N_{\rm ratio}>3/\sqrt{2}$) galaxies.  The dashed
  line shows the $L\OIII/M_*>29.5$ empirical limit for AGN
  selection.  Most poorly-measured galaxies have $L\OIII/M_*<29.5$
  and so are typically less likely to host AGNs than well-measured
  galaxies.
\label{fig:weaklines}}
\end{figure}

In Section 3.1 we demonstrated that the VO87 AGN classification
(Equations 2 and 5) tends to select AGNs with $L\OIII/M_*>29.5$
(units of erg~s$^{-1}$ and $M_\odot$), and we impose the same limit to
select non-LINER BPT AGNs (Equation 6).  Figure \ref{fig:weaklines}
presents the $L\OIII/M_*$ distributions for galaxies with both
poorly-measured (black) and well-detected emission-line ratios (gray),
alongside the AGN fraction (red) measured from the well-detected
galaxy sample.  Some of the $\OIII$ measurements for poorly-measured
galaxies are likely to have large error bars, especially at the lowest
$L\OIII/M_*$ ratios.  Still, it is clear that most galaxies with
poorly-measured line ratios typically have $\log(L\OIII/M_*)<<29.5$,
suggesting that they are unlikely to host AGNs.  There is only a small
tail of poorly-measured galaxies with $\log(L\OIII/M_*)>29.5$ which
might have a significant number of ``missed'' AGNs.

We account for the potentially missed AGNs by assuming that
poorly-measured galaxies have the same AGN fraction as a function of
$L\OIII/M_*$ as observed in the well-measured galaxies, in each bin of
galaxy properties.  Since poorly-measured galaxies have lower typical
$L\OIII/M_*$ than well-measured galaxies, bins with significant
numbers of missed galaxies (i.e., high-mass, red, high-sSFR, and
concentrated galaxies) are likely to have lower AGN fractions than
those presented in Section 3.2 and Figure \ref{fig:obsfrac} for the
well-measured galaxy sample.

\begin{figure*}[ht]   
\epsscale{1.15}
{\plotone{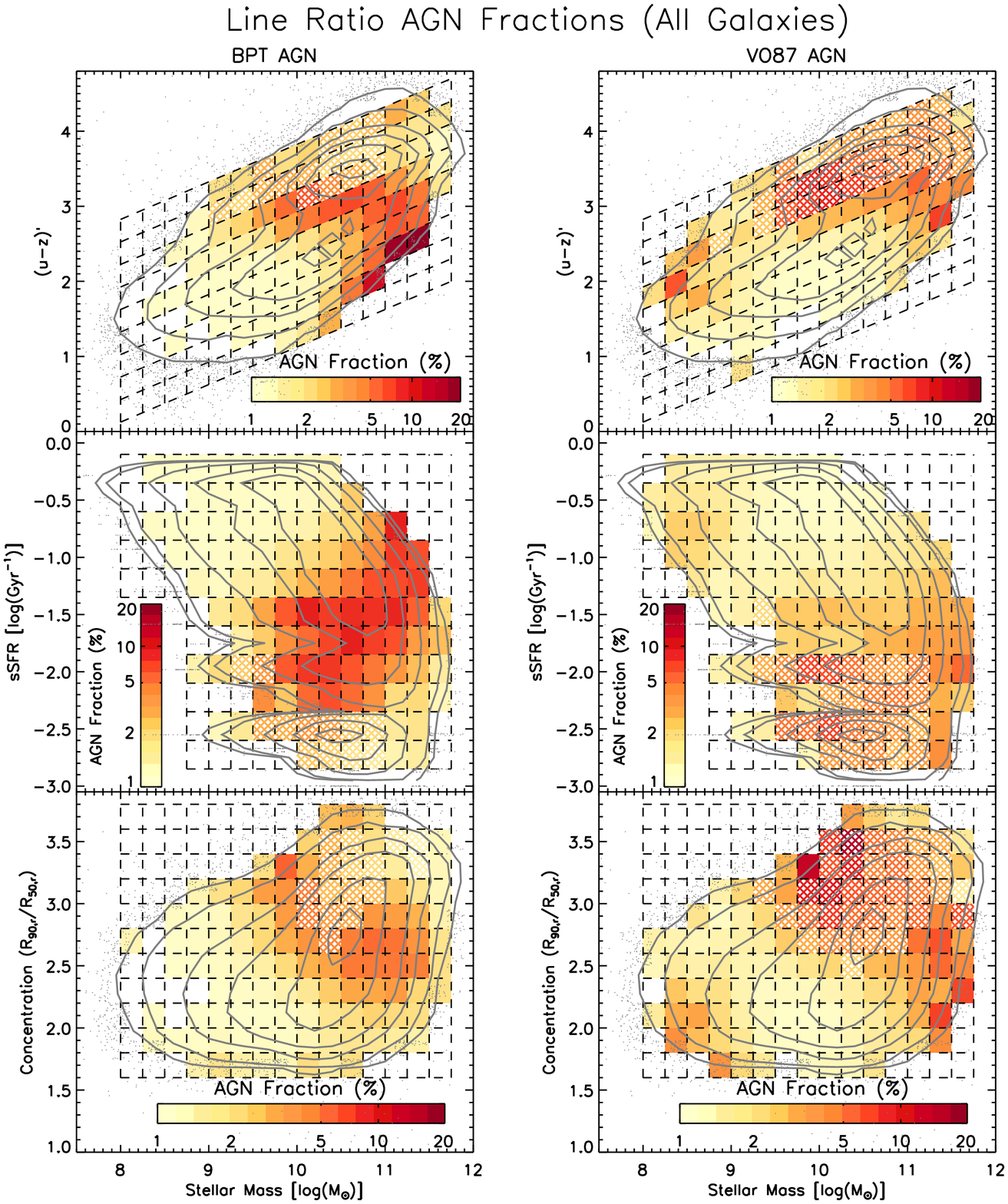}}
\figcaption{Total fractions of observed BPT and VO87 AGNs estimated
  for the entire parent sample of 317,192 galaxies, in bins of galaxy
  color (k-corrected $u-z$), specific star formation rate, and
  concentration ($R_{90,r}/R_{50,r}$) versus stellar mass.  The AGN
  fractions are determined from the well-measured sample with a
  correction for the AGNs missed among galaxies with poorly-detected
  lines, assuming the same AGN fraction with $L\OIII/M_*$ as outlined
  in Section 3.3.  Gray contours show the overall galaxy population
  from the full parent sample.  Bins where the correction is $>$25\%
  of the total AGN fraction are marked with crossed hashes: this only
  occurs among massive, red, low-sSFR, concentrated galaxies.  The
  corrected AGN fractions are very similar to the well-measured AGN
  fractions in Figure \ref{fig:obsfrac}.  The observed BPT and VO87
  AGN fractions are highest among massive green-valley hosts, with
  lower AGN fractions for both massive red (low-sSFR, concentrated)
  and low-mass blue (high-sSFR, low-concentration) galaxies.
\label{fig:obsfrac_weakfix}}
\end{figure*}   

Figure \ref{fig:obsfrac_weakfix} presents the total BPT and VO87 AGN
fractions for the entire parent sample of 317,192 galaxies.  In each
bin, we begin with the measured AGN fraction among well-measured
galaxies, then add a correction by assuming the same AGN fraction as a
function of $L\OIII/M_*$ in the poorly-measured galaxies.  These
corrections are only significant (with $>$25\% as many ``missed'' AGNs
as well-measured AGNs) in high-mass red, low-sSFR, and concentrated
galaxies, and the correction in these galaxies results in a smaller
AGN fraction than found in the well-measured sample.  The lack of
non-LINER AGNs among massive quenched galaxies agrees with previous
AGN host studies for line-ratio AGNs \citep{kau03, hec04, kau09,
  schaw10, tan12}, broad-line AGNs \citep{tru13a, mat14}, and X-ray
AGNs \citep{mul12, har12, ros13a, ros13b, chen13, aza15}.  Otherwise
the estimated total BPT and VO87 AGN fractions are very similar to the
well-measured AGN fractions (Figure \ref{fig:obsfrac}), and the
correction for poorly-detected galaxies leads to a qualitatively
identical preference for AGNs to be observed in massive green-valley
(moderate-sSFR, bulge-dominated) galaxies.

\section{Biases of Line-Ratio Classification}

The AGN line-ratio signature can be hidden by $\HII$ region line
ratios in galaxies with a significant level of star formation.  Below
we investigate how this ``star formation dilution'' changes with
galaxy properties and influences the observed AGN fractions.
We begin by computing basic AGN detection limits, quantified by
$L\OIII_{\rm AGN}/M_*$, for galaxies to host BPT and VO87 AGNs.
These detection limits implicitly account for the changing
spectroscopic aperture by using observed $\HII$ region emission within
the same aperture, although we also directly investigate aperture
effects by comparing the observed AGN fractions in different redshift
bins.  Finally, we compute the intrinsic AGN fractions across host
galaxy properties using Monte Carlo simulations of
physically-motivated distributions of Eddington ratio and black hole /
galaxy mass ratios.

\subsection{Basic Limits for AGN Detection}


\begin{figure*}[ht]   
\epsscale{1.15}
{\plotone{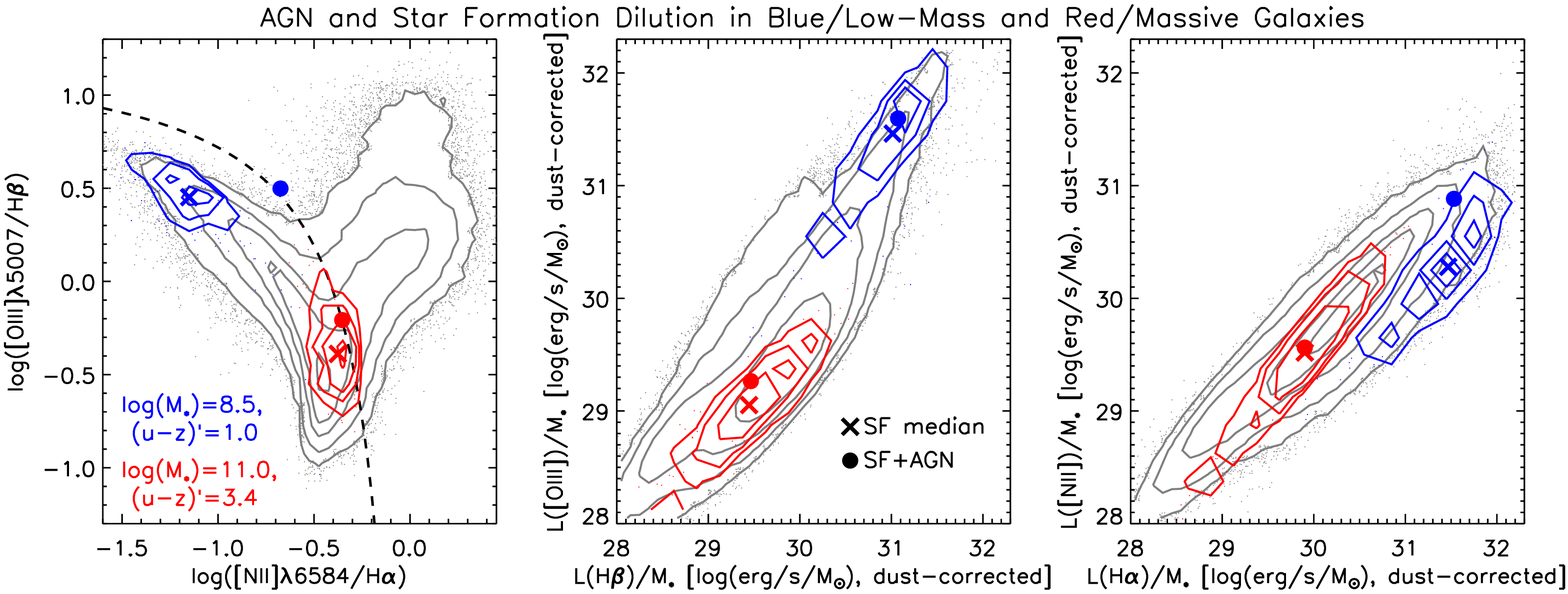}}
\figcaption{An illustration of star formation dilution and AGN
  detectability for galaxies with different properties.  The contours
  show the line ratios and strengths of BPT-classified star-forming
  galaxies: in blue, low-mass blue galaxies with
  $8.5<\log(M_*/M\odot)<8.75$ and
  $-3.275<(u-z)'-0.5\log(M_*/M\odot)<=-2.975$; in red, high-mass
  red galaxies with $11<\log(M_*/M\odot)<11.25$ and
  $-2.075<(u-z)'-0.5\log(M_*/M\odot)<=-1.775$; and in gray, all
  galaxies with $(S/N)_{\rm ratio}>3/\sqrt{2}$.  Crosses represent the
  median values of the star-forming galaxies, and filled circles
  indicate the combined ratios after adding enough ``pure AGN NLR''
  emission to reach the \citet{kau03} AGN/SF division line.  There is
  more star formation dilution in low-mass blue (high-sSFR,
  low-concentration) galaxies, requiring a relatively more powerful
  AGN to be selected as a BPT AGN.
\label{fig:colorbpt}}
\end{figure*}   

The biases of BPT and VO87 AGN selection can be characterized by
estimating the threshold at which AGN emission exceeds star formation
dilution within a 3$\arcsec$ fiber for galaxies of different
properties.  We quantify this threshold using the intrinsic
(dust-free) $\OIII$ luminosity of the AGN divided by the galaxy
stellar mass: $L\OIII_{\rm AGN}/M_*$.  This quantity is related to the
Eddington ratio $\eddratio$ by a $\OIII$ bolometric correction and the
mass ratio $\Mbh/M_*$.  For now, we simply assume that $L\OIII_{\rm
  AGN}/M_*$ is monotonically related to $\eddratio$, reserving a
detailed discussion of the functional form until Section 5.

We estimate the changes in star formation dilution with galaxy
properties using the typical $\HII$ region line strengths in the bins
of the various panels in Figure \ref{fig:obsfrac}.  That is, for each
bin in color--mass, sSFR--mass, and concentration--mass, we calculate
the median dust-corrected (assuming an intrinsic $\Ha/\Hb=2.86$)
luminosity of each emission line for galaxies with line ratios
classified as star-forming in the BPT or VO87 diagrams.  The typical
$\HII$ region emission depends both on spectroscopic aperture and
galaxy properties, and we implicitly account for both by using the
observed line fluxes of SF galaxies in each bin.  The AGN detection
limit is given by the AGN contribution needed to push these line
ratios onto the AGN/SF diagnostic lines: \citet{kau03} on the BPT
(Equation 1), or our empirical line on the VO87 (Equation 2).  For the
moment we neglect the AGN/LINER divisions (Equations 5 and 6),
although we include these requirements in our detailed models of AGN
selection biases in Section 5.  We assume a ``pure AGN NLR'' has
$\log(\OIII/\Hb)=0.5$, $\log(\NII/\Ha)=0.0$, and $\Ha/\Hb=3.1$.
Because we are starting from dust-corrected $\HII$ region emission
lines, the derived AGN detection limits are similarly dust-free,
assuming that both NLR and $\HII$ region gas have the same dust
extinction.  We demonstrate that this assumption is valid in the
Appendix.  Using dust-free $L\OIII_{\rm AGN}/M_*$ makes it easier to
relate to an Eddington ratio, although it does mean that our detection
limits are not strictly observed quantities.

Our method of computing AGN detection limits is illustrated by Figure
\ref{fig:colorbpt}, which shows the line ratios and strengths (in
$L_{\rm line}/M_*$) for two bins in galaxy color--mass.  Low-mass
blue galaxies have higher $L_{\rm line}/M_*$ from integrated $\HII$
regions than high-mass red galaxies.  Thus low-mass blue galaxies
(which also tend to have high-sSFR and low-concentration) require an
AGN with higher $L\OIII_{\rm AGN}/M_*$ to be classified as a BPT or
VO87 AGN compared to massive red (low-sSFR, high-concentration)
galaxies.

\begin{figure*}[t]  
\epsscale{1.15}
{\plotone{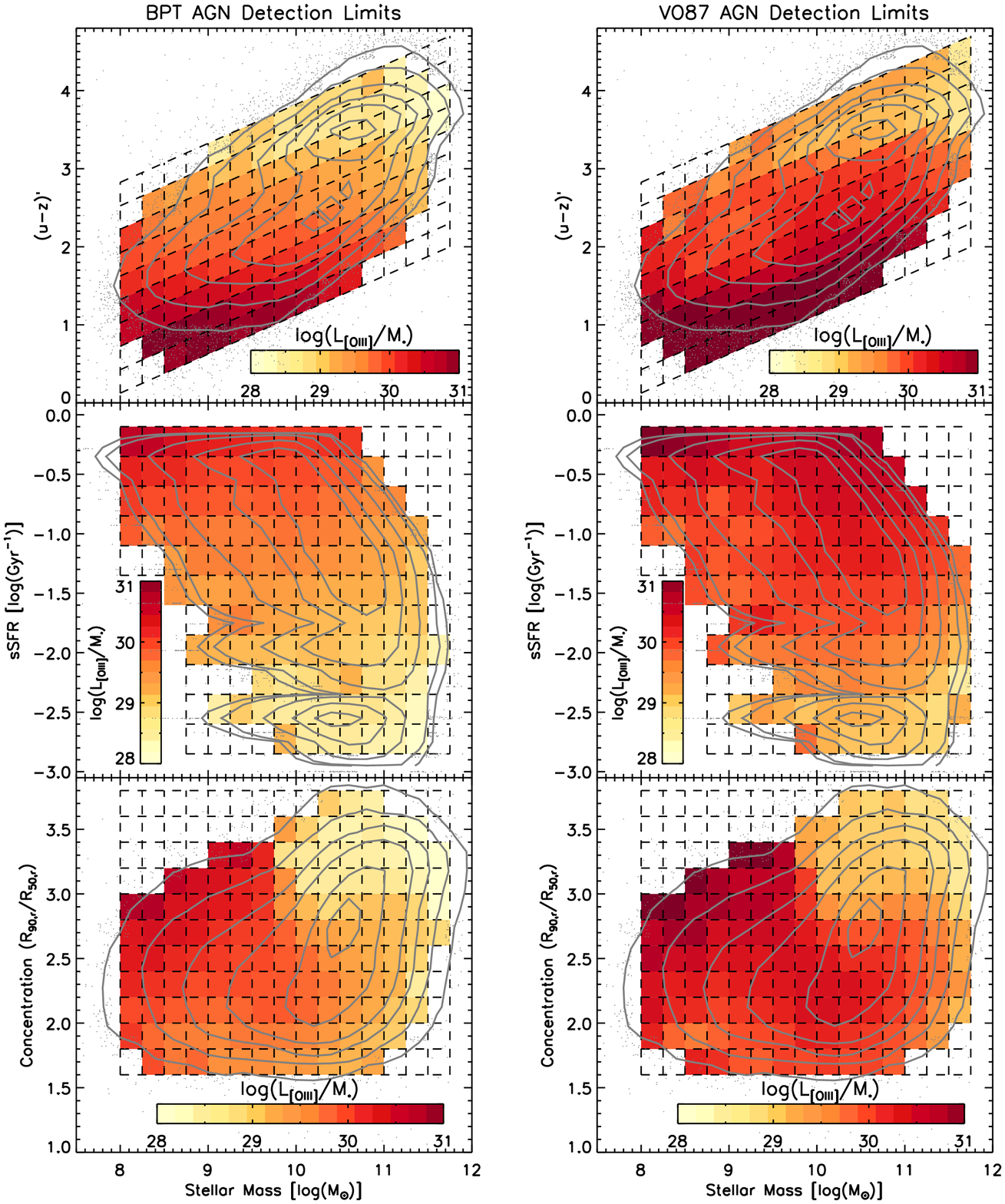}}
\figcaption{The detection limits for AGN line ratios to exceed star
  formation dilution, as a function of galaxy color, specific star
  formation rate, concentration, and stellar mass.  Gray contours show
  the full parent sample of galaxies.  The AGN detection limit is
  quantified as $L\OIII_{\rm AGN}/M_*$ (in units of
  erg~s$^{-1}$~$M_\odot^{-1}$), the dust-corrected AGN $\OIII$
  luminosity divided by the stellar mass.  These limits do not
  consider the AGN/LINER classification.  The $\HII$ region emission
  in low-mass blue, high-sSFR, and low-$R_{90,r}/R_{50,r}$ galaxies
  tends to dilute AGN-like line ratios, leading to a $L\OIII_{\rm
    AGN}/M_*$ detection threshold a factor of $\sim$500 higher than in
  high-mass red galaxies.
\label{fig:eddlimits}}
\end{figure*}  

The $L\OIII_{\rm AGN}/M_*$ detection limits for an AGN to exceed the
star formation dilution are shown in bins of color--mass, sSFR--mass,
and concentration--mass in Figure \ref{fig:eddlimits}.  At a basic
level the detection limits mirror the observed AGN fractions: the bins
of lowest AGN fraction (low-mass, blue, high-sSFR, low-concentration)
tend to also have the highest AGN detection thresholds.  If we assume
$\gamma=\Mbh/M_* \sim 0.001$ \citep[][assuming $M_* \sim 1.5M_{\rm
  bulge}$]{hr04} and $k_{\rm [OIII]} \sim 100$ \citep[][for
$\log(L\OIII_{\rm AGN}) \sim 40$]{lam09}, then the $L\OIII_{\rm
  AGN}/M_*$ limits translate to $\log(\eddratio) \sim -2$ for blue
low-mass (high-sSFR, mixed-concentration) galaxies and
$\log(\eddratio) \sim -5$ for red high-mass (low-sSFR,
high-concentration) galaxies.  Since the AGN Eddington ratio
distribution is steeply declining \citep[e.g.,][]{kau09, aird12,
  hic14}, this behavior leads to a significantly lower observed AGN
fraction in low-mass galaxies.  We use more detailed models of
Eddington ratio, bolometric correction, and $\Mbh/M_*$ to investigate
the AGN selection biases in Section 5 below.

Interpreting the biases of BPT and VO87 AGN selection crucially
depends on the metric used to estimate AGN power.  \citet{kau03}
\citep[and later,][]{schaw10} used $L\OIII$ as an indicator of AGN
strength, concluding that ``powerful'' ($L\OIII>10^7L_\odot$) AGNs are
present only in high-mass galaxies.  These luminous AGNs make up the
bulk of the total cosmic accretion history of AGNs, since mass
accretion rate $\dot{M}_{\rm BH}$ is related to $L\OIII$ by a
bolometric correction $k_{\rm [OIII]}$ and an efficiency $\eta$,
$\dot{M}_{\rm BH} = k_{\rm [OIII]}L\OIII/({\eta}c^2)$).  However, the
Eddington ratio $\eddratio$ is a more effective indicator of AGN
accretion structure and wind/jet outflow properties
\citep[e.g.,][]{nar08, ho08, tru11a, hec14}.  In this sense, our
Eddington ratio proxy $L\OIII/M_*$ is an effective way to quantify how
line-ration selection biases relate to AGN-galaxy coevolution.

The ``pure AGN NLR'' line ratios assumed here are estimated from the
locus of AGN-dominated galaxies on the BPT diagram.  Figure
\ref{fig:obsfrac} demonstrates that AGN-dominated line ratios are most
often seen in massive hosts.  Massive galaxies also tend to have high
(near-solar) metallicities, which means that the pure AGN NLR line
ratios are appropriate only for metal-rich NLR emission.  In the
Appendix we demonstrate that the observed line-ratio AGN occupation
fractions are better described by the presence of low-metallicity AGN
NLRs in low-metallicity galaxies.  This effect further worsens the AGN
detection bias in low-mass galaxies beyond that already present in
Figure \ref{fig:eddlimits}.

\subsection{Aperture Effects on AGN Detection}

\begin{figure}[ht]
\epsscale{1.15}
{\plotone{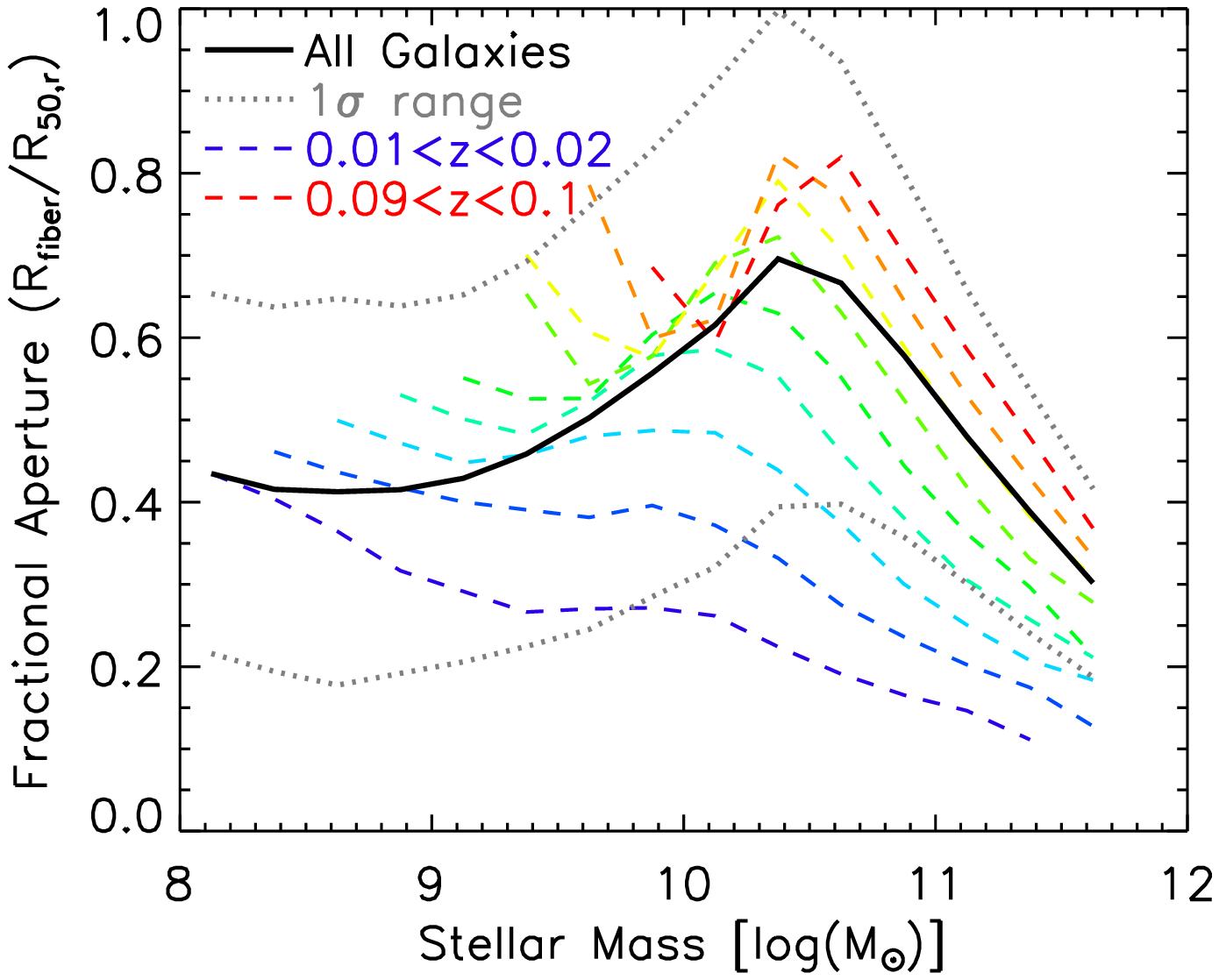}}
\figcaption{The fractional aperture covered by the SDSS fiber,
  quantified as the ratio between fiber radius and $r$-band half-light
  radius, with stellar mass.  The thick black line shows the median
  aperture coverage for all galaxies, and the gray dotted lines
  represent the $1\sigma$ range at each stellar mass.  The dashed
  lines display the median fractional aperture in redshift bins of
  width $\Delta{z}=0.01$ (i.e., 9 bins from $0.01<z<0.02$ to
  $0.09<z<0.10$).  Although there are significant aperture changes
  with redshift, the average fractional aperture for the whole sample
  changes by less than a factor of two with stellar mass.
\label{fig:aperture}}
\end{figure}

The size of the spectroscopic aperture affects the amount of star
formation dilution in the observed emission-line ratios.  Smaller
apertures include less star formation relative to the nuclear AGN: for
example, \citet{mor02} demonstrated that many nearby narrow-line AGNs
are hidden by star formation dilution in larger spectroscopic
apertures.  Our objects were all observed in 3$\arcsec$ fibers with
the SDSS spectrograph.  Figure \ref{fig:aperture} demonstrates that
the 3$\arcsec$ aperture covers a different fraction of the galaxy
light at different redshifts and different galaxy sizes.  The median
ratio of fiber radius to $r$-band half-light radius, $R_{\rm
  fiber}/R_{50,r}$, ranges from $\sim0.3$ to $\sim0.6$ (and $1\sigma$
bounds extending to $\sim0.2$ and $\sim0.8$).  The fractional aperture
changes are smaller than the changes in absolute aperture size because
low-mass galaxies, which are smaller in physical size, are only
observed at low redshift, corresponding to smaller physical aperture
sizes.  Larger high-mass galaxies are mostly in the higher-redshift
volume and consequently have larger physical aperture sizes.  These
$r$-band half-light radii are not necessarily equivalent to the
emission-line half-light radii, but Figure \ref{fig:aperture} remains
a useful demonstration of the changes in aperture effects with
redshift.

\begin{figure*}[ht]   
\epsscale{1.15}
{\plotone{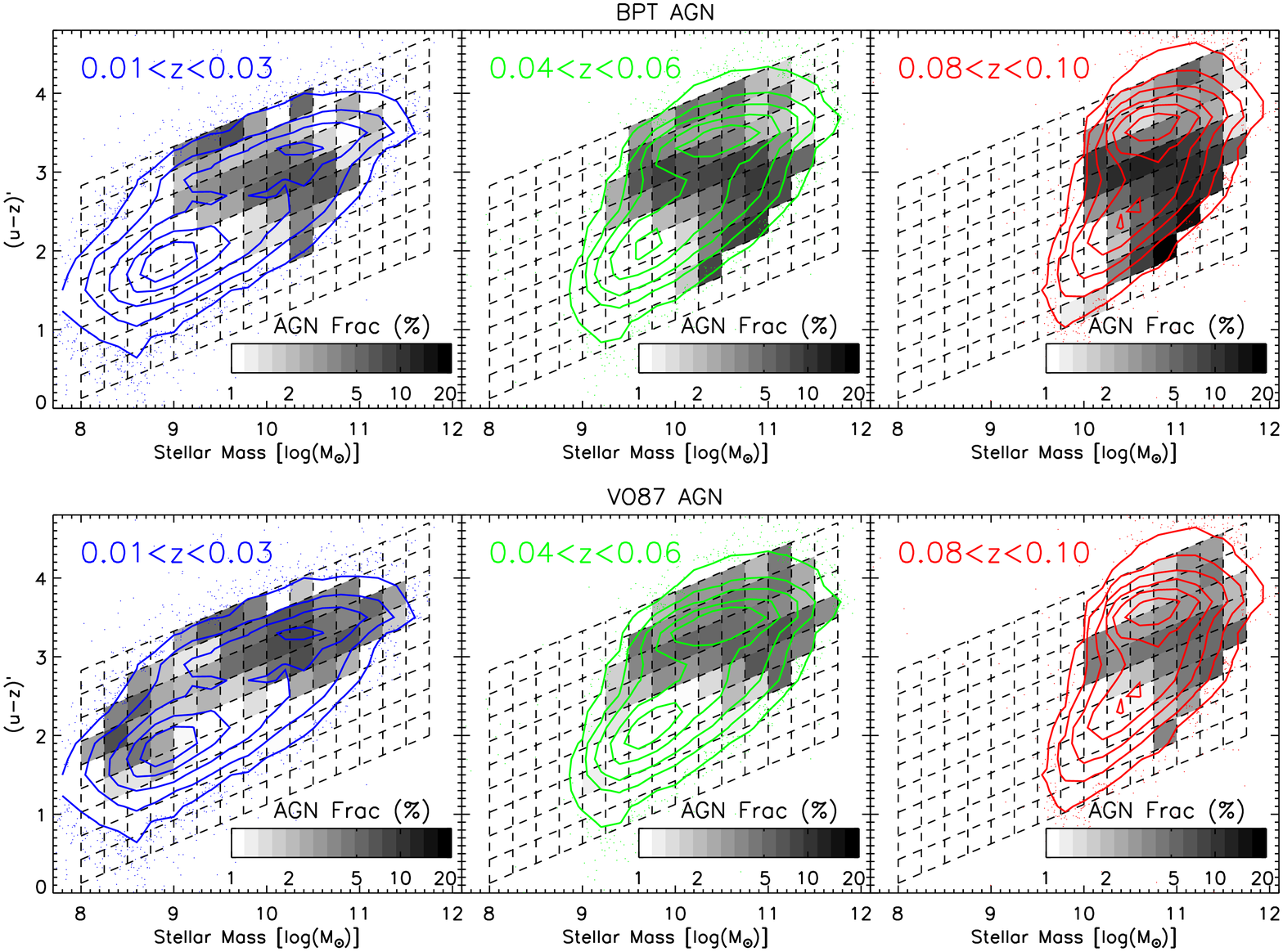}}
\figcaption{The fraction of galaxies classified as BPT AGN and VO87
  AGN in three bins of redshift, from low to high.  Gray contours show
  the distribution of galaxies in each redshift bin.  Where galaxies
  are present in multiple redshift bins, there is little difference in
  the observed AGN fraction, suggesting that AGN selection is
  minimally affected by the changing aperture.  Essentially the
  increase in galaxy size with stellar mass is balanced by a
  corresponding increase in the typical physical aperture size, since
  low-mass galaxies are observed only at low redshift and high-mass
  galaxies are mostly at high redshift.
\label{fig:obsfrac_hilo}}
\end{figure*}   

The lack of aperture effects with redshift is similarly seen in Figure
\ref{fig:obsfrac_hilo}, which compares observed fractions of galaxies
classified as BPT AGN and VO87 AGN at $0.01<z<0.03$, $0.04<z<0.06$,
and $0.08<z<0.1$.  The ``break'' from high to low AGN fraction is at
$\log(M_*) \sim 10$ in all three redshift bins.  As seen in Figure
\ref{fig:aperture}, there are no significant aperture effects causing
differential star formation dilution and AGN identification as a
function of stellar mass.

It should be emphasized again that both the AGN limits from Section
4.1 and the Monte Carlo simulations of Section 5 implicitly account
for aperture effects.  We empirically quantify star formation dilution
using line fluxes of $\HII$ regions only within the spectroscopic
aperture, rather than total-galaxy measurements.  Thus even if there
were aperture effects missed by Figures \ref{fig:aperture} and
\ref{fig:obsfrac_hilo}, they would be accounted for in our models of
AGN limits and in the resultant intrinsic AGN fraction with host
properties.

\section{Modeling the Intrinsic AGN Population}

In Section 4 we demonstrated that line-ratio AGN selection is biased
against host galaxies with low stellar mass, blue color, and high
sSFR.  Such galaxies have more ``star formation dilution'' from
optical emission lines produced in $\HII$ regions.  But it remains
unclear if the absence of observed AGNs in low-mass blue disk hosts is
caused purely by bias, or there is also some change in AGN occupation
with galaxy properties.

In this Section we directly model the intrinsic AGN population as a
function of galaxy properties.  Separate simulations are run for both
the BPT and VO87 AGN selection methods, comparing the modeled and
observed fraction of galaxies meeting each set of AGN selection
criteria.  Within each bin of galaxy properties (color--mass,
sSFR--mass, or concentration--mass), the simulations follow these
basic steps:

\begin{enumerate}

  \item A set of ``non-AGN'' line ratios is randomly drawn (with
    replacement) from the dust-corrected lines of observed galaxies
    with well-measured line ratios not classified as BPT AGNs or VO87
    AGNs.  We require that these galaxies be face-on
    ($\mathbf{b/a>0.5}$) and remove outliers in velocity dispersion
    (using iterative 2$\sigma$ clipping).  The number of simulated
    galaxies is the same as the total number of galaxies within each
    bin.

  \item Each simulated ``non-AGN'' galaxy is assigned an AGN with an
    Eddington ratio $\eddratio$ randomly drawn from a probability
    distribution.  The average Eddington ratio is either fixed or
    allowed to vary as a free parameter.

  \item The Eddington ratio is translated to AGN emission-line
    luminosities in three steps.  First, black hole mass is estimated
    from galaxy velocity dispersion (or allowed to be a free
    parameter).  $\OIII$ luminosity is then derived from $\Lbol$ using
    a (luminosity-dependent) bolometric correction.  From $L\OIII_{\rm
      AGN}$, we then compute the other line luminosities assuming
    ``pure AGN NLR'' line ratios.

  \item We compute the number of simulated galaxies with resultant
    AGN+galaxy line luminosities classified as BPT AGNs (Equations 1
    and 6) or VO87 AGNs (Equations 2 and 5).  The simulated AGN
    fraction is given by the median from 200 Monte Carlo simulations
    of steps 1--3 above, with an associated error given by the
    normalized median absolute deviation (NMAD).

\end{enumerate}

The assumptions inherent in these steps are motivated and discussed in
Section 5.1, with detailed tests of the assumptions also given in the
Appendix.  Inspired by the work of \citet{aird12} for X-ray AGN hosts,
in Section 5.2 we test if the observed AGN fractions are consistent
with uniform SMBH growth regardless of host properties.  In Sections
5.3 and 5.4, respectively, we instead fit the Eddington ratio and
black hole mass as free parameters, demonstrating that AGN accretion
is correlated with galaxy sSFR and low-mass galaxies may have
undermassive black holes.

\subsection{Basic Modeling Assumptions}

In each bin of galaxy properties, initial sets of ``non-AGN'' line
ratios are empirically drawn from well-measured ($(S/N)_{\rm
  ratio}>3/\sqrt{2}$) galaxies that have reliable velocity dispersions
(described below) and do not meet the BPT or VO87 AGN selection
criteria.  The empirical sets of galaxies implicitly reflect the
distribution of non-AGN line ratios in each bin of galaxy properties.
The breadth of the line-ratio distribution comes from the scatter in
the mass-metallicity relation \citep[e.g.,][]{tre04}, a range of
ionization conditions from $\HII$ regions and old star winds
\citep[e.g.,][]{san15}, and fiber aperture effects for different
galaxy sizes and redshifts.  Beginning the simulations with empirical
non-AGN line ratios accurately reproduces the observed line-ratio
distributions (see Section 5.5).  Since we do not classify LINERs as
BPT or VO87 AGNs, they are included in the ``non-AGN'' galaxy sets.
LINER galaxies are likely to host weakly accreting AGNs
\citep[e.g.,][]{ho08,ho09}, but \citet{yan12} demonstrated that the
bulk of their emission-line flux is extended rather than due to a
nuclear AGN.  Thus we include LINERs among the empirically-drawn
``non-AGN'' line ratios, simulating the addition of more powerful AGNs
with Seyfert-like line ratios to such galaxies.

Because the simulations begin with empirical line ratios of non-AGN
galaxies, they cannot predict the AGN fraction in poorly-measured
galaxies (which, by definition, cannot be reliably classified as
star-forming or AGN).  For this reason we only compare (or fit) the
simulations to the observed AGN fractions among well-measured
galaxies.  The inability to simulate poorly-measured line ratios has
consequences for massive red (low-sSFR, concentrated) galaxies, which
tend to be excluded in the well-measured sample.  We investigate this
issue in Section 5.3 by assuming that poorly-measured galaxies have
the same median ratio of $\eddratio / (L\OIII_{\rm total}/M_*)$ as the
well-measured galaxies.

The Eddington ratio distribution of our simulated AGNs is
parameterized as a Schechter function, following \citet{hh09}:
\begin{equation}
  P(\lambda) = \frac{dt}{d\log\lambda} = P_0 \left(\frac{\lambda}{\lambda_*}\right)^{-\alpha} \exp{\left(-\frac{\lambda}{\lambda_*}\right)}.
\end{equation}
This functional form is effectively a power-law distribution, with a a
``turnover'' Eddington ratio of $\lambda_*=0.4$ set to impose the
Eddington limit $\lambda<1$.  The Eddington ratio $\lambda \equiv
\eddratio$, with $L_{\rm Edd} \simeq 1.3\times10^{38}
(\Mbh/M_\odot)$~erg~s$^{-1}$, and $P_0$ is a normalization constant.
Equation 7 is, at best, an approximation for the true Eddington ratio
distribution, which is poorly known and not well-constrained by
observations \citep[e.g.,][]{kel10}.  However, adopting a functional
form allows us to effectively compare \textit{relative} Eddington
ratios across galaxy properties.

We use $\alpha=0.6$ for the power-law slope of Equation 7, consistent
with the observed Eddington ratio distribution \citep{hh09,kau09} and
similar to the distribution used by \citet{aird12}.  In the Appendix
we demonstrate that shallower Eddington ratio distributions (e.g.,
$\alpha=0.2$ from \citealp{hic14} or $\alpha=0.05$ from
\citealp{schulze10}) result in simulations with worse fits to the
observed AGN fractions.  We also prefer a steeper slope in $\eddratio$
because it qualitatively reflects the decreasing efficiency $\eta$ for
radiatively inefficient accretion at low $\dot{m}$ \citep{xie12}.
With a fixed number of Eddington ratios in the random distribution, we
set $P_0=1$ and numerically control the normalization (and average
Eddington ratio) by changing the lower bound $\log(\lambda_{\rm min})$
of the distribution.

To convert Eddington ratio to luminosity, we estimate black hole mass
using the \citet{gul09} $\Mbh-\sigma$ relation:
\begin{equation}
  \log(\Mbh/M_\odot) = \alpha + \beta\log(\sigma/200~{\rm km~s}^{-1}) + \epsilon_0
\end{equation}
Here $\alpha=8.12$, $\beta=4.24$, and the intrinsic scatter
$\epsilon_0=0.44$.  \citet{kor13} demonstrate that the $\Mbh-\sigma$
relation for classical bulges has significantly higher normalization
($\alpha=8.49$) and lower scatter ($\epsilon_0=0.29$), suggesting that
the \citet{gul09} normalization is lower due to the inclusion of
pseudobulges.  However nearly all of our galaxies have insufficient
data to reliably separate classical bulges and pseudobulges, so we
retain the \citet{gul09} relation.  In the Appendix we also test a
constant $\Mbh/M_*$ ratio, finding that it results in a significantly
worse fit to the data.

Equation 8 is appropriate only for velocity dispersions corresponding
to a galaxy bulge.  We remove velocity dispersions from unresolved
rotation or disordered kinematics by requiring that the
empirically-drawn galaxies be face-on ($b/a>0.5$), have well-measured
velocity dispersion (error in $\sigma$ $<$60~km~s$^{-1}$), and using
iterative 2$\sigma$ clipping to remove outliers.  \citet{kas12} find
that most $z \sim 0$ galaxies with $\log(M_*/M_\odot) \gtrsim 8.5$ are
settled disks, suggesting that unsettled disks with
dispersion-dominated kinematics are likely to be removed by the
2$\sigma$ outlier rejection.  The 3'' fiber aperture means that these
velocity dispersions are typically measured over a radius smaller than
$r_e$ (see Figure \ref{fig:aperture}).  However bulges typically have
flat velocity dispersion profiles over $0.1<r/r_e<10$: \citet{cap06}
measure a typical $\sigma(r)=\sigma_e(r/r_e)^{-0.066}$ over this
range, and so we assume $\sigma\simeq\sigma_e$ for our galaxies.
Galaxies with upper limits in velocity dispersion are included using
survival analysis, assuming a normal parent distribution (before
clipping) described by the median $\sigma$ and its NMAD.

To convert bolometric luminosity into the extinction-corrected $\OIII$
line luminosity we use the following bolometric correction:
\begin{equation}
  \frac{L_{\rm bol}}{10^{40}~{\rm erg~s}^{-1}} = 112\left(\frac{L\OIII}{10^{40}~{\rm erg~s}^{-1}}\right)^{1.2}
\end{equation}
This equation is a power-law fit to the luminosity-dependent
bolometric corrections found by \citet{lam09}.  Equation 9 is roughly
consistent with the findings of \citet{sl12} for observed (not
corrected for extinction) $L\OIII$, given the 1--2~mag of extinction
increasing from low to high $L\OIII$ observed in our galaxies.  We
adopt an intrinsic scatter of 0.4~dex in the $L\OIII$ bolometric
correction which reflects the range of AGN spectral energy
distributions and details of the emission geometry.  This intrinsic
scatter dominates over the measurement errors in line flux and dust
correction.  \citet{sl12} measure a slightly larger scatter of 0.6~dex
for $L\OIII$, but some of their measured scatter is due to the range
of AGN extinction.  A smaller scatter of 0.4~dex is probably more
appropriate for the extinction-corrected $L\OIII$ used in this work.
Using a scatter of 0.6~dex in $\Lbol/L\OIII$ instead of the adopted
0.4~dex systematically decreases the inferred Eddington ratios by
$\sim$0.1~dex, with no change in the relative Eddington ratios across
galaxy properties.

From $L\OIII_{\rm AGN}$, all of the (extinction-corrected) emission
lines in the BPT and VO87 diagrams can be estimated using line ratios
for ``pure AGN NLRs'':
\begin{equation}
  \Ha/\Hb=3.1.
\end{equation}
\begin{equation}
  \log(\OIII/\Hb)=0.5 \pm 0.3,
\end{equation}
\begin{equation}
  \log(\SII/\Ha)=-0.2 \pm 0.2,
\end{equation}
\begin{equation}
  \log(\NII/\Ha) = \langle \log(\NII/\Ha)_{\rm gal} \rangle + 0.45 \pm
  0.2.
\end{equation}
Equation 10 is the intrinsic Balmer decrement that best describes
typical AGN NLR conditions \citep{ost06}, with ionization from the AGN
causing a value slightly exceeding the $\Ha/\Hb=2.86$ of $\HII$
regions.  Equations 11 and 12 are empirically drawn from the AGN locus
in the observed BPT and VO87 diagrams.  Equation 13 similarly reflects
the range of observed AGNs, but also allows for a
metallicity-dependent AGN NLR.  The scatter in each equation was
chosen to empirically reproduce the observed line-ratio distributions
in the BPT and VO87 diagrams, and effectively accounts for a range in
NLR geometry and ionization conditions (see Section 5.5).

\begin{figure*}[t]  
\epsscale{1.15}
{\plotone{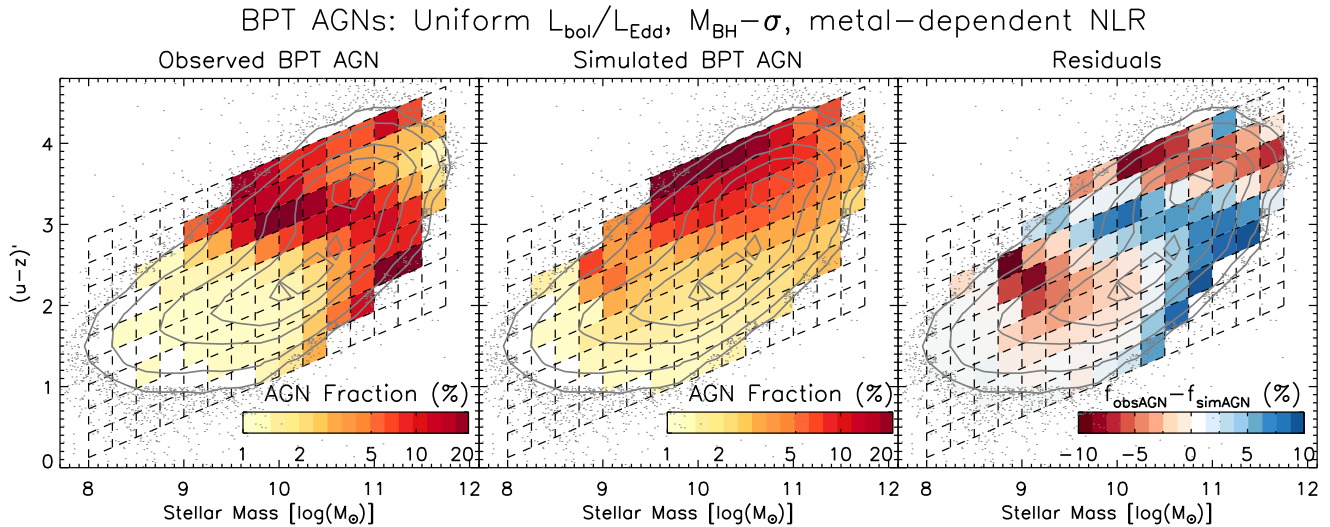}}
\figcaption{The fraction of BPT AGNs with host galaxy color and
  stellar mass from both observations and the ``uniform-$\eddratio$''
  simulation.  The simulation uses the assumptions outlined in
  Equations 7--13: a Schechter function $\eddratio$, $\Mbh-\sigma$
  relation, luminosity-dependent $\Lbol/L\OIII$ bolometric correction,
  and metallicity-dependent AGN NLR line ratios.  AGN fraction is
  indicated by the bin shading and gray contours represent the
  distribution of well-measured galaxies.  Similar to the simple AGN
  limits of Section 4.1 and Figure \ref{fig:eddlimits}, star formation
  dilution results in fewer AGNs identified in blue galaxies.  However
  the uniform-$\eddratio$ model does a poor job of reproducing the
  data, with too many simulated AGNs in red galaxies and too few in
  high-mass blue/green galaxies.
\label{fig:simbpt_colorzmsig}}
\end{figure*}  

\begin{figure*}  
\epsscale{1.15}
{\plotone{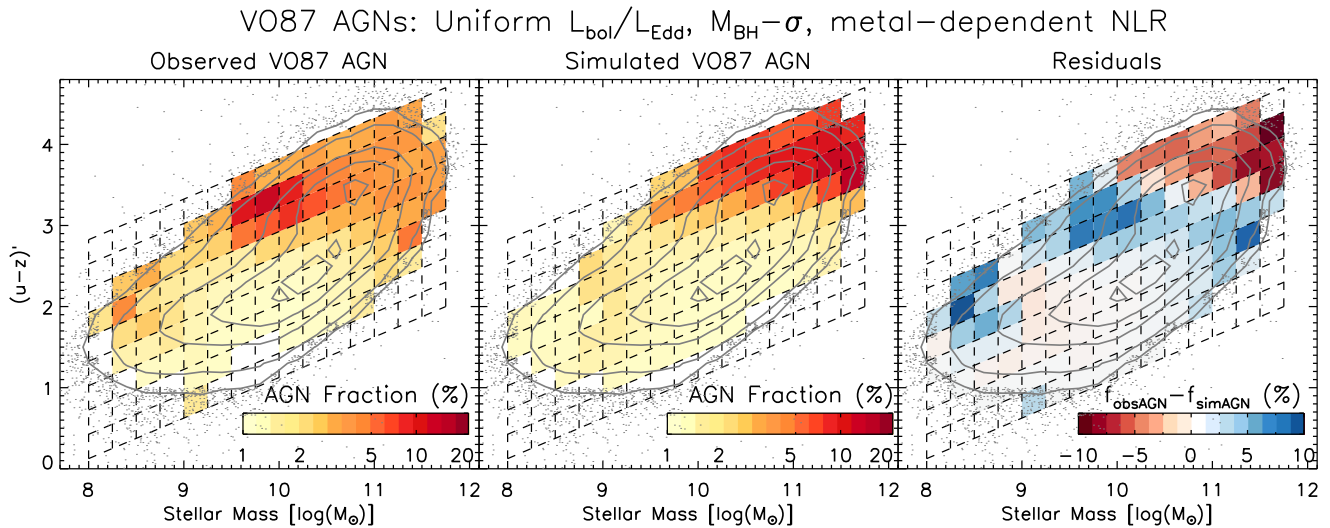}}
\figcaption{The fraction of VO87 AGNs with host galaxy color and
  stellar mass from the ``uniform-$\eddratio$'' simulation compared to
  the observations.  Bin shading indicates the AGN fraction and gray
  contours represent the well-measured galaxy sample.  As in Figure
  \ref{fig:simbpt_colorzmsig}, the simulation overpredicts the number
  of AGNs in massive red galaxies and underpredicts the AGN fraction
  in massive blue galaxies.  The uniform-$\eddratio$ simulation also
  fails to reproduce the AGNs in low-mass hosts identified by the VO87
  but not the BPT.
\label{fig:simvo87_colorzmsig}}
\end{figure*}  

\begin{figure*}[t]  
\epsscale{1.15}
{\plotone{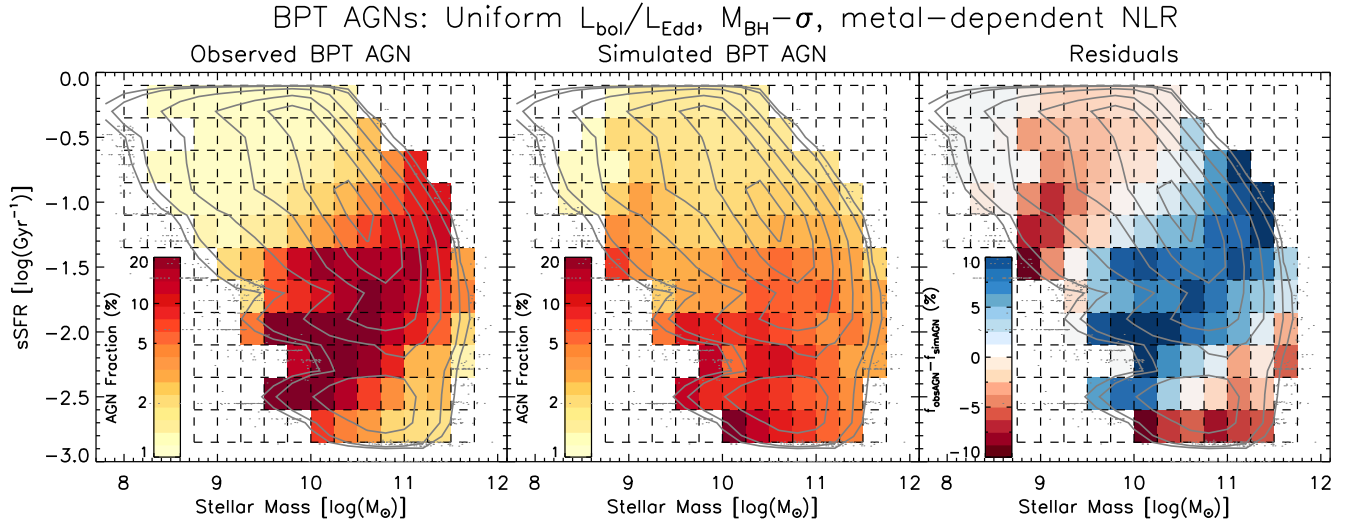}}
\figcaption{The observed and simulated BPT AGN fractions with host
  galaxy specific star formation rate and stellar mass, using the
  ``uniform-$\eddratio$'' model.  The simulation underpredicts the
  number of AGNs in moderate-sSFR and moderate-mass galaxies,
  underpredicts the observed AGN fraction at both high and low sSFR
  and stellar mass.  In Sections 5.3 and 5.4 we show that, rather than
  a uniform Eddington ratio, the observations imply a correlation
  between Eddington ratio and sSFR in massive galaxies, with some hint
  of undermassive black holes in low-mass hosts.
\label{fig:simbpt_sfrzmsig}}
\end{figure*}  

\begin{figure*}[t]  
\epsscale{1.15}
{\plotone{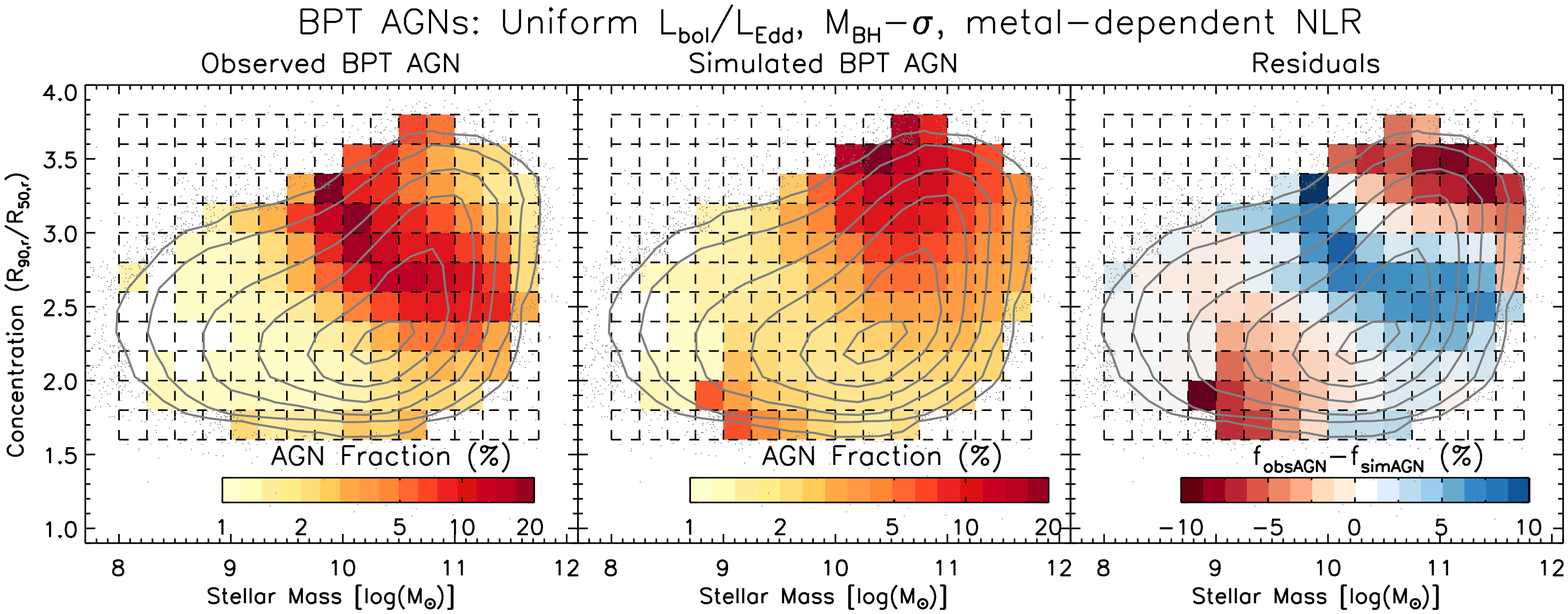}}
\figcaption{The observed and simulated BPT AGN fractions with host
  galaxy specific concentration ($R_{90,r}/R_{50,r}$) and stellar
  mass, using ``uniform-$\eddratio$'' model.  The uniform Eddington
  ratio model has the same zebra-like mismatch to the data as in
  Figures \ref{fig:simbpt_colorzmsig} and \ref{fig:simbpt_sfrzmsig},
  underpredicting the observed AGN fraction in
  moderate-mass/moderate-concentration galaxies and overpredicting at
  low-mass/low-concentration and high-mass/high-concentration.
\label{fig:simbpt_conczmsig}}
\end{figure*}  

To model the metallicity-dependent NLR line ratios, we follow the
models of \citet{kew01}, which show that lower metallicity decreases
the $\NII/\Ha$ ratio of the AGN NLR and galaxy by the same degree,
while the AGN $\OIII/\Hb$ and $\SII/\Ha$ ratios change little.
Similar metallicity effects on the AGN narrow-line region and host
galaxy are plausible due to the large (few-kpc) size of the typical
AGN narrow-line region \citep[e.g.,][]{ben02}\footnote{The large size
  and potential metallicity variation of the AGN NLR is in contrast to
  the much smaller broad-line region, which is observed to have
  super-solar metallicity over a wide range in mass and redshift
  \citep[e.g.,][]{nag06,jua09,mat11}.}.  Most galaxies have fairly
flat metallicity gradients with $\alpha_{\rm O/H}=-0.1$~dex/$R_e$
\citep{zar94,san14}, further suggesting that the NLR metallicity
should be similar to the galaxy metallicity measured within the SDSS
fiber (which typically covers 0.3--0.6$R_e$, see Figure
\ref{fig:aperture}).  The scatter in Equation 13 effectively includes
the potential effects of scatter in metallicity between the galaxy and
the AGN NLR.  In the Appendix we also test a simulation with a
uniform-metallicity NLR (i.e., a constant NLR $\NII/\Ha$ ratio),
finding that it results in a significantly worse fit to the
observations.  Section 5.5 demonstrates that the line ratios and
scatter assumed in Equations 10--13 result in simulated BPT and VO87
diagrams which are similar to the observed diagrams.

\subsection{Testing a Uniform Eddington Ratio Distribution}

We begin by testing a ``uniform-$\eddratio$'' model similar to that
advocated by \citet{aird12}: the same Eddington ratio distribution in
every bin of galaxy properties (color, sSFR, and concentration versus
stellar mass).  The simulation uses the Eddington ratio distribution,
$\Mbh-\sigma$ relation, $\Lbol/L\OIII$ bolometric correction, and
metallicity-dependent AGN NLR line ratios outlined in Equations 7--13
above.  The normalization of Equation 7 is set to be the same over all
galaxy properties: $\log(\lambda_{\rm min})=-5.0$ for BPT AGNs and
$\log(\lambda_{\rm min})=-5.5$ for VO87 AGNs, with each value set to
minimize the summed $\chi^2$ values from comparing to the
observations:
\begin{equation}
  \chi^2=\sum\limits_{i=1}^{N_{\rm bins}} \left(\frac{(f_{\rm
        obsAGN}-f_{\rm simAGN})^2}{\sigma^2(f_{\rm obsAGN})}\right).
\end{equation}
Here $f_{\rm obsAGN}$ and $f_{\rm simAGN}$ are the observed and
simulated AGN fractions among well-measured galaxies, and
$\sigma(f_{\rm obsAGN})$ is the binomial error in $f_{\rm obsAGN}$
\citep[calculated following][]{binomerr}.

The ``uniform-$\eddratio$'' simulation is compared to the observed BPT
AGN fractions across the color--mass diagram in Figure
\ref{fig:simbpt_colorzmsig}.  A uniform Eddington ratio distribution
does a poor job of reproducing the observations, with significant
residuals across the galaxy population.  The simulations similarly do
a similarly poor job of reproducing the observed VO87 AGN population,
as seen in Figure \ref{fig:simvo87_colorzmsig}.  The simulated and
observed BPT AGN fractions in the sSFR--mass and concentration--mass
diagrams are also shown in Figures \ref{fig:simbpt_sfrzmsig} and
\ref{fig:simbpt_conczmsig}, respectively.  In the Appendix we show
that the mismatch with observations is even worse when assuming a
constant $\Mbh/M_*$ ratio or a metal-rich NLR.  In all cases the
``uniform-$\eddratio$'' model overpredicts the observed AGN fraction
in massive red, low-sSFR, concentrated galaxies, underpredicts in the
bluest, highest-sSFR, least-concentrated massive galaxies, and
overpredicts again in most low-mass galaxies.

The overprediction of AGNs in massive red (low-sSFR, concentrated)
galaxies is especially pronounced if we consider the difference in AGN
fractions between the well-measured and full samples of galaxies.  In
Section 3.3 we showed that the apparent AGN fraction for massive red
(low-sSFR, concentrated) galaxies for the full sample is lower than in
the well-measured sample.  Thus the simulated AGN fraction in such
galaxies is likely to be even more overpredicted when compared to the
full dataset.

From Figures \ref{fig:simbpt_colorzmsig}--\ref{fig:simbpt_conczmsig}
we conclude that, with plausible model assumptions for a bolometric
correction, the $\Mbh-\sigma$ relation, and pure AGN line ratios, the
observed fractions of line-ratio AGNs are inconsistent with a uniform
Eddington ratio distribution.  In the next subsections we instead
allow Eddington ratio and black hole mass to be free parameters with
galaxy properties.  Rather than a uniform Eddington ratio
distribution, the observed line-ratio AGN fractions are best
reproduced by a model where (a) AGN accretion rate is correlated with
host sSFR in massive ($\log(M_*/M_\odot)>10$) host galaxies and (b)
low-mass ($\log(M_*/M_\odot)<10$) galaxies may have a lower black hole
occupation function.

\subsection{Non-Uniform AGN Duty Cycle}

Given the poor fit of the ``uniform-$\eddratio$'' simulations, in this
subsection we instead allow the normalization of the Eddington ratio
(i.e., the average Eddington ratio) to be a free parameter with galaxy
properties.  Varying the normalization of the Eddington ratio
distribution effectively changes the AGN duty cycle (i.e., the number
of AGNs above some accretion rate threshold).  Functionally, this
simulation follows steps 1--4 described in the beginning of the
Section.  We retain the same bolometric correction (Equation 9),
$\Mbh-\sigma$ relation (Equation 8), and metallicity-dependent NLR
(Equations 10--13) as in the uniform-$\eddratio$ simulation.  We also
use the Schechter function to describe the AGN Eddington ratio
distribution given in Equation 7 (with the same declining power-law
slope of $\alpha=0.6$), numerically controlling its normalization
using the lower bound $(\eddratio)_{\rm min}$.  However instead of a
uniform lower bound, we fit over a grid starting from
$\log(\eddratio)_{\rm min}=-3$ and decreasing by
$\Delta\log(\eddratio)_{\rm min}=0.05$ until the simulations produce
an AGN fraction equal to the observations.  Bins with observed AGN
fractions consistent with zero (using their 1-$\sigma$ binomial error)
result in upper limits for the Eddington ratio normalization, since
lower Eddington ratios would similarly result in zero simulated AGNs.
We report the best-fit Eddington ratios using the distribution's
average: for our Schechter function parameterization, $\langle
\log(\eddratio) \rangle \simeq 0.91\log(\eddratio)_{\rm min} + 0.25$.
A higher average Eddington ratio can also be interpreted as a higher
AGN duty cycle (i.e. a higher fraction of galaxies hosting an AGN
above some accretion rate).

\begin{figure*}  
\epsscale{1.15}
{\plotone{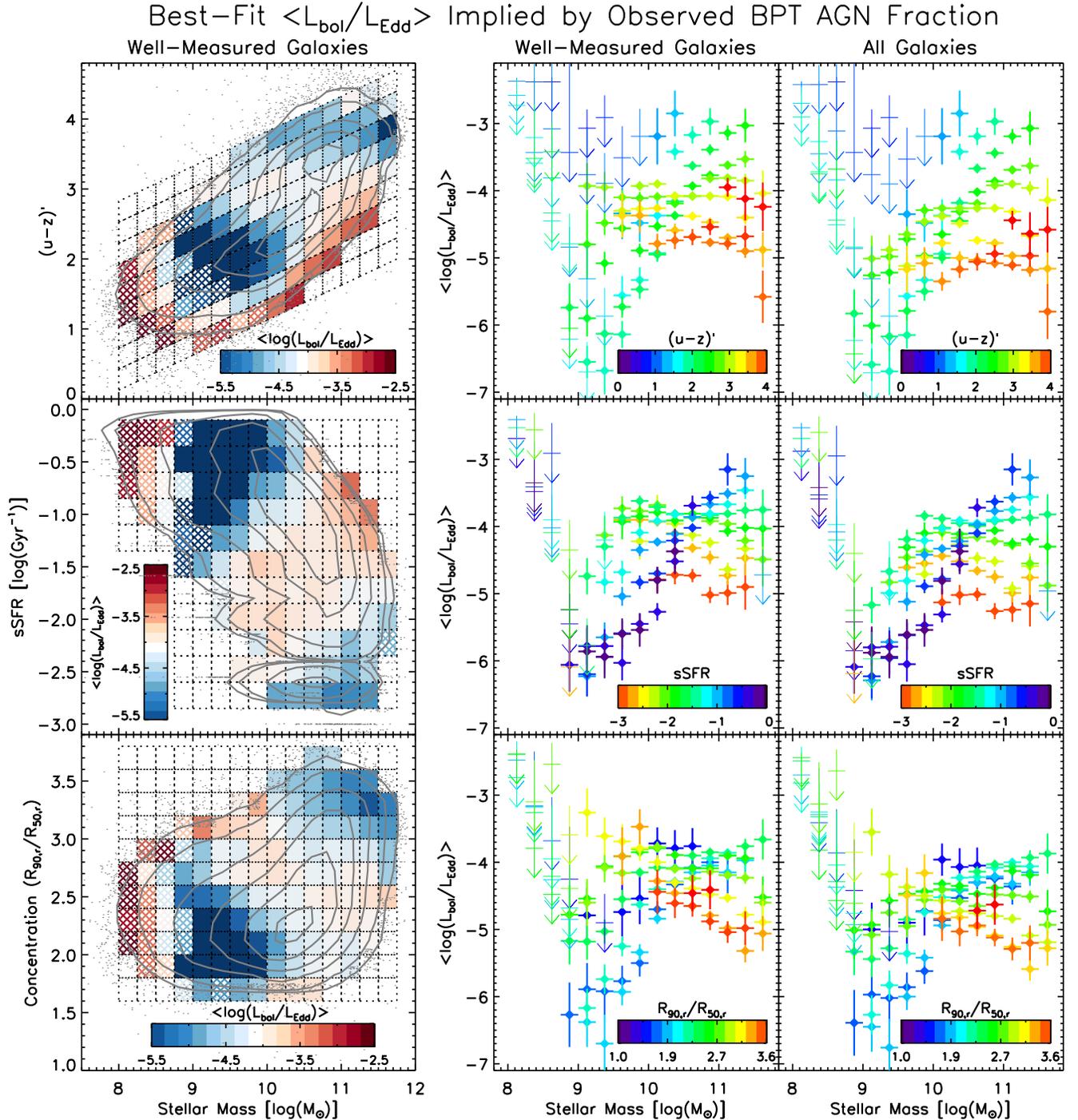}}
\figcaption{\textit{Left:} The average Eddington ratios in bins of
  galaxy color--mass, sSFR--mass, and concentration--mass from the
  ``non-uniform'' $\eddratio$ model fits that reproduce the observed
  BPT AGN fraction for well-measured galaxies in each bin.  Gray
  contours represent the well-measured galaxy sample, and hashed bins
  denote Eddington ratio upper limits where the observed AGN fraction
  is consistent with zero (given its binomial error).
  \textit{Center:} The same average Eddington ratios versus stellar
  mass in each bin, color-coded by $(u-z)'$, sSFR, or
  $R_{90,r}/R_{50,r}$.  Vertical error bars represent the NMAD of the
  Monte Carlo simulation results.  Upper limits due to an observed AGN
  fraction consistent with zero (the hashed bins in the left panel)
  are shown as thinner lines.  \textit{Right:} Average Eddington
  ratios for the full galaxy sample, assuming that poorly-measured
  galaxies have the same median $L\OIII_{\rm AGN} / L\OIII_{\rm
    total}$ as the well-measured galaxies.  In massive galaxies, the
  AGN Eddington ratio tends to be highest in hosts with blue color,
  high sSFR, and low concentration.  The connection between AGN
  accretion and galaxy properties is murkier in low-mass hosts.
\label{fig:fiterbpt}}
\end{figure*}  

\begin{figure*}  
\epsscale{1.15}
{\plotone{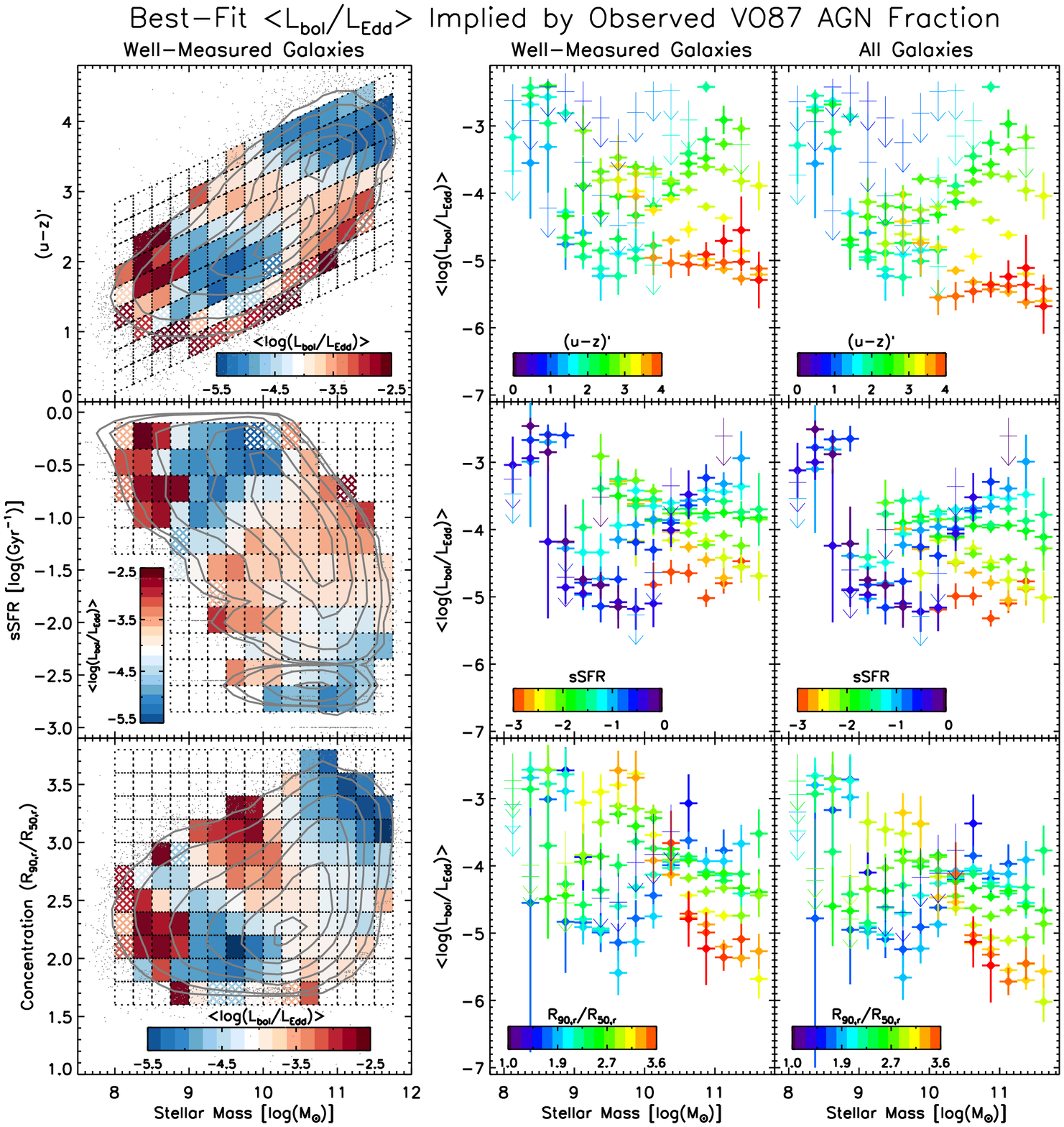}}
\figcaption{The average Eddington ratios from the ``non-uniform''
  $\eddratio$ model fits that match the observed VO87 AGN fractions in
  bins of galaxy color--mass, sSFR--mass, and concentration--mass.
  The well-measured galaxy sample is shown by gray contours.  Upper
  limits in Eddington ratio are marked as hashed bins (left panel) and
  thinner lines (center and right panels).  Just as was the case for
  BPT AGNs in Figure \ref{fig:fiterbpt}, there is a connection between
  Eddington ratio and host galaxy blue color, sSFR, and low
  concentration in high-mass ($\log(M_*/M_\odot) \gtrsim 10$)
  galaxies.  In low-mass ($\log(M_*/M_\odot) \lesssim 10$) galaxies
  the connection between host properties and AGN accretion is much
  less clear.
\label{fig:fitervo87}}
\end{figure*}  

Figures \ref{fig:fiterbpt} and \ref{fig:fitervo87} present the average
Eddington ratios from this ``non-uniform $\eddratio$'' model in bins
of color--mass, sSFR--mass, and concentration--mass for both BPT AGNs
and VO87 AGNs.  The left and middle columns of each figure show the
average Eddington ratios calculated from the well-measured set of
galaxies.  In the right column we make a correction to estimate the
average Eddington ratio for the full galaxy sample, assuming that
poorly-measured galaxies have the same median ratio of $\eddratio /
(L\OIII_{\rm total}/M_*)$ as the well-measured galaxies.  Within the
narrow ranges of $M_*$ and $\sigma$ in each bin of galaxy properties,
this assumption is essentially the same as assuming the same
fractional AGN contribution to $L\OIII_{\rm total}$ in both
well-measured and poorly-measured galaxies.  Correcting for
poorly-measured galaxies tends to slightly decrease (by up to 0.3~dex)
the average Eddington ratio in high-mass red, low-sSFR, concentrated
galaxies (i.e., those galaxies most likely to be weak-lined with
poorly-constrained line ratios).  In all three panels, the
  observed AGN fractions in $M_* \lesssim 10^9 M_\odot$ galaxies are
  frequently consistent with zero, resulting in poorly-constrained
  upper limits for the intrinsic Eddington ratios.

\begin{figure}[ht]
\epsscale{1.15}
{\plotone{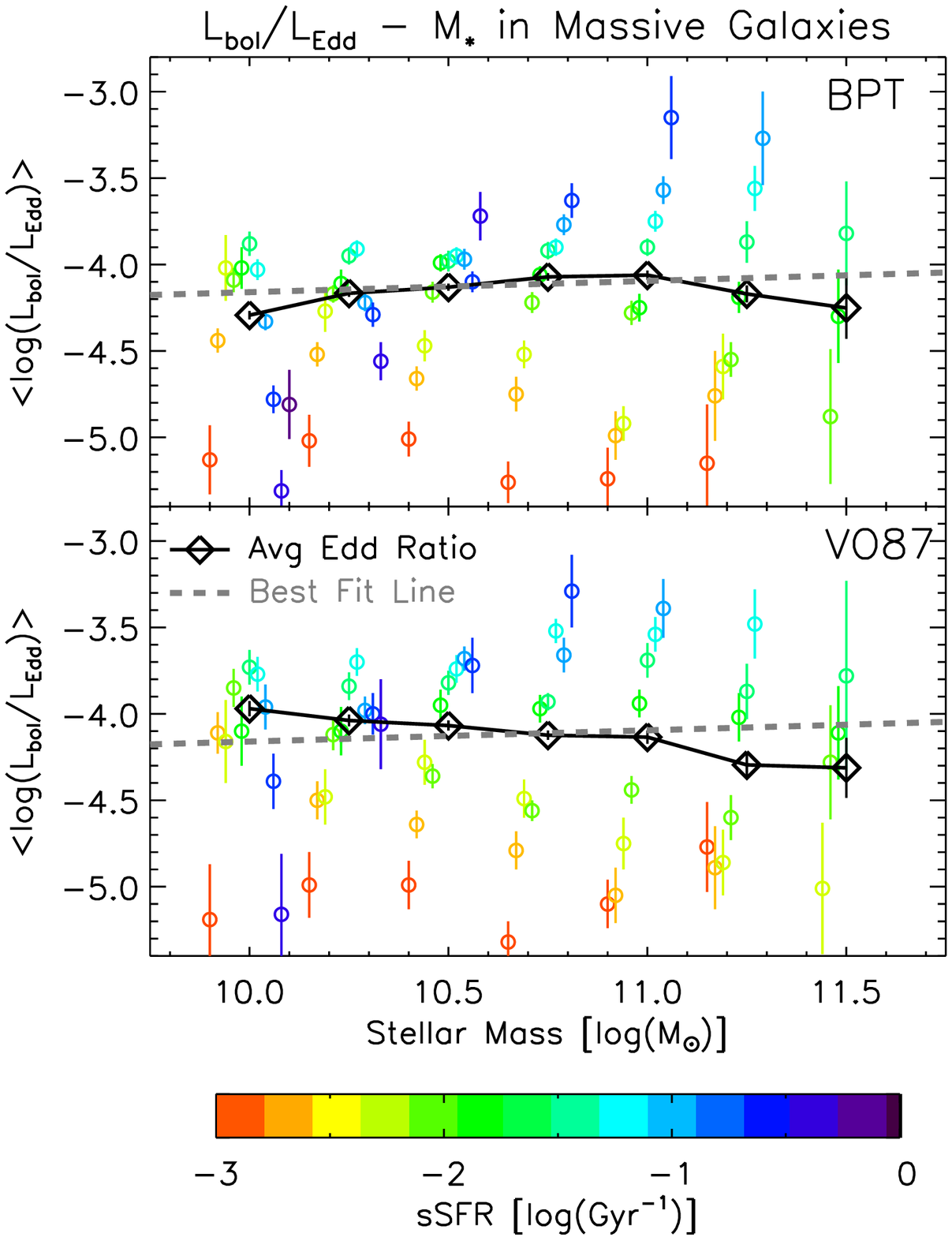}}
\figcaption{The average AGN Eddington ratio implied by the
  ``non-uniform'' $\eddratio$ model fits to the observed BPT (top) and
  VO87 (bottom) AGN fractions, as a function of galaxy stellar mass.
  Only massive ($\log(M_*/M_\odot) \geq 10$) galaxies are shown.  Bins
  in sSFR are shown by the colored points (with small horizontal
  offsets for clarity) and the black points represent the weighted
  mean at each stellar mass (along with the error of the weighted
  mean).  The best-fit line (dashed gray line) is consistent with
  zero, indicating that, in massive galaxies, AGN Eddington ratio is
  independent of stellar mass.
\label{fig:ermass}}
\end{figure}

\begin{figure}[ht]
\epsscale{1.15}
{\plotone{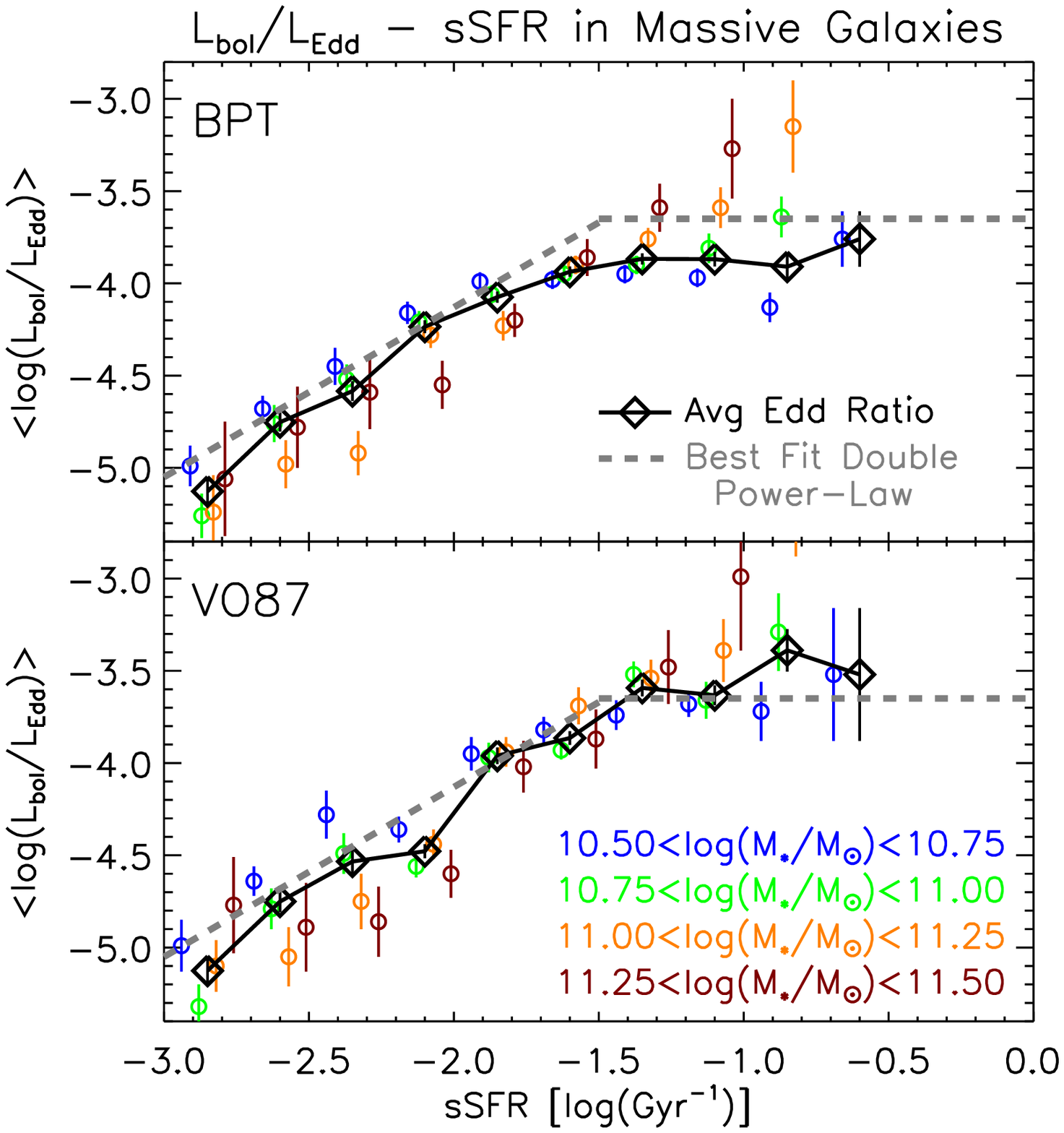}}
\figcaption{The connection between AGN Eddington ratio and host galaxy
  sSFR for high-mass ($\log(M_*/M_\odot) \geq 10.5$) galaxies, as
  implied by the BPT AGN (top) and VO87 AGN (bottom) fractions.
  Individual mass bins are shown by the colored points (offset
  horizontally from one another for clarity) and the black points
  represent the weighted mean at each sSFR (along with the error of
  the weighted mean).  The gray lines are a double power-law fit to
  the combined BPT and VO87 data: $y=0.92x-2.29$ at $x<1.5$ and
  $y=-3.65$ at $x>1.5$, where $x=\log(\rm sSFR/Gyr^{-1})$ and
  $y=\log(\eddratio)$.
\label{fig:ersfr}}
\end{figure}

The relationship between AGN Eddington ratio and galaxy properties is
best understood separately for high-mass ($\log(M_*/M_\odot) \gtrsim
10$) and low-mass ($\log(M_*/M_\odot) \lesssim 10$) galaxies.  We
begin with massive galaxies, examining the average AGN Eddington ratio
as a function of stellar mass and specific star formation rate.

Figure \ref{fig:ermass} presents the average Eddington ratio in
massive ($\log(M_*/M_\odot) \geq 10$) galaxies as a function of
stellar mass, for both individual bins of sSFR (colored points) and
the weighted average of the full massive galaxy population (black
lines).  The best-fit line to the combined BPT and VO87 models has a
slope of $0.065 \pm 0.023$: consistent with zero.  In other words, in
massive ($\log(M_*/M_\odot) \geq 10$) galaxies, we find that AGN
accretion is independent of stellar mass.  This result agrees with the
uniform $\eddratio$ with stellar mass proposed by \citet{aird12} for
X-ray AGN hosts.

However, Figure \ref{fig:ersfr} demonstrates that the average
Eddington ratio in massive galaxies is strongly \textit{non-uniform}
with galaxy sSFR, increasing by $\sim$2~dex with increasing sSFR.
Previous work similarly found that rapidly accreting AGNs are much
more common in massive star-forming hosts than in massive quiescent
hosts \citep[e.g.,][]{kau03, hec04, kau09, tru13a, ros13a, ros13b,
  mat14, aza15}.  The solid line in Figure \ref{fig:ersfr} shows the
weighted average Eddington ratio over all mass bins, at a given sSFR.
We fit a double power-law to the Eddington ratios in each sSFR bin for
both the BPT and VO87 model results.  With $x=\log(\rm sSFR/Gyr^{-1})$
and $y=\log(\eddratio)$, the best-fit (minimum-$\chi^2$) relationship
is given by:
\begin{enumerate}
  \item For $x=\log({\rm sSFR/Gyr^{-1}}) < -1.5$,
      \begin{equation}
        y= (0.92 \pm 0.07) x - (2.29 \pm 0.03).
      \end{equation}
  \item For $x=\log({\rm sSFR/Gyr^{-1}}) > -1.5$,
      \begin{equation}
        y = (-0.10 \pm 0.11) x - (3.79\pm 0.09).
      \end{equation}
\end{enumerate}
Error bars in slope and y-intercept are the 1$\sigma$ errors of the
best-fit values.  In other words, $\eddratio$ increases as a power-law
with sSFR until $\log(\rm sSFR/Gyr^{-1})=-1.5$, at which point
$\log(\eddratio)$ becomes flat with $\log(\rm sSFR)$.  The slope of
the log-log line (exponent of the power-law) is well-constrained to be
close to unity, in striking agreement with the constant ratio of
$\eddratio$ to sSFR in the ensemble of higher-redshift X-ray hosts
presented by \citet{mul12}.  The turnover and lack of dependence of
$\eddratio$ on sSFR at $\log(\rm sSFR/Gyr^{-1})>-1.5$ might be due to
AGN contamination in the broad-band photometry at high Eddington
ratios: contribution from a blue AGN continuum would effectively
overestimate sSFR \citep[e.g., Appendix A of][]{bon12}.  If sSFR is
accurately measured and there is negligible AGN contamination, the
flattening of $\eddratio$ with sSFR may represent a ceiling in AGN
accretion for host galaxies on the ``star formation mass sequence,''
described by $\log(\rm sSFR/Gyr^{-1}) \simeq -1.1 \pm 0.3$ at $z \sim
0$ \citep{whi12}.  Similarly, \citet{ros13b} and \citet{aza15} argue
that AGNs are more common in star-forming hosts, but otherwise there
is no enhancement of AGN activity with additional star formation.



Interpreting the relationship between AGN accretion and galaxy
properties from Figures \ref{fig:fiterbpt} and \ref{fig:fitervo87} is
more difficult in low-mass galaxies.  At the lowest masses
($\log(M_*/M_\odot)<9$) the Eddington ratios are essentially
unconstrained due to zero observed AGNs and extreme star formation
dilution.  Even when the average Eddington ratios are well-constrained
at $9<\log(M_*/M_\odot)<10$, such low-mass galaxies may have black
hole masses that are not well-described by the $\Mbh-\sigma$ relation.
Applying the $\Mbh-\sigma$ relation might overestimate black hole mass
either because $\sigma$ is poorly measured or results from something
other than bulge kinematics, or due to a black hole occupation
function that is less than 100\%.  In the next section we investigate
a non-uniform $\Mbh-\sigma$ relation.

\subsection{Non-Uniform Black Hole Occupation}

\citet{gre10} demonstrated that low-mass ($\sim$10$^9M_\odot$)
galaxies with accurate megamaser-disk $\Mbh$ estimates are
undermassive by $\sim$0.5~dex with respect to the $\Mbh-\sigma$
relation of massive galaxies.  This sort of non-uniform $\Mbh-\sigma$
relation, with undermassive black holes in low-mass galaxies, can be
explained as a consequence of black hole seeding mechanisms.  Direct
collapse of primordial gas is likely to form the most massive (up to
$\sim$10$^6M_\odot$) SMBH seeds \citep{bromm03,beg06,john12}, but this
process becomes inefficient in low-mass halos \citep{van10}.  Instead,
low-mass galaxies likely have black holes initially formed as
lower-mass (up to $\sim$200$M_\odot$) Pop III remnants
\citep{fry01,mad01}.  Thus low-mass galaxies, with low-mass initial
seeds and not many mergers over their lifetime, might be expected to
have a lower black hole occupation function compared to massive
galaxies \citep{vol09}.

In addition to the physics of black hole seed formation, it is also
possible that problematic $\sigma$ measurements might lead to a
non-uniform $\Mbh-\sigma$ relation.  First, the measured velocity
dispersions at $\log(M_*/M_\odot)<10$ are frequently below the SDSS
instrumental resolution ($\sim$70~km~s$^{-1}$).  Even when
well-measured, the velocity dispersion of a low-mass galaxy may
describe disordered kinematics rather than a settled bulge.  If the
velocity dispersion is overestimated for either of these reasons it
will lead to an overestimate of black hole mass.  However, the
velocity dispersions of low-mass galaxies already imply very low-mass
black holes: a velocity dispersion of 35~km~s$^{-1}$, typical of the
survival analysis, results in only $\Mbh \simeq 10^{5}M_\odot$.  This
value already approaches the ``intermediate-mass'' black hole regime
of direct-collapse SMBH seeds.

\begin{figure*}  
\epsscale{1.15}
{\plotone{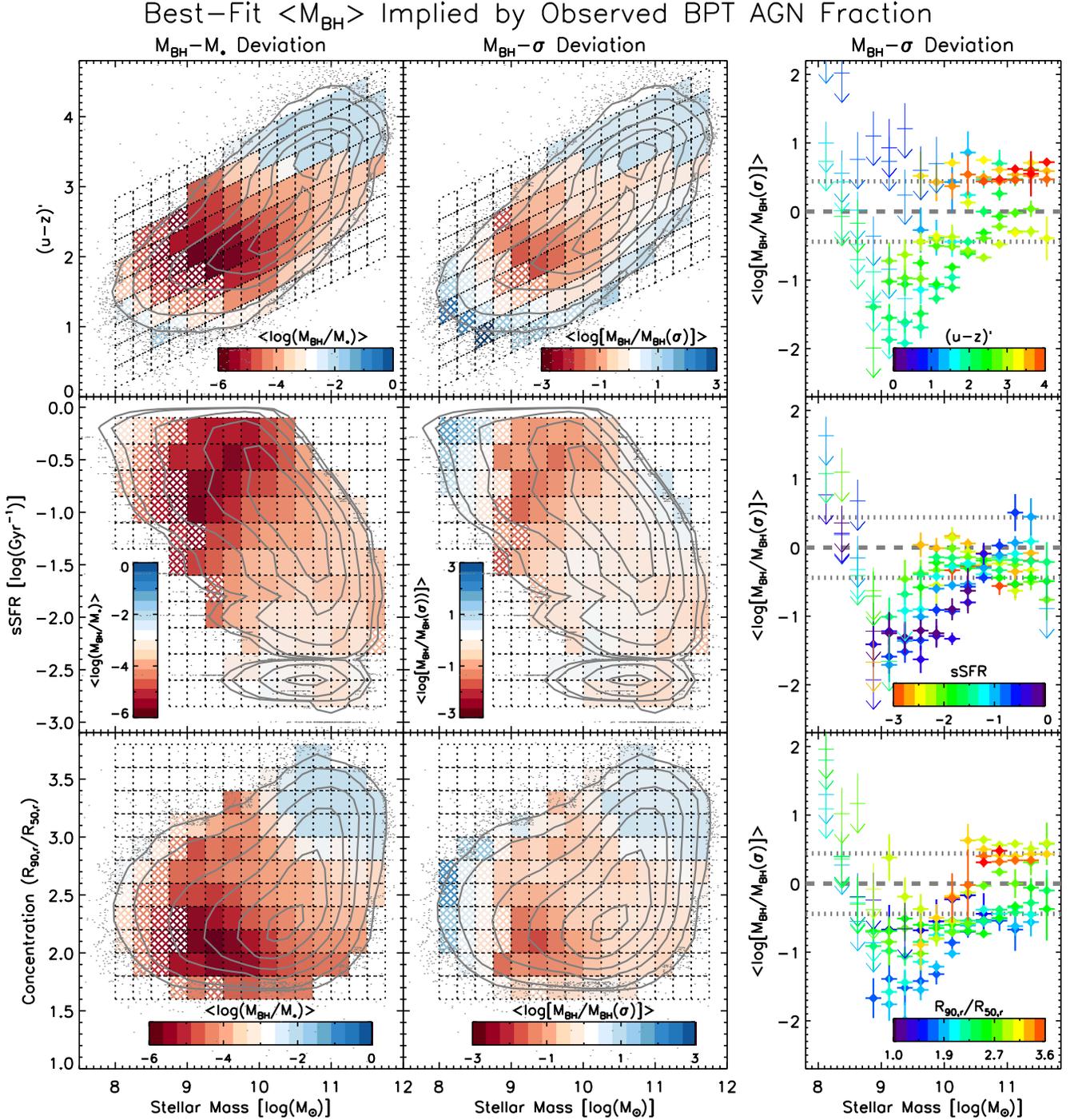}}
\figcaption{The best-fit average $\Mbh$ implied by the observed BPT
  AGN fraction in bins of galaxy color--mass, sSFR--mass, and
  concentration--mass.  \textit{Left:} Average black hole mass
  quantified as $\log(\Mbh/M_*)$, where bins colored white match
  $\Mbh/M_*=0.001$ (the \citet{hr04} relation with $M_*=1.4M_{\rm
    bulge}$).  Gray contours show the distribution of well-measured
  galaxies, and hashed bins denote upper limits where the observed AGN
  fraction is consistent with zero.  \textit{Center:} Average black
  hole mass quantified as the deviation from $\Mbh-\sigma$.  Upper
  limits with no or very few detected AGNs (the hashed bins at left)
  are shown as thinner lines.  \textit{Right:} The same deviation from
  $\Mbh-\sigma$ plotted versus stellar mass in each bin, color-coded
  by $(u-z)'$, sSFR, or $R_{90,r}/R_{50,r}$.  The \citet{gul09}
  $\Mbh-\sigma$ relation and its 1-$\sigma$ scatter are shown by the
  dashed and dotted gray lines.  Low-mass ($10^9--10^{10}M_\odot$)
  galaxies may have black hole masses that are 0.5--1.5~dex lower than
  expected from the $\Mbh-\sigma$ relation and their velocity
  dispersions.
\label{fig:fitmbhbpt}}
\end{figure*}  

\begin{figure*}  
\epsscale{1.15}
{\plotone{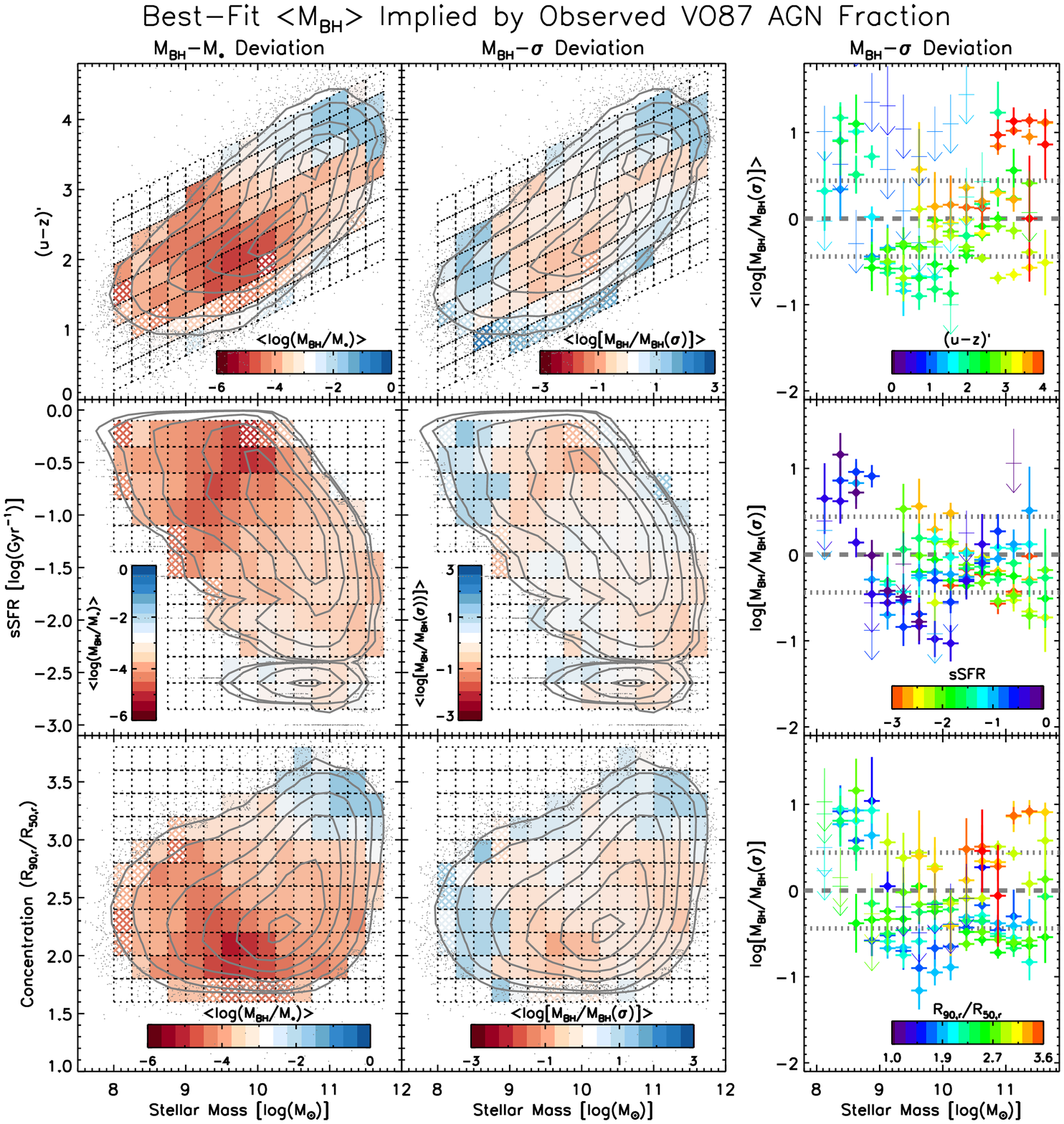}}
\figcaption{The best-fit average $\Mbh$, quantified as
  $\log(\Mbh/M_*)$ (left) or deviation from $\Mbh-\sigma$ (center and
  right), implied by the observed VO87 AGN fraction in bins of galaxy
  color--mass, sSFR--mass, and concentration--mass.  As before, hashed
  bins (left panel) and thinner lines (center and right panels) denote
  upper limits where the observed AGN fraction is consistent with
  zero.  Low-mass ($10^9--10^{10}M_\odot$) galaxies may have slightly
  undermassive black holes that are $\sim$0.5--1~dex lower than
  expected from the $\Mbh-\sigma$ relation and their velocity
  dispersions.
\label{fig:fitmbhvo87}}
\end{figure*}  

Fitting the AGN occupation with $\Mbh$ as a free parameter avoids
problematic $\sigma$ measurements and allows us to probe the black
hole occupation function.  We use the same approach as our previous
simulations, following steps 1--4 outlined in Section 5.  The
normalization of the Eddington ratio distribution is fixed by the
galaxy sSFR according to Equations 14 and 15 above, with $\langle
\log(\eddratio) \rangle \simeq 0.91\log(\eddratio)_{\rm min} + 0.25$.
We assume that low-mass galaxies follow the same accretion--sSFR
relation as massive galaxies, with any deviations caused by
non-uniform black hole mass.

Figure \ref{fig:fitmbhbpt} and \ref{fig:fitmbhvo87} show the best-fit
black hole masses with galaxy color--mass, sSFR--mass, and
concentration--mass, implied by both BPT and VO87 AGNs.  We quantify
this $\Mbh$ in two ways.  The left panels present $\log(\Mbh/M_*)$,
such that bins colored white match the \citet{hr04} relation with
$M_*=1.4M_{\rm bulge}$, $\Mbh/M_*=0.001$.  The center and right panels
display the deviation from $\Mbh-\sigma$, measured as the log-ratio of
the best-fit $\Mbh$ and the $\Mbh(\sigma$ expected from the
$\Mbh-\sigma$ relation.  Negative log-ratios and redder bins denote
undermassive black holes, while higher log-ratios and bluer bins
denote overmassive black holes.  Significantly overmassive black holes
are likely due to errors in the assumptions, for example, where the
$\eddratio$--sSFR relation fails to adequately describe the data.

The ``non-uniform $\Mbh$'' model does not provide any real constraints
on $\Mbh$ for the least-massive ($\log(M_*/M_\odot) \lesssim 9$)
galaxies, since their lack of observed AGNs and extreme star formation
dilution allow almost any $\Mbh-\sigma$ normalization.  The model fits
do, however, provide interesting results for galaxies in the range $9
\lesssim \log(M_*/M_\odot) \lesssim 10$.  The BPT and VO87 AGN
fractions both imply undermassive black holes by 0.5--1.5~dex in these
low-mass hosts.  This result is particularly true for the
least-concentrated and highest-sSFR galaxies, which are also most
likely to have poorly-understood velocity dispersions (due to
disordered kinematics or below the instrumental resolution) and/or
lower black hole occupation.  We further discuss how the $\Mbh-\sigma$
deviations of Figures \ref{fig:fitmbhbpt} and \ref{fig:fitmbhvo87}
relate to black hole occupation and the SMBH seed distribution in
Section 6.5.

\subsection{Comparing the Observed and Simulated Line Ratio
  Distributions}

\begin{figure}[t]
\epsscale{1.15}
{\plotone{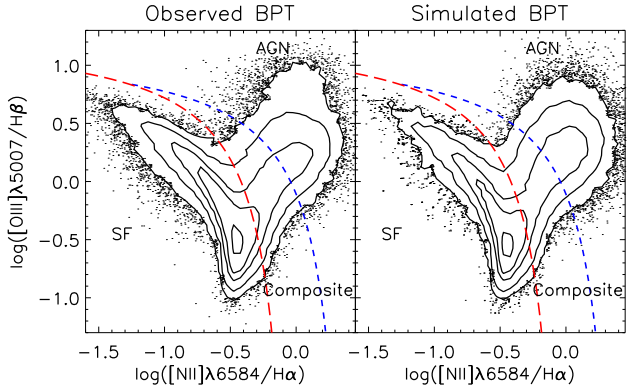}}
\figcaption{The observed and simulated line ratio distributions in the
  BPT diagram.  The dashed red line represents the empirical line of
  \citet{kau03} that we use for AGN/SF classification, and the
  short-dashed blue line also shows the \citet{kew01} maximal
  starburst line.  The simulated AGN line ratios are broadly
  consistent with the observed data, with minor differences likely
  arising from details of NLR physical conditions.
\label{fig:simbpt}}
\end{figure}

\begin{figure}[t]
\epsscale{1.15}
{\plotone{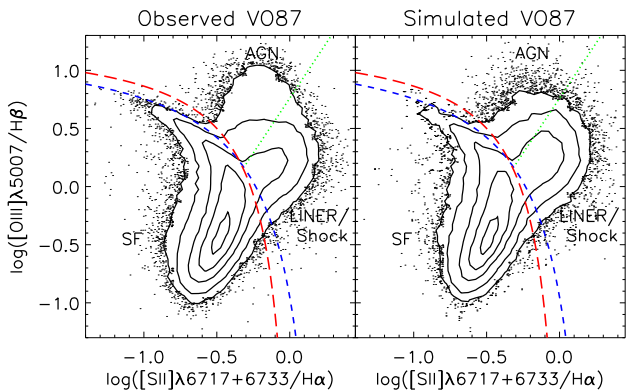}}
\figcaption{The observed and simulated VO87 diagrams.  The dashed red
  and dotted green lines show our adopted AGN/SF and AGN/LINER
  classification, respectively: we empirically create the AGN/SF line
  and the AGN/LINER line is from \citet{kew06}.  The blue short-dashed
  line also shows the \citet{kew01} maximal starburst line.  As in the
  BPT diagram in Figure \ref{fig:simbpt}, there is broad agreement
  between the observed and simulated VO87 AGN line ratio
  distributions.
\label{fig:simvo87}}
\end{figure}

The previous two simulations are fit to match the observed AGN
fraction as a function of galaxy properties.  It is also useful to see
if they reproduce the observed line ratio distribution: this functions
as a consistency check on our assumed ``pure AGN NLR'' line ratios.

Figure \ref{fig:simbpt} compares the observed and simulated BPT
diagrams, and Figure \ref{fig:simvo87} compares the observed and
simulated VO87 diagrams.  In both cases the simulated line ratios are
drawn from a single realization of the Monte Carlo simulations in the
non-uniform $\eddratio$ model fit to the galaxy color--mass
distribution.  The line ratio distributions are similar for fits to
bins in galaxy sSFR--mass and concentration-mass, and for the
non-uniform $\Mbh$ model, differing only due to the randomly drawn
nature of each realization.  Because the simulations were fit to match
the observed AGN fractions, the number of galaxies classified as AGNs
or SF-dominated is identical for both observed and simulated line
ratios.  The simulated SF sequence is also nearly identical to the
observed ratios, since the simulations begin from a randomly-drawn
subset of observed non-AGN line ratios.

Meanwhile the simulated AGN line ratios are broadly similar to the
observations.  The agreement is especially good in the $\NII/\Ha$ and
$\SII/\Ha$ ratios, suggesting that our assumed metallicity-dependent
NLR model (Equations 10--13) is reasonably accurate.  There are some
differences in the shapes of the observed and simulated $\OIII/\Hb$
distributions.  In particular, the observed line ratios probably do
not follow a log-normal distribution, but instead depend on physical
NLR conditions such as ionization or cloud structure
\citep[e.g.,][]{gro04,ric14}.  The NLR physical conditions may change
with AGN Eddington ratio, suggesting that it is inaccurate to assume
the same normally-distributed pure line ratios for every AGN.  Still,
our simple ``evolving NLR'' description broadly captures the locus and
range of observed line ratios, and our assumed log-normal
distributions are reasonable in the absence of a better understanding
of AGN NLR physical conditions.

\section{Implications for AGN Host Galaxy Properties}

The previous sections showed that both the BPT and VO87 AGN selection
methods suffer from significant bias in low-mass, blue, high-sSFR,
low-concentration galaxies due to star formation dilution.  After
carefully modeling this bias, we found that the intrinsic AGN
population is not well-described by a uniform Eddington ratio
distribution over all galaxy properties.  Instead there is likely to
be a connection between AGN accretion and host galaxy sSFR in massive
galaxies.  In low-mass galaxies there is also some potential for
slightly undermassive black holes compared to the $\Mbh-\sigma$
relation.

Below we discuss the ramifications of the line-ration selection bias
and the implied AGN occupation function with galaxy properties.

\subsection{Limitations of Line-Ratio AGN Selection}

As noted in Section 4.1, the biases of line-ratio AGN selection depend
on how AGN power is quantified.  The total mass accretion of black
holes (integral of $\dot{M}$ over time) is best understand by AGN
luminosity, since $L_{\rm bol}=\eta\dot{M}c^2$.  Meanwhile the
Eddington ratio $\eddratio$ governs the structure of the accretion
flow and feedback mode \citep{nar08,ho08,tru11a,hec14}.

\begin{figure}[t]
\epsscale{1.15}
{\plotone{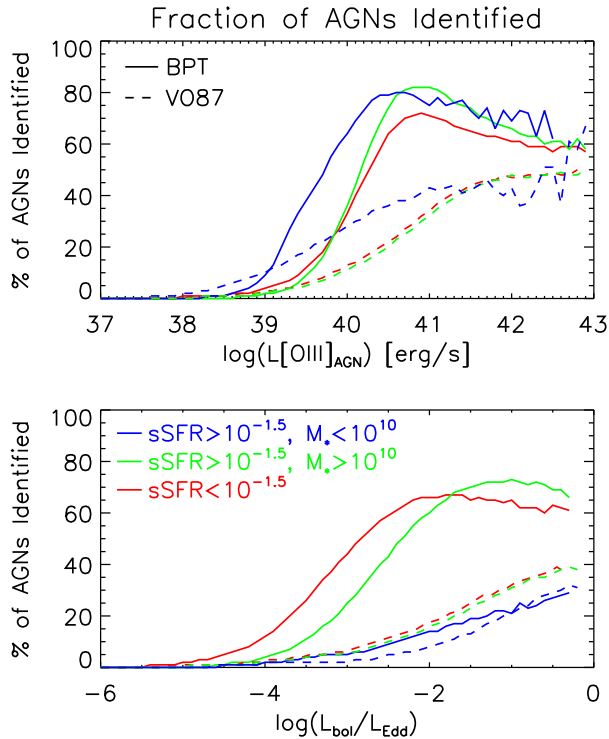}}
\figcaption{The fraction of simulated AGNs (from the ``non-uniform
  $\eddratio$'' simulation) identified by the BPT and VO87 AGN
  classification methods.  Detection fractions are presented for three
  different bins of galaxy properties: low-mass and high-sSFR (blue
  lines), high-mass and high-sSFR (green lines), and low-sSFR (which
  also tend to be high-mass, red lines).  In the legend, sSFR is in
  units of Gyr$^{-1}$ and $M_*$ is in $M_\odot$.  In terms of AGN
  luminosity, AGNs are reliably identified at $L\OIII \gtrsim
  10^{40.5}$~erg~s$^{-1}$ regardless of host galaxy properties.  The
  selection function in $\eddratio$, on the other hand, is a strong
  function of galaxy properties, with fewer weakly-accreting AGNs
  identified in low-mass star-forming galaxies.  Similar trends are
  seen for both the BPT and VO87 classification methods, although VO87
  AGN selection has an overall lower efficiency.
\label{fig:detfrac}}
\end{figure}

Figure \ref{fig:detfrac} presents the fraction of AGNs identified by
the BPT and VO87 methods as a function of both AGN luminosity and
Eddington ratio.  The detection fraction is determined from the
``non-uniform $\eddratio$'' simulation presented in Section 5.3, and
is the number of simulated AGNs identified by the BPT or VO87 AGN
selection.  As found by \citet{kau03}, the AGN detection fraction is
high at $L\OIII \gtrsim 10^{40.5}$~erg~s$^{-1}$ ($L\OIII \gtrsim
10^{7}L_\odot$) regardless of galaxy properties.  In other words, the
use of line-ratio AGN selection is not significantly biased for
estimates of \textit{total black hole accretion} with galaxy
properties.  However, galaxy properties cause a strong bias for
line-ratio AGN identification at a given accretion rate, similar to
that discussed in Section 4.1 and Figure \ref{fig:eddlimits}.  The
strong ``star formation dilution'' in low-mass star-forming hosts
means that, at fixed Eddington ratio, many fewer AGNs are identified
than in massive and low-sSFR host galaxies.  Estimates of AGN
\textit{feedback and fueling efficiency} with galaxy properties are
thus strongly biased when using line-ratio AGN selection.  Figure
\ref{fig:detfrac} also demonstrates that the VO87 AGN selection method
has an overall lower efficiency than BPT AGN selection.


The bias against line-ratio AGN selection is worst in galaxies with
high star formation rates and low metallicity: i.e., the conditions
typical of $z \gtrsim 1$ galaxies \citep[e.g.,][]{mad14}.  $\HII$
regions associated with rapid star formation may also have harder
ionization conditions \citep{liu08, bri08, stei14, shap15}, further
decreasing the contrast between AGN and star formation.  Making
matters even worse, the BPT locus at $z \gtrsim 1$ is suggestive of
lower-metallicity AGN NLRs \citep{kew13, jun14, coil15}.  Thus it is
likely that $\HII$ regions dominate the emission lines of $z \gtrsim
1$ galaxies, biasing against AGN line-ratio detection to a greater
degree than in our low-redshift SDSS sample \citep[see
also][]{coil15}.

Other AGN selection methods that use emission-line ratios will be
biased in a qualitatively similar fashion to the star formation
dilution affecting BPT and VO87 selection.  The widely-used
``mass-excitation'' \citep[MEx,][]{jun11} and ``color-excitation''
\citep[CEx,][]{yan11} methods, which retain $\OIII/\Hb$ but replace
the $\NII/\Ha$ or $\SII/\Ha$ line ratios with stellar mass or color,
will have similar (though more explicit) biases against identifying
AGNs in blue and low-mass galaxies.  The CEx method is likely to be
more affected by lower-metallicity AGN NLRs than MEx AGN selection,
since lower metallicity makes a galaxy bluer without changing its
mass.

Line ratios with higher contrast between AGNs and $\HII$ regions are
likely to be less affected by star formation dilution.
High-ionization lines such as $\NeV$ or $\HeII$ and (semi-)permitted
lines such as $\CIII$ or $\CIV$ are likely to result in less biased
AGN selection since the AGN NLR typically has harder ionization and
higher density than $\HII$ regions.
Among ratios of strong lines, it is possible that the $\NeIII/\OII$
ratio used in the ``TBT'' method \citep{tro11} offers better
diagnostic power for AGN selection since $\NeIII$ has a critical
density $\sim$10 times higher than $\OIII$.  However there are many
modes of $\NeIII$ emission in galaxies, most of which are not well
understood \citep{zei15}.  Spatially-resolved line ratios also result
in higher contrast between extended star formation and a nuclear AGN,
potentially revealing SMBH accretion even when $\HII$ regions dominate
the integrated emission lines \citep[e.g.,][]{wri10,tru11b}.

\subsection{AGN Contribution to Emission Lines}

\begin{figure*}  
\epsscale{1.15}
{\plotone{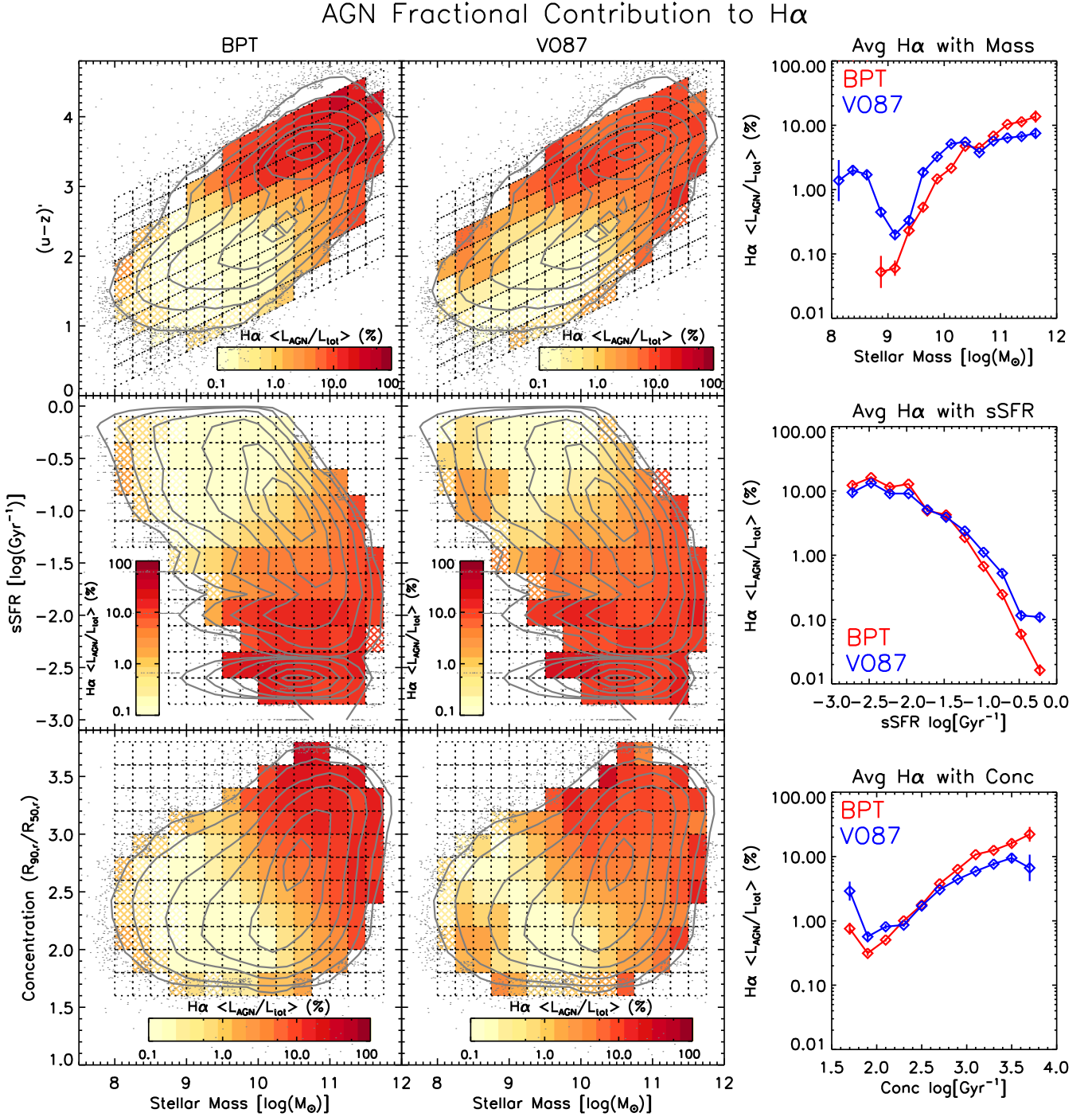}}
\figcaption{The average fraction of $\Ha$ line emission coming from
  AGNs in bins of galaxy color--mass, sSFR--mass, and
  concentration--mass.  In the left and center panels, gray contours
  show the full parent sample of galaxies, and bin shading represents
  the average AGN contribution.  The right panels show the
  weighted-average AGN contribution with stellar mass, sSFR, and
  concentration, weighting by the Monte Carlo error in each bin and
  excluding upper limits.  The AGN fraction of $\Ha$ emission is
  highest (up to 15\%) in massive red/quiescent galaxies, and is
  lowest ($<$1\%) in low-mass blue/star-forming galaxies.  In most
  galaxies, AGNs contribute $\lesssim$10\% to the total $\Ha$
  emission.
\label{fig:hafrac}}
\end{figure*}  

\begin{figure*}  
\epsscale{1.15}
{\plotone{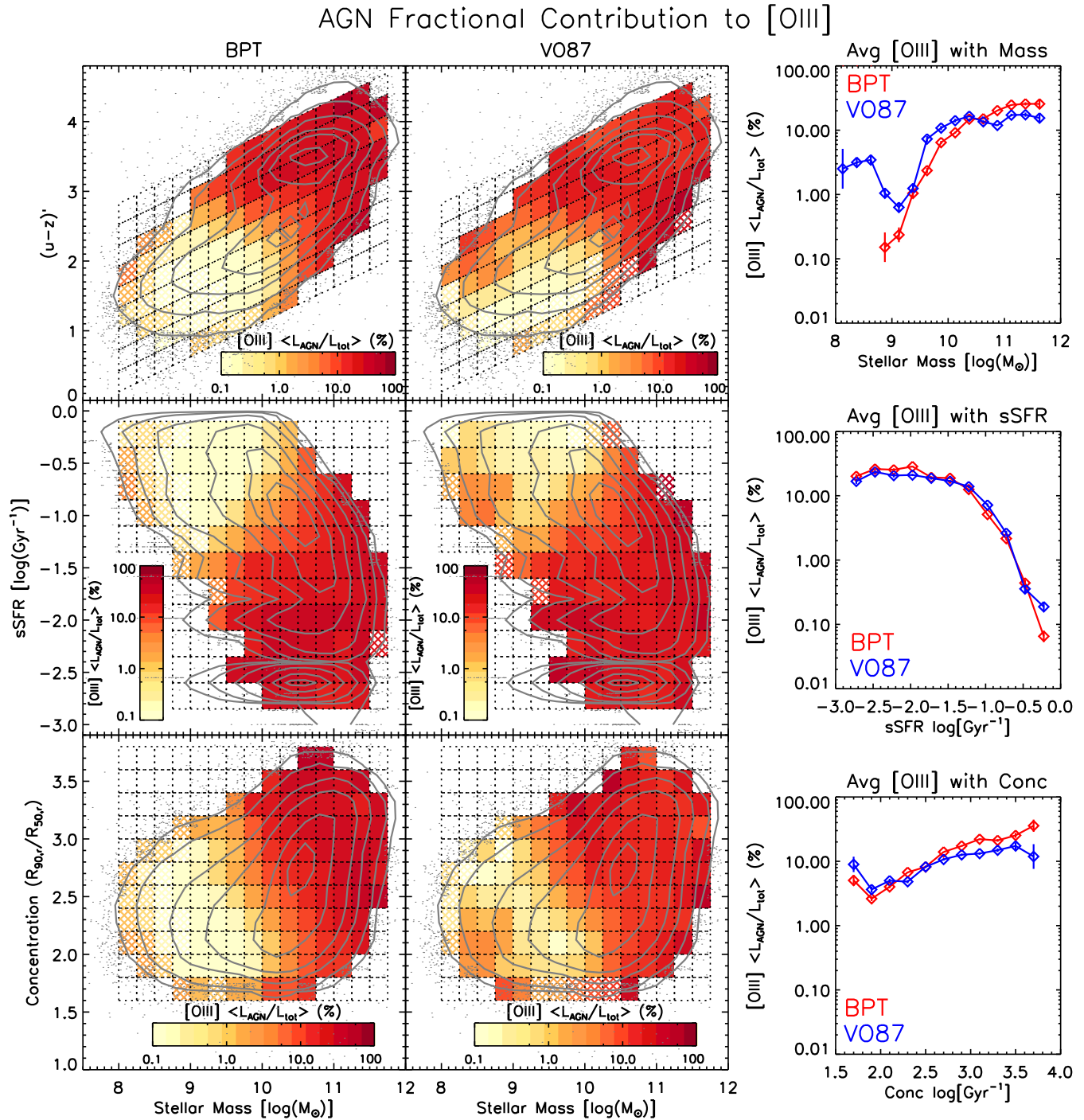}}
\figcaption{The weighted-average fractional AGN contribution to the
  total $\OIII$ emission in galaxies binned by color--mass,
  sSFR--mass, and concentration--mass.  As in Figure \ref{fig:hafrac},
  the averages in the right panels are weighted by the Monte Carlo
  errors in each bin excluding upper limits.  AGNs are a significant
  portion ($\sim$20--30\%) of the average $\OIII$ emission in massive
  red galaxies with low sSFR and high concentration, suggesting that
  the $\OIII$ line is not a good indicator of galaxy properties in
  such systems.
\label{fig:o3frac}}
\end{figure*}  

\begin{figure}[h]
\epsscale{1.15}
{\plotone{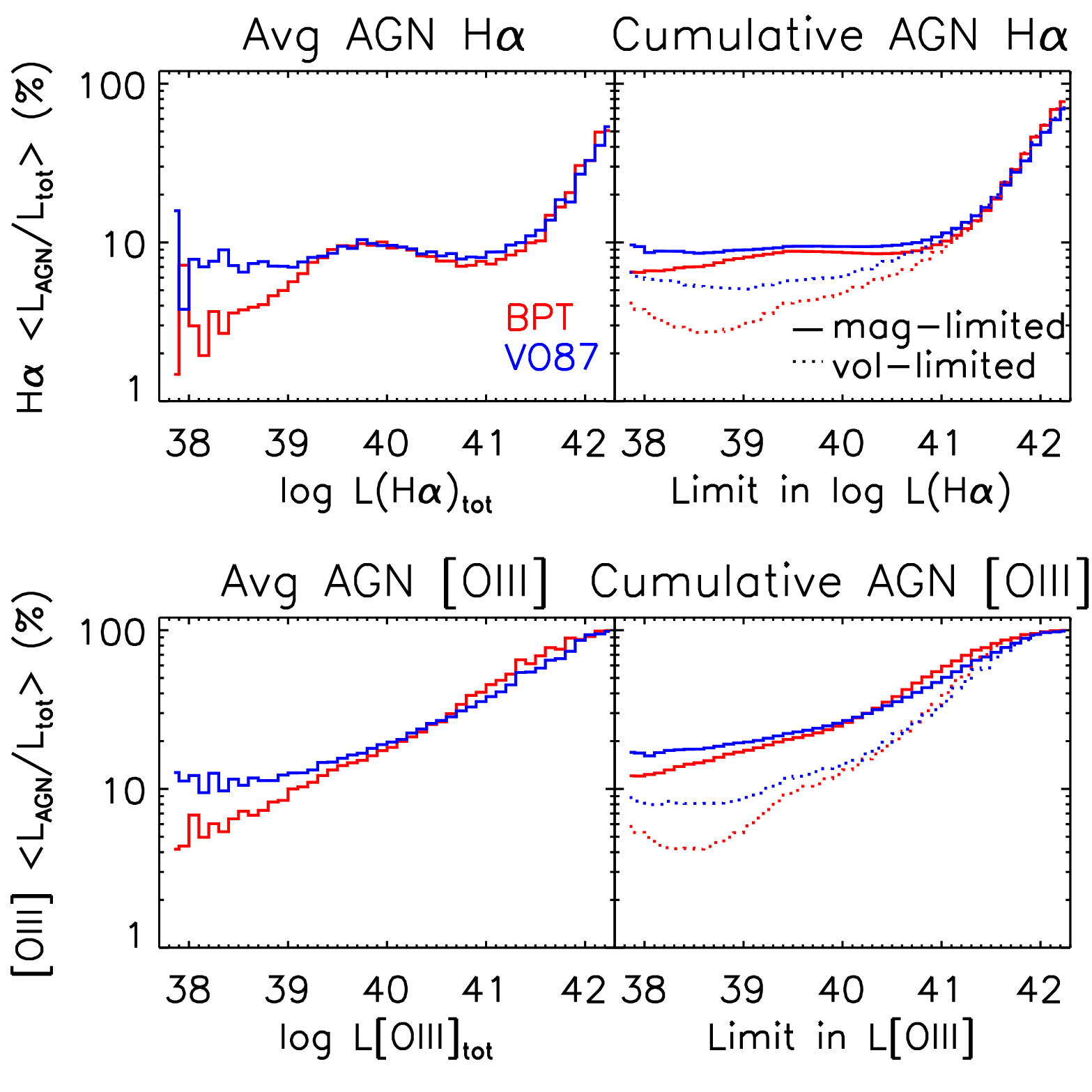}}
\figcaption{The average fractional AGN contribution to $\Ha$ (top
  panels) and $\OIII$ (bottom panels), as a function of total line
  luminosity (left panels) and limiting line luminosity (right
  panels).  We account for the line-flux limit of the SDSS in the
  right panels, with the solid line describing the AGN fraction for
  the magnitude-limited ($r<17.77$) SDSS and the dotted line using
  $1/V_{\rm max}$ weighting to estimate the AGN fraction for a
  volume-limited survey.  As in Figure \ref{fig:hafrac}, the AGN
  contribution to $\Ha$ is small ($\lesssim$10\%) in most galaxies,
  and only luminous ($L(\Ha)>10^{42}$~erg~s$^{-1}$) galaxies are
  AGN-dominated.  In contrast, the AGN contribution to $\OIII$ is
  significant (10--20\%) even in weak-lined galaxies and
  magnitude-limited surveys, and galaxies with
  $L\OIII>10^{41}$~erg~s$^{-1}$ are dominated by AGNs in $\OIII$
  emission.
\label{fig:linefracs}}
\end{figure}

The main goal of this work has been to characterize the population of
emission-line AGN, including those hidden by star formation dilution.
But by definition, the presence of significant dilution means that
$\HII$ regions dominate the observed emission lines over AGN NLR
emission.  Many of the AGNs identified in this work have little effect
on the emission lines from the point of view of characterizing galaxy
properties (such as star formation rate or gas-phase metallicity).

We use the ``non-uniform $\eddratio$'' model of Section 5.3 to measure
the typical AGN contribution to the observed $\Ha$ and $\OIII$
emission lines.  Quantifying AGN luminosity is simpler than inferring
the Eddington ratio or black hole occupation function, since AGN
luminosity is directly related to AGN detection above the $\HII$
region emission.  The assumed $\Mbh$ distribution does not actually
matter for determining the typical AGN luminosity: for example, a
lower black hole occupation would simply require an accordingly higher
Eddington ratio normalization to maintain the same number of detected
AGNs.

The fractional AGN contributions to the $\Ha$ and $\OIII$ emission
lines are shown in Figures \ref{fig:hafrac} and \ref{fig:o3frac},
respectively.  Both the BPT and VO87 AGN selection methods result in
similar estimates of AGN emission-line fractions.  The average AGN
contribution is highest in massive, low-sSFR, concentrated galaxies,
but is $\lesssim$15\% in all galaxies.  The low average AGN
contribution to most galaxies' $\Ha$ emission suggests that, for
example, $\Ha$-derived star formation rates are generally unbiased by
AGN emission in similar low-redshift galaxy samples.  Meanwhile AGNs
contribute a significant fraction ($\sim$10--30\%) of the overall
$\OIII$ emission in massive, low-sSFR, concentrated galaxies.  In
low-mass, high-sSFR, and low-concentration galaxies the AGN fraction
of both $\Ha$ and $\OIII$ emission is negligible ($\lesssim$1\%),
matching our conclusion that such galaxies have the largest amount of
star formation dilution.

We also measure the fractional AGN contribution as a function of line
luminosity, shown in the left panels of Figure \ref{fig:linefracs}.
The right panels of Figure \ref{fig:linefracs} present the average
emission-line fraction of AGNs for galaxies above a given line
luminosity limit, useful for estimating the typical AGN contribution
in emission-line surveys.  Each cumulative AGN fraction is limited to
the redshift range at which the the SDSS flux limit ($1 \times
10^{-16}$~erg~s$^{-1}$~cm$^{-2}$) is above the luminosity limit.  The
solid line is appropriate for the magnitude-limited ($r<17.77$) SDSS,
while the dotted line shows AGN fractions weighted by the inverse
maximum detection volume, $1/V_{\rm max}$ (calculated following
\citealp{hogg99}), giving the AGN fraction for a volume-limited
survey.  The larger number of low-mass galaxies (which tend to be
star-forming and metal-poor) in volume-limited surveys leads to
typically lower AGN fractions compared to magnitude-limited surveys.
In general AGNs are an insignificant contribution to $\Ha$ in all but
the most luminous ($L(\Ha) \gtrsim 10^{41}$~erg~s$^{-1}$) galaxies and
shallowest surveys.  However, AGNs are a significant fraction of a
galaxy's $\OIII$ emission even in low-luminosity galaxies and deep
magnitude-limited surveys, and AGNs dominate galaxy emission at line
luminosities of $L\OIII \gtrsim 10^{41}$~erg~s$^{-1}$.

\subsection{AGN Fueling}

Section 5 demonstrated that the Eddington ratio distributions implied
by both BPT and VO87 AGNs is not uniform across all galaxy properties.
At high mass ($\log(M_*/M_\odot) \gtrsim 10$), Section 5.3 and Figure
\ref{fig:ersfr} indicated that AGN accretion is connected to specific
star formation rate, and otherwise has no dependence on stellar mass.
Here we agree with \citet{aird12}, who found a uniform Eddington ratio
distribution with stellar mass for X-ray AGN hosts, along with a
moderate enhancement of AGN activity with blue galaxy color.  The
slope of unity we derive for the $\eddratio$ -- sSFR correlation at
$\log({\rm sSFR}) < -1.5$~Gyr$^{-1}$ is very similar to that of
\citet{mul12} for X-ray AGN hosts.  We also find that the correlation
flattens at $\log({\rm sSFR}) > -1.5$~Gyr$^{-1}$.  This might be
caused by AGN contamination affecting the estimated sSFR at the
highest AGN accretion rates.  It is also possible that, as suggested
by \citet{ros13b} and \citet{aza15}, AGNs prefer hosts on the
``star-forming mass sequence'' but are otherwise uncorrelated with
star-forming excess.

Our estimates of the Eddington ratio distribution are less robust at
$\log(M_*/M_\odot) \lesssim 10$, since these galaxies have less
reliable velocity dispersion measurements used to estimate $\Mbh$.
Thus we cannot confidently claim any non-uniformity in AGN accretion
at low stellar masses.  In the Appendix we do, however, find that the
observed BPT and VO87 AGN fractions are much better described by the
\citet{gul09} $\Mbh-\sigma$ relation than by a constant $\Mbh/M_*$
ratio.  In a galaxy of given sSFR, Eddington ratio seems to be uniform
no matter the bulge fraction, but galaxies with smaller bulges tend to
have smaller black holes.  This contradicts the constant $\Mbh/M_*$
ratio used by \citet{aird12}, although some high-redshift observations
suggest an evolving $\Mbh/M_{\rm bulge}$ but a constant $\Mbh/M_*$
fraction \citep[e.g.,][]{jah09, schramm13, sun15}.


\subsection{AGN Feedback}

The high energy output of AGNs suggests the potential for powerful
feedback to quench host galaxy star formation.  Such feedback could
occur either through blowout of the star-forming gas by
radiatively-driven winds \citep{sil98,fab02,dim05}, or by radio jets
mechanically heating the gas \citep{cro06}.  The feedback mode likely
depends on Eddington ratio, with powerful radiative winds likely only
at high accretion rates and jets dominating the outflows of weakly
accreting AGNs \citep{nar08, ho08, tru11a, hec14}.  But in general
both feedback modes are likely to increase with AGN luminosity.

We define ``feedback timescale'' as the ratio of a galaxy's
gravitational binding energy to the AGN bolometric luminosity,
$\tau_{\rm fb} = U_{\rm gal}/L_{\rm AGN}$.  The inverse of this
timescale $\tau_{\rm fb}^{-1}$ quantifies the efficiency of AGN
feedback:
\begin{equation}
  \tau_{\rm fb}^{-1} \simeq 40 \frac{L_{42}R_{10}}{M_{10}^2}~{\rm Gyr}^{-1}.
\end{equation}
Here $L_{42}=L_{\rm AGN}/(10^{42}~{\rm erg~s^{-1}})$,
$R_{10}=R_{50}/(10~{\rm kpc})$, and $M_{10}=M_*/(10^{10}M_\odot)$.
Very roughly, at a fixed Eddington ratio and fixed black hole
occupation defined by $\Mbh \sim \sigma^4$, $L_{\rm AGN} \sim
(\eddratio)\Mbh \sigma^4 \sim (f_bM_*)^2$ (where $f_b$ is the
bulge-to-total ratio).  Meanwhile, galaxy mass and size are correlated
as $M \sim R^\alpha$, with $0.15<\alpha<0.55$ increasing from low-mass
disks to high-mass ellipticals \citep{shen03}.  For a uniform
Eddington ratio distribution this means that AGN feedback efficiency
increases with bulge fraction and stellar mass, $\tau_{\rm fb} \sim
f_b^2M_*^\alpha$.  Of course, the actual situation is somewhat more
complicated, as Sections 5 and 6 showed that the Eddington ratio
distribution is not uniform.

We use the ``non-uniform $\eddratio$'' simulation of Section 5.3 to
estimate the AGN feedback efficiency $\tau_{\rm fb}^{-1}$ across
galaxy properties.  Just as in Section 6.2, the details of the $\Mbh$
distribution do not matter, since AGN luminosity (with respect to the
empirically measured star formation dilution) is directly constrained
by the observed AGN fraction.  Our feedback estimates depend only on
the assumed $\Lbol/L\OIII$ bolometric correction and ``pure AGN'' line
ratios.

\begin{figure*}  
\epsscale{1.15}
{\plotone{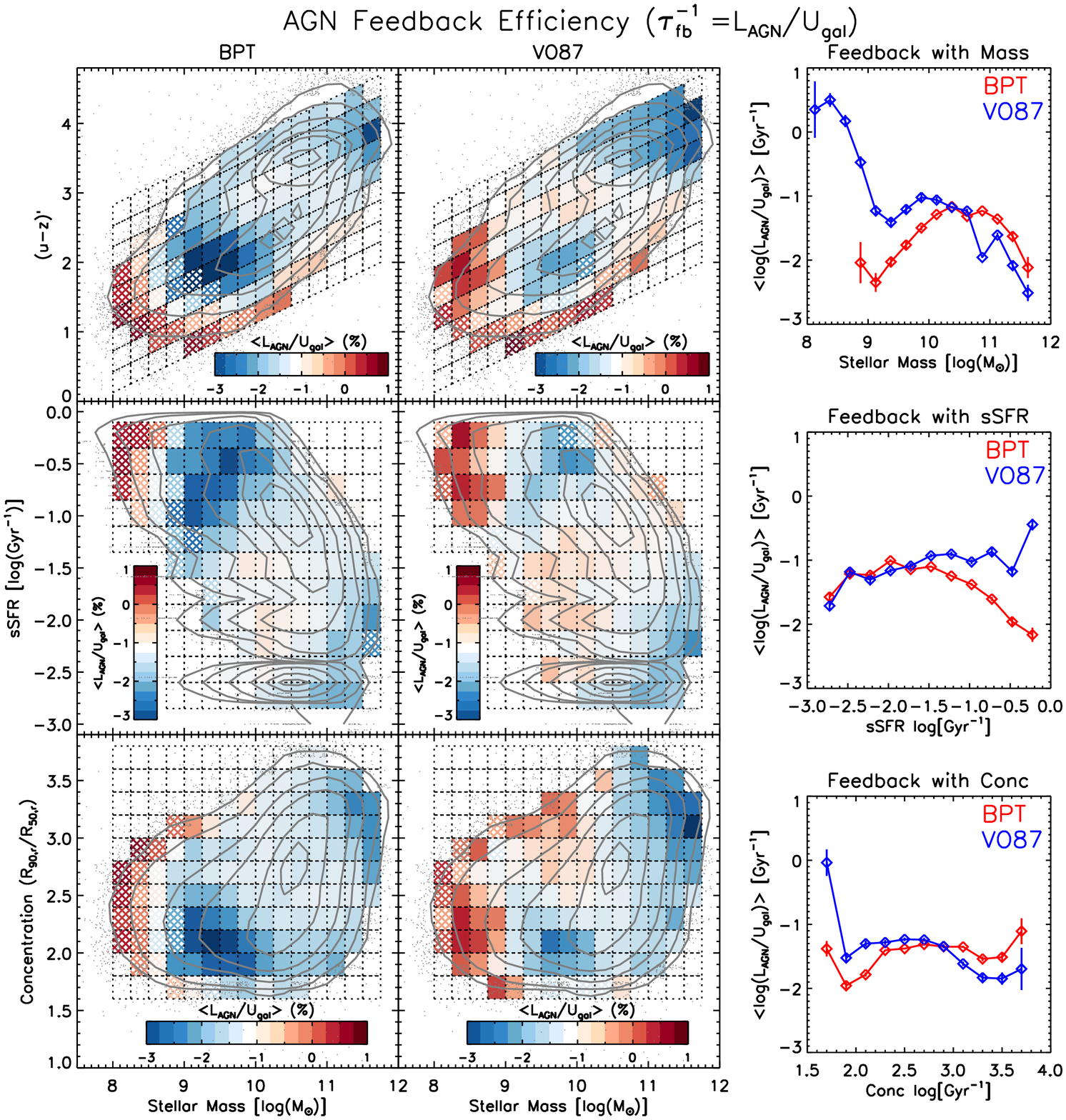}}
\figcaption{The average AGN feedback efficiency $\tau_{\rm fb}^{-1} =
  L_{\rm AGN}/U_{\rm gal}$ with galaxy color--mass, sSFR--mass, and
  concentration--mass, inferred by both the BPT and VO87 AGN
  selection.  The average feedback efficiency is given by the bin
  shading in the left and center panels, with the full parent sample
  of galaxies shown by gray contours.  The feedback efficiency is
  shown with stellar mass, sSFR, and concentration in the right
  panels, computed as weighted averages using the Monte Carlo error in
  each bin and excluding upper limits.  AGN feedback efficiency is
  fairly uniform with galaxy properties, with no evidence that AGNs
  dominate the quenching of star formation for most galaxies of any
  type.
\label{fig:feedback}}
\end{figure*}  

Figure \ref{fig:feedback} shows the average AGN feedback efficiency
$\tau_{\rm fb}^{-1} = L_{\rm AGN}/U_{\rm gal}$ in bins of galaxy
color--mass, sSFR--mass, and concentration--mass.  In each bin the
feedback efficiency is averaged over the Eddington ratio distribution
(which has a declining power-law slope $\alpha=0.6$), with a small
number of high-$\eddratio$ AGNs and a large population of
weakly-accreting AGNs.  There are slight differences in the feedback
inferred from the observed BPT and VO87 AGN populations, largely at
low masses and high sSFR where the BPT selection results in upper
limits due to the heavy bias.  By either method there is little
connection between the average AGN feedback efficiency and galaxy
properties at $z<0.1$: feedback by typical AGNs is likely to operate
just as efficiently in low-mass and high-mass galaxies.  The
uniformity of $\tau_{\rm fb}^{-1}$ at high and low sSFR also implies
that feedback from line-ratio AGNs is not the dominant mode of star
formation quenching in most galaxies of any mass or morphology.  This
is in agreement with recent simulations showing that AGN winds, even
when present at high velocity, have little effect on the larger gas
reservoirs within their host galaxies \citep{gab14, roos15}.

\subsection{Black Hole Seeds}

\begin{figure*}  
  \epsscale{1.15}
  {\plotone{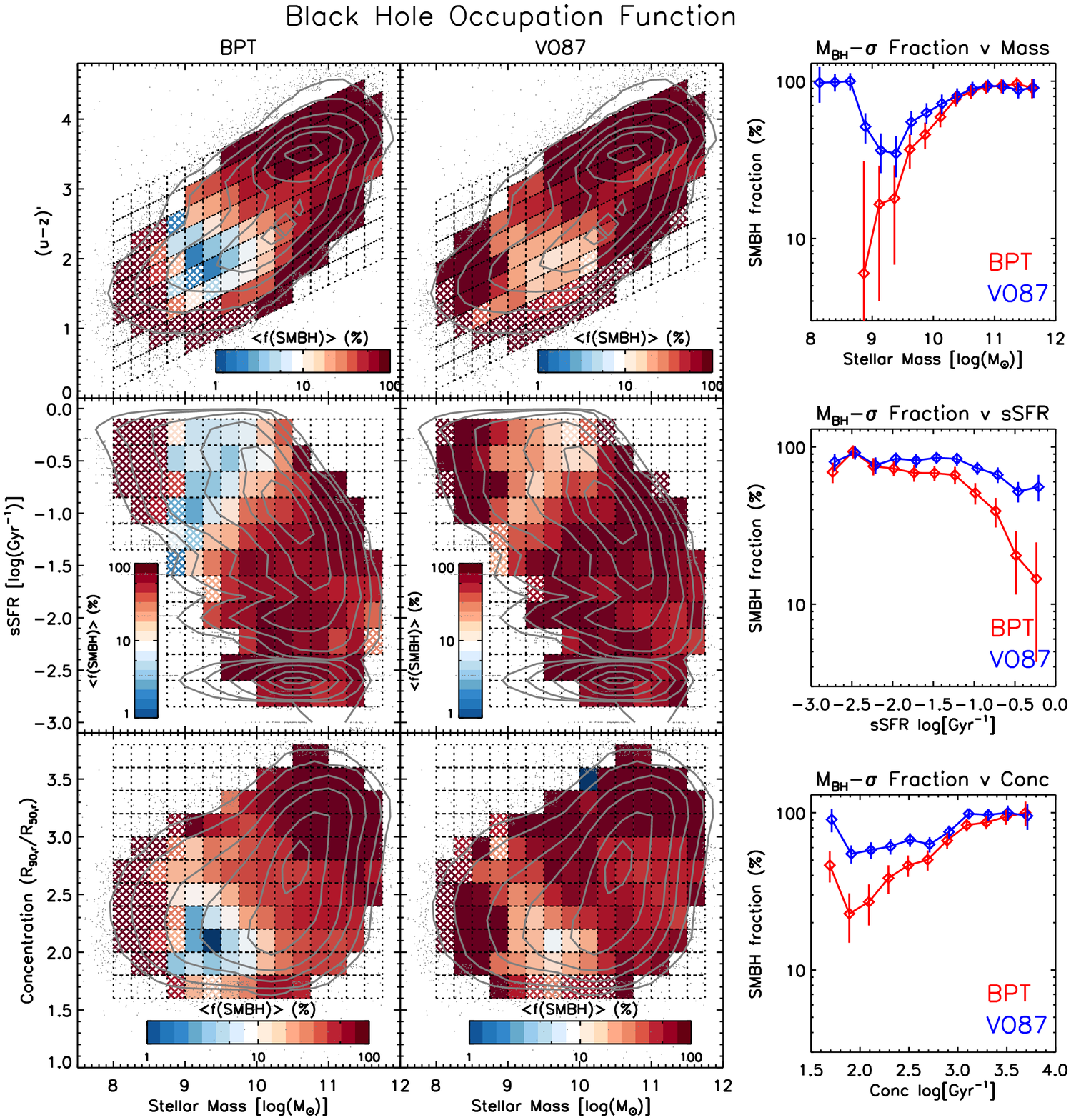}}
  \figcaption{The SMBH occupation function derived from our model fit
    to the observed BPT AGN and VO87 AGN fractions.  Here we assume
    that the black hole population is bimodal with two log-normal
    peaks, one from Pop III star remnants at 150$M_\odot$ (with
    $\pm$0.5~dex scatter) and one from massive seed ``fossils''
    described by the \citet{gul09} $\Mbh-\sigma$ relation (Equation
    14).  With the caveat that $\sigma$ may be inaccurate (in both
    measurement and interpretation as bulge kinematics) in our
    low-mass galaxies, there is some hint of a lower SMBH occupation
    function in $\log(M_*/M_\odot) \lesssim 10$ hosts.
\label{fig:seedfrac}}
\end{figure*}  

The SMBH occupation function in low-redshift galaxies carries an
imprint of the initial seeding mechanism in the early universe.  In
particular, low-mass ($\lesssim$150~$M_\odot$) SMBH seeds from Pop III
remnants are likely to be common in most early halos, while massive
($10^4$--$10^6$~$M_\odot$) seeds from direct collapse are rare in
low-mass halos due to their higher spin parameter \citep{beg06, vol09,
  van10}.  Subsequent accretion and mergers of these seeds is thought
to eventually form the local $\Mbh-\sigma$ correlation
\citep[e.g.,][]{vol11, vol12}.  Due to the high direct-collapse
seeding rate and high merger rate, essentially every massive galaxy is
likely to end up on the $\Mbh-\sigma$ relation.  Both the merger rate
and the fraction of massive seeds are thought to be lower for low-mass
($\log(M_*/M_\odot) \lesssim 10$) hosts, and so many of these
galaxies may have undermassive black holes compared to the
$\Mbh-\sigma$ relation.

We determine the ``SMBH occupation function'' as the fraction of
galaxies with black hole masses well-described by the $\Mbh-\sigma$
relation.  Our galaxies are assumed to contain a bimodal population of
black holes, with one peak of massive seeds described by the
\citet{gul09} $\Mbh-\sigma$ relation (Equation 8), and a second
log-normal distribution of lower-mass Pop III seeds centered at
$150M_\odot$~$\pm0.5$~dex.  From this distribution the SMBH occupation
function is fit using a simulation similar to that in Section 5.4,
with the same bolometric correction (Equation 9),
metallicity-dependent NLR (Equations 10--13), and sSFR-dependent
Eddington ratio distribution (Equations 14 and 15), while using the
bimodal distribution for $\Mbh$.  From this distribution, the SMBH
occupation function is described by the fraction of galaxies hosting
black holes drawn from the higher-mass $\Mbh-\sigma$ peak.

Figure \ref{fig:seedfrac} shows the SMBH occupation function from our
model fits to the observed BPT and VO87 AGN fractions.  There are some
differences between the two selection methods, likely due to the
uncertainties associated with NLR metallicity in low-mass galaxies.
In addition, the velocity dispersions in $\log(M_*/M_\odot) \lesssim
10$ galaxies are poorly understood, due to both the low SDSS
resolution ($\sim$70~km/s) and the difficulty in interpretation
(bulges versus disordered disks).  Nonetheless both the BPT and VO87
AGN fractions imply a slightly lower black hole occupation of
$\sim$30--50\% in low-mass and disk-dominated hosts.  This result is
roughly consistent with the independent study of \citet{mil15}, who
similarly found a marginally lower SMBH occupation function of a few
tens of percent for $\log(M_*/M_\odot) \lesssim 10$ X-ray AGN hosts.
Confirming this marginal evidence and robustly determining the SMBH
black hole occupation function at low stellar mass requires higher
resolution spectroscopy than available from our SDSS sample.

\section{Summary}

We constructed a sample of over 300,000 galaxies from the Sloan
Digital Sky Survey to investigate the biases of line-ratio AGN
selection and recover the intrinsic AGN population across a range of
galaxy properties.  While line-ratio AGNs are observed to be most
common in massive green-valley hosts, we demonstrated that this result
is a selection effect caused by the bias from ``star formation
dilution'' in low-mass and star-forming galaxies.  After accounting
for this bias, we find that AGNs are most common in massive galaxies
with high specific star formation rates, implying that SMBH accretion
and star formation are fueled by the same gas reservoir.  AGNs
contribute little to the overall $\Ha$ and $\OIII$ emission lines in
low-mass galaxies, but their feedback effects are likely to be just as
efficient at all stellar masses and host morphologies.  There is
marginal evidence that the black hole occupation function may be a
factor of a few lower in $\log(M_*/M_\odot) \lesssim 10$ hosts,
although this result is not robust due to poorly-understood velocity
dispersions in these galaxies.  Higher resolution spectroscopy is
needed to better constrain SMBH occupation function, while selection
methods via spatially resolved line ratios or high-ionization lines
suffer less bias and can better reveal the connection between AGN
accretion and star formation.

\acknowledgements

We thank Jenny Greene and Renbin Yan for valuable discussions which
contributed to this work.  JRT acknowledges support from NASA through
Hubble Fellowship grant \#51330 awarded by the Space Telescope Science
Institute, which is operated by the Association of Universities for
Research in Astronomy, Inc., for NASA under contract NAS 5-26555.  MS
acknowledges support from the China Scholarship Council
(No. [2013]3009) and SJ acknowledges support from the European
Research Council through grant ERC-StG- 257720.

Funding for the SDSS and SDSS-II has been provided by the Alfred
P. Sloan Foundation, the Participating Institutions, the National
Science Foundation, the U.S. Department of Energy, the National
Aeronautics and Space Administration, the Japanese Monbukagakusho, the
Max Planck Society, and the Higher Education Funding Council for
England. The SDSS Web Site is {\tt http://www.sdss.org/}.

\appendix

\section{Testing Other Modeling Assumptions}

In this Appendix we justify the modeling assumptions used in the
simulations of Section 5 (and outlined in Equations 7--13) by testing
other plausible assumptions.  The models used the same dust extinction
for both $\HII$ and AGN NLR emission lines, and we test that
assumption here.  The original ``uniform-$\eddratio$'' simulation is
also compared with additional simulations that replace the
metallicity-dependent NLR ratios with metal-rich (constant) NLR
ratios, or replace the $\Mbh-\sigma$ relation with a constant
$\Mbh/M_*$ ratio.  We also test different power-law slopes for the
Schechter function describing the Eddington ratio distribution.

\begin{figure*}  
\epsscale{1.15}
{\plotone{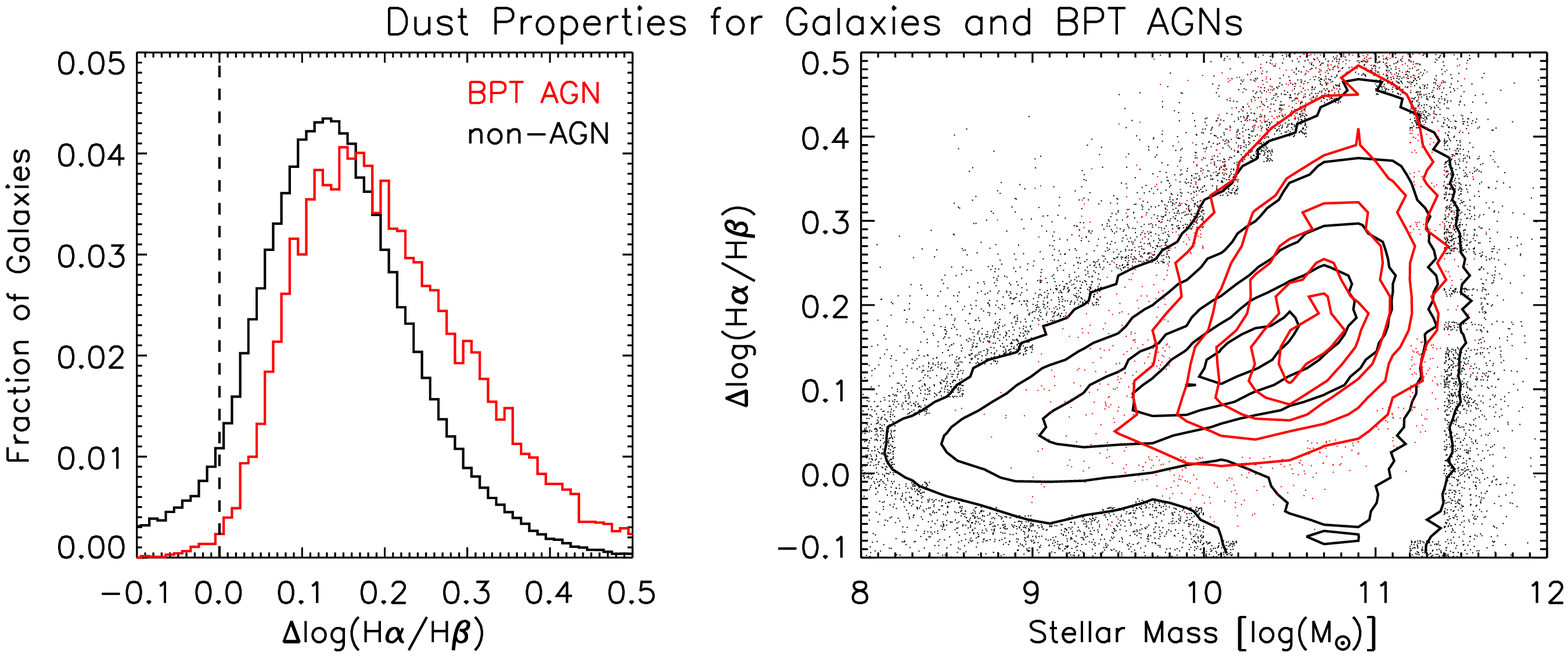}}
\figcaption{The distribution of dust extinction for galaxies
  classified as BPT AGNs and non-AGNs (SF galaxies and LINERs).  We
  quantify the dust extinction as $\Delta\log(\Ha/\Hb) \equiv
  \log(\Ha/\Hb)_{\rm obs}-\log(\Ha/\Hb)_{\rm i}$, where the intrinsic
  Balmer decrement $(\Ha/\Hb)_{\rm i}$ is 3.1 for AGNs and 2.86 for
  non-AGNs \citep{ost06}.  AGNs have marginally more dust extinction
  than the full set of non-AGNs (left panel), but this is because BPT
  AGNs are identified only in massive galaxies.  AGNs have nearly
  identical dust extinction compared to non-AGNs of the same stellar
  mass (right panel).
\label{fig:comparedust}}
\end{figure*}  

\begin{figure*}  
\epsscale{1.15}
{\plotone{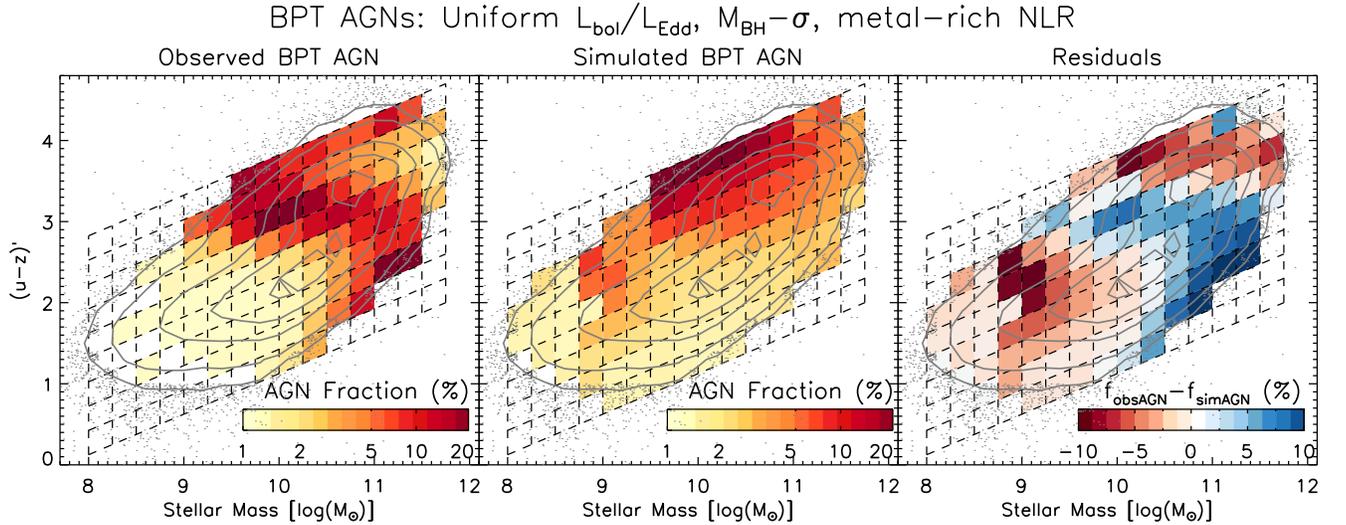}}
\figcaption{The observed and simulated fractions of BPT AGNs with host
  galaxy color and stellar mass for the ``uniform $\eddratio$''
  simulation with metal-rich NLR line ratios.  Bin shading indicates
  the AGN fraction and the gray contours represent the distribution of
  well-measured galaxies.  Using a metal-rich NLR results in a larger
  overprediction of BPT AGNs in low-mass galaxies compared to the
  metallicity-dependent NLR used in Section 5.2.
\label{fig:simbpt_colormsig}}
\end{figure*}  

\begin{figure*}  
\epsscale{1.15}
{\plotone{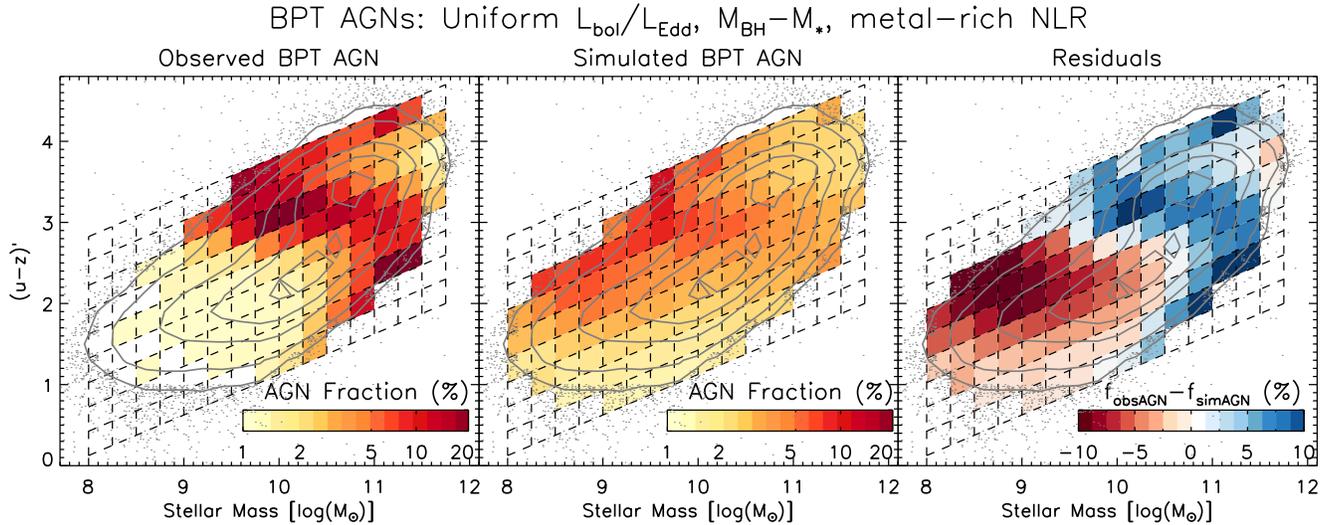}}
\figcaption{The observed and simulated fractions of BPT AGNs with host
  galaxy color and stellar mass for the ``uniform $\eddratio$''
  simulation with metal-rich NLR line ratios and a constant $\Mbh/M_*$
  ratio.  As before, the AGN fraction is indicated by the bin shading
  and the well-measured galaxy distribution is given by the gray
  contours.  Assuming a constant $\Mbh/M_*$ ratio results in a
  significantly worse match to the observations than the simulation
  using the $\Mbh-\sigma$ relation in Figure
  \ref{fig:simbpt_colormsig} (and in Section 5.2).
\label{fig:simbpt_colorbpt}}
\end{figure*}  

\begin{figure*}  
\epsscale{1.15}
{\plotone{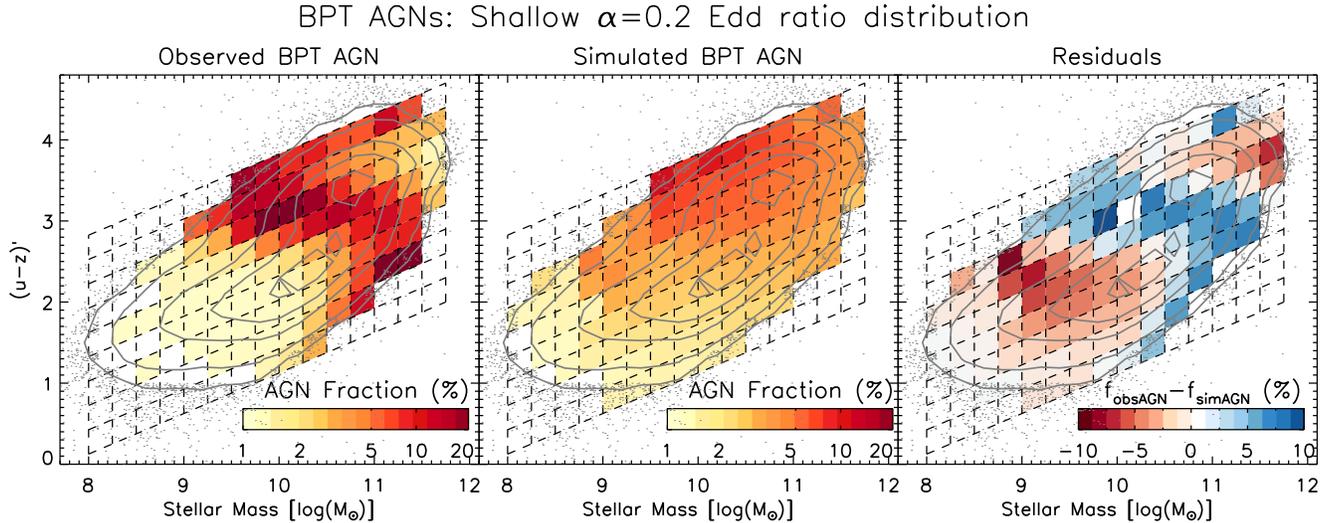}}
\figcaption{A comparison of the observed and simulated BPT AGN
  fractions in the galaxy color--mass diagram, using a simulation with
  a ``shallow'' Eddington ratio distribution of power-law slope
  $\alpha=0.2$.  The shallow $\eddratio$ distribution results in
  larger residuals than the distribution with $\alpha=0.6$ adopted in
  the remainder of this work (e.g., Figure
  \ref{fig:simbpt_colorzmsig}).
\label{fig:simbpt_shallower}}
\end{figure*}  

\begin{figure*}  
\epsscale{1.15}
{\plotone{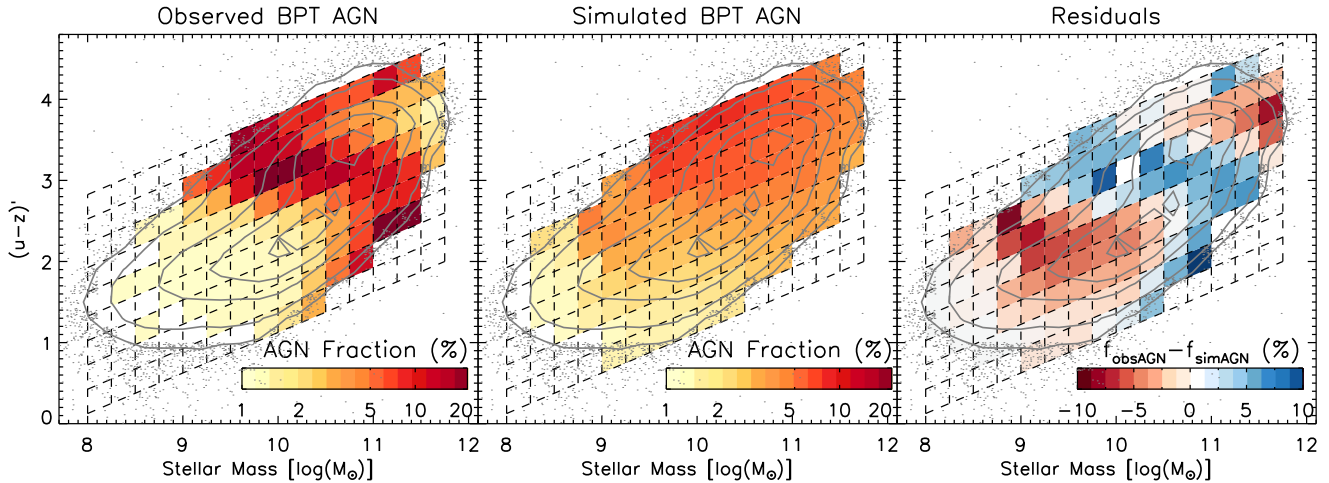}}
\figcaption{A comparison of the observed and simulated BPT AGN
  fractions in the galaxy color--mass diagram for a simulation with a
  ``flat'' Eddington ratio distribution of power-law slope
  $\alpha=0.05$.  The simulation with a flat $\eddratio$ distribution
  produces a significantly poorer fit to the data than the steeper
  $\alpha=0.6$ $\eddratio$ distribution.
\label{fig:simbpt_flater}}
\end{figure*}  

\begin{deluxetable*}{l|ccc|ccc}
  \tablecolumns{7}
  \tablecaption{$\chi^2/1000$ Values for Simulated AGN Fractions
    \label{tbl:simchi2}}
  \tablehead{
    \colhead{Model} & 
    \colhead{} & 
    \colhead{BPT} & 
    \colhead{} & 
    \colhead{} & 
    \colhead{VO87} & 
    \colhead{} \\
    \colhead{} & 
    \colhead{$(u-z)'-M_*$} & 
    \colhead{sSFR $-M_*$} & 
    \colhead{$C_r-M_*$} & 
    \colhead{$(u-z)'-M_*$} & 
    \colhead{sSFR $-M_*$} & 
    \colhead{$C_r-M_*$} }
  \startdata
    Uniform $\eddratio$, $\Mbh-\sigma$, metal-dependent NLR & 12.8 & 10.7 & 9.83 & 2.62 & 2.32 & 2.37 \\
    Uniform $\eddratio$, $\Mbh-\sigma$, metal-rich NLR & 23.1 & 19.2 & 15.7 & 2.62 & 2.32 & 2.37 \\
    Uniform $\eddratio$, $\Mbh-M_*$, metal-rich NLR & 46.0 & 33.2 & 34.0 & 4.59 & 4.34 & 4.97 \\
    \hline
    Shallow ($\alpha=0.2$) $\eddratio$ distribution & 24.5 & 18.6 & 16.7 & 4.59 & 4.34 & 4.97 \\
    Flat ($\alpha=0.05$) $\eddratio$ distribution & 28.2 & 21.4 & 18.9 & 5.48 & 4.09 & 4.27
  \enddata
\end{deluxetable*}

\subsection{Dust Corrections}

Our estimates of the intrinsic AGN occupation fraction used dust-free
emission lines, beginning from dust-corrected emission lines observed
for star-forming galaxies, and then adding dust-free model AGN NLR
emission lines.  Thus the models implicitly assume the same dust
extinction for both star-forming galaxies and AGNs: a plausible
assumption, given that the AGN NLR gas is at $>$kpc scales from the
galaxy center \citep{ben02}, similar to locations of $\HII$ regions.
We directly compare the dust extinction of BPT-classified AGNs and
non-AGNs in Figure \ref{fig:comparedust}, quantified as the excess
Balmer decrement $\Delta\log(\Ha/\Hb) \equiv \log(\Ha/\Hb)_{\rm
  obs}-\log(\Ha/\Hb)_{\rm i}$.  Here the intrinsic Balmer decrement
$(\Ha/\Hb)_{\rm i}=2.86$ for $\HII$ regions and $(\Ha/\Hb)_{\rm
  i}=3.1$ for AGNs \citep{ost06}.  Compared to galaxies of the same
(high) stellar mass, AGNs have nearly identical Balmer decrements,
justifying our assumption that both $\HII$ region and AGN NLR gas are
affected by the same dust extinction.

\subsection{Constant-Metallicity NLR and Constant $\mathbf{\Mbh/M_*}$
  Ratio}

To test a constant, metal-rich NLR we replace Equation 13 with:
\begin{equation}
  \log(\NII/\Ha)=0.0 \pm 0.2,
\end{equation}
In other words, we fix the AGN $\NII/\Ha$ ratio to a normal
distribution rather than having it depend on the $\NII/\Ha$ ratio of
the host galaxy.  The metal-rich NLR assumption uses the same
equations as the Section 5 simulations (Equations 10--12) for the
other AGN NLR line ratios: this means there is no difference between
the metallicity-dependent and metal-rich VO87 AGN simulations.  We
follow the same steps outlined in Section 5.2 to create a
``uniform-$\eddratio$'' simulation with a metal-rich AGN NLR, using
the same $\log(\lambda_{\rm min})=-5.0$ for BPT AGNs and
$\log(\lambda_{\rm min})=-5.5$ for VO87 AGNs to normalize the
Eddington ratio distribution and minimize the total $\chi^2$ in
comparing to observations.  This uniform-$\eddratio$, metal-rich NLR
simulation for BPT AGNs is compared to observations in the color--mass
diagram in Figure \ref{fig:simbpt_colormsig}.

We additionally test a different relationship between black hole mass
and galaxy properties by replacing Equation 8 with:
\begin{equation}
  \Mbh/M_*=0.001,
\end{equation}
with an intrinsic scatter of 0.5~dex.  This is consistent with the
$\Mbh/M_{\rm bulge}$ relation of \citet{hr04}, assuming $M_{\rm
  bulge}=M_*/1.4$.  \citet{aird12} used a similar constant $\Mbh/M_*$
ratio in their study showing that X-ray AGNs have a uniform Eddington
ratio distribution over a wide range of host galaxy stellar mass.  In
the local universe velocity dispersion correlates with black hole mass
much better than does total stellar mass \citep[e.g.,][]{kor13}, but
several $z \gtrsim 1$ studies suggest that $\Mbh/M_{\rm bulge}$
evolves with redshift while $\Mbh/M_*$ is constant
\citep[e.g.,][]{jah09, schramm13, sun15}.  Thus there is some
motivation for comparing a constant $\Mbh/M_*$ ratio with the
$\Mbh-\sigma$ assumption used in Section 5.  We create a
``uniform-$\eddratio$'' simulation with both a metal-rich NLR and a
constant $\Mbh/M_*$ ratio following the same steps as in Section 5.2,
minimizing the total $\chi^2$ by setting $\log(\lambda_{\rm
  min})=-5.75$ in both the BPT and VO87 simulations.  Figure
\ref{fig:simbpt_colorbpt} displays this uniform-$\eddratio$,
metal-rich NLR, constant-$\Mbh/M_*$ simulation for BPT AGNs in the
color--mass diagram.

Table \ref{tbl:simchi2} compares the $\chi^2/1000$ values from each
simulation for BPT and VO87 AGN fractions in the color--mass,
sSFR--mass, and concentration--mass host galaxy diagrams.  As seen in
Figures \ref{fig:simbpt_colormsig} and \ref{fig:simbpt_colorbpt},
using a metal-rich AGN NLR or a constant $\Mbh/M_*$ relation results
in significantly worse matches between the simulated and observed AGN
fractions.  Thus we are justified in using the metallicity-dependent
NLR and $\Mbh-\sigma$ relation for our simulations in Section 5.

\subsection{Different Power-Law Slopes for the Eddington Ratio
  Distribution}

The simulations in Sections 5 and 6 used a Schechter form of the
Eddington ratio distribution (Equation 7) with a power-law slope
$\alpha=0.6$.  Observations of X-ray and narrow-line AGNs, as well as
models for AGN fueling, are consistent with this parameterization
\citep{hh09, kau09, aird12}, but neither are strongly constraining due
to their considerable uncertainties.  Other observations suggest a
somewhat shallower power-law slope: \citet{hic14} argue that
$\alpha=0.2$ better describes the observed $L_{AGN}$--SFR correlation,
and \citet{schulze10} measure $\alpha=0.05$ for low-redshift
broad-line AGNs.  Here we test the effects of adopting shallower
Eddington ratio distributions on the fit between the simulated and
observed AGN fractions for the uniform-$\eddratio$ model.

Figure \ref{fig:simbpt_shallower} compares the uniform-$\eddratio$
simulation with an Eddington ratio distribution of slope $\alpha=0.2$
with the observed AGN fraction.  This ``shallow'' $\eddratio$
distribution was normalized (minimizing the $\chi^2$) using a lower
$\log(\lambda_{\rm min})=-7$ for BPT AGNs and $\log(\lambda_{\rm
  min})=-6.5$ for VO87 AGNs.  The fit to the observed AGN fractions is
significantly worse than the fiducial $\alpha=0.6$ used in Sections 5
and 6, with $\chi^2$ values $\sim$1.5$\times$ higher, as shown in
Table \ref{tbl:simchi2}.

The uniform-$\eddratio$ simulation with a ``flat'' $\alpha=0.05$
Eddington ratio distribution is compared to the observations in Figure
\ref{fig:simbpt_flater}.  Normalizing the flat $\eddratio$
distribution requires a very low minimum Eddington ratio:
$\log(\lambda_{\rm min})=-10$ for BPT AGNs and $\log(\lambda_{\rm
  min})=-9$ for VO87 AGNs.  The $\chi^2$ values for the $\alpha=0.05$
distribution are roughly double those of the fiducial $\alpha=0.6$
Eddington ratio distribution (see Table \ref{tbl:simchi2}), with a
higher-amplitude ``striping'' pattern of residuals in the color-mass
diagram.  For a uniform Eddington ratio distribution in all host
galaxy types, the steeper $\alpha=0.6$ Eddington ratio distribution
provides the best (lowest-$\chi^2$) match to the observed AGN
fractions.

\end{document}